\documentclass[onecolumn,english,smallextended]{svjour3}
\usepackage[T1]{fontenc}
\usepackage[latin9]{inputenc}
\usepackage{array}
\usepackage{booktabs}
\usepackage{amsmath}
\usepackage{graphicx}
\usepackage{amssymb}
\usepackage[authoryear]{natbib}

\newcommand{\noun}[1]{\textsc{#1}}
\providecommand{\tabularnewline}{\\}

\usepackage{url}
\smartqed

\begin{document}

\title{Exact Indexing for Massive Time Series Databases under Time Warping
Distance}

\author{Vit Niennattrakul \and Pongsakorn Ruengronghirunya \and Chotirat
Ann Ratanamahatana}

\institute{V. Niennattrakul \at Department of Computer Engineering, Chulalongkorn
University \\ \email{g49vnn@cp.eng.chula.ac.th} \and P. Ruengronghirunya
\at Department of Computer Engineering, Chulalongkorn University
\\ \email{g51prn@cp.eng.chula.ac.th} \and C.A. Ratanamahatana \at
Department of Computer Engineering, Chulalongkorn University \\ \email{ann@cp.eng.chula.ac.th}}

\maketitle

\abstract{Among many existing distance measures for time series data, Dynamic
Time Warping (DTW) distance has been recognized as one of the most
accurate and suitable distance measures due to its flexibility in
sequence alignment. However, DTW distance calculation is computationally
intensive. Especially in very large time series databases, sequential
scan through the entire database is definitely impractical, even with
random access that exploits some index structures since high dimensionality
of time series data incurs extremely high I/O cost. More specifically,
a sequential structure consumes high CPU but low I/O costs, while
an index structure requires low CPU but high I/O costs. In this work,
we therefore propose a novel indexed sequential structure called TWIST
(Time Warping in Indexed Sequential sTructure) which benefits from
both sequential access and index structure. When a query sequence
is issued, TWIST calculates lower bounding distances between a group
of candidate sequences and the query sequence, and then identifies
the data access order in advance, hence reducing a great number of
both sequential and random accesses. Impressively, our indexed sequential
structure achieves significant speedup in a querying process by a
few orders of magnitude. In addition, our method shows superiority
over existing rival methods in terms of query processing time, number
of page accesses, and storage requirement with no false dismissal
guaranteed.\keywords{Time Series, Indexing, Dynamic Time Warping}}

\section{Introduction}

\begin{figure}[h]
\noindent \begin{centering}
\begin{tabular}{c}
\includegraphics[width=6cm]{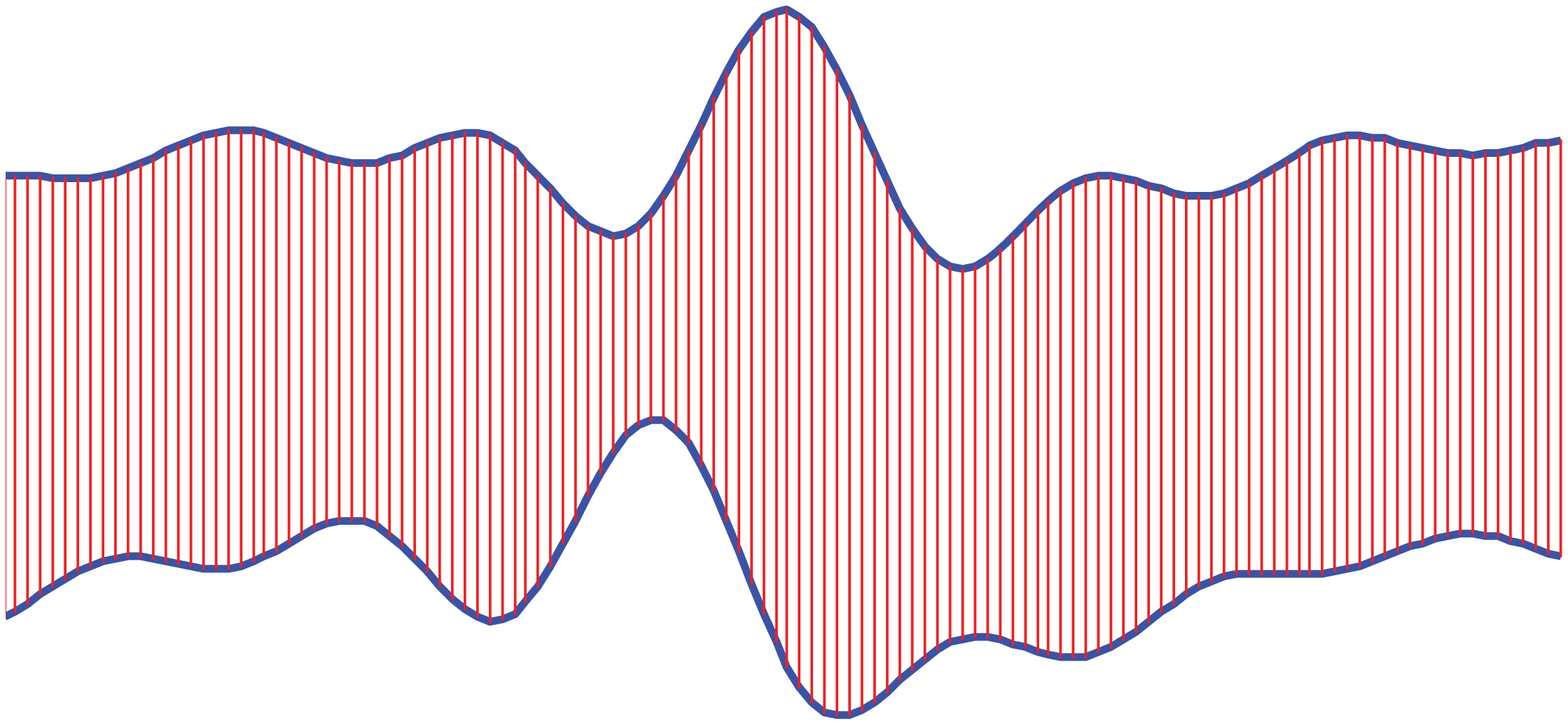}\tabularnewline
(a)\tabularnewline
\tabularnewline
\includegraphics[width=6cm]{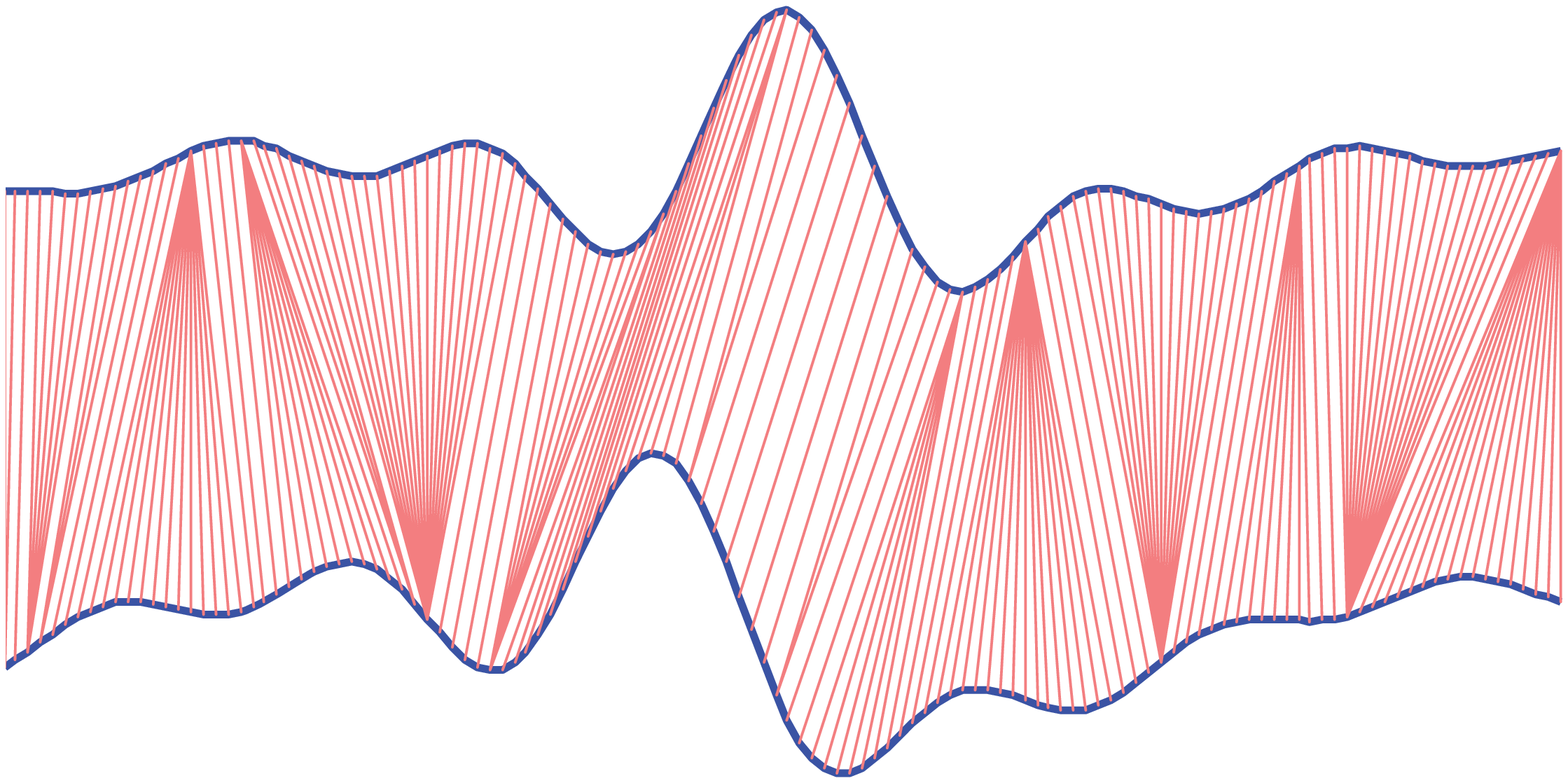}\tabularnewline
(b)\tabularnewline
\end{tabular}
\par\end{centering}

\caption{The comparison of sequence alignments between a) Euclidean distance
and b) DTW distance}
\label{Flo:introduction1}
\end{figure}

Dynamic Time Warping (DTW) distance \citep{BerndtC94,RatanamahatanaK04,RatanamahatanaK05,Sakurai2007}
has been known as one of the best distance measures \citep{ding08,KeoghK03}
suited for time series domain over the traditional Euclidean distance
because DTW distance has much more flexibility in sequence alignment.
In addition, DTW distance tries to find the best warping, while Euclidean
distance is calculated in one-to-one manner, as shown in Figure \ref{Flo:introduction1}.
However, DTW distance has a major drawback, i.e., it requires extremely
high computational cost, especially when DTW distance is used in similarity
search problems, including top-$k$ query. More specifically, in top-$k$
querying problem, after a query sequence has been issued, a set of
$k$ candidate sequences most similar to the query sequence ranked
by DTW distance is returned. Traditionally, the naïve approach needs
to calculate DTW distances for all candidate sequences. As a result,
its query processing time mainly depends on distance calculation and
the number of data accesses. 

So far, many speedup techniques have been proposed including lower
bounding functions and index structures. Lower bounding functions
\citep{YiJF98,KimPC01,keogh2005eid,ZhuS03b,sakurai2005ffs}, whose
complexity is typically much lower than that of a DTW distance measure,
are used for a lower bounding distance calculation which guarantees
that DTW distance must be equal to or larger than the lower bounding
distance. Additionally, in sequential scan, before calculating DTW
distance between the query sequence and a candidate sequence, a lower
bounding function is utilized to approximate and prune off the candidate
sequence which has larger lower bounding distance than the current
best-so-far distance. And in indexing, the lower bounding distance
is also used to guide the similarity search. Currently, many lower
bounding functions have been proposed to reduce computational costs
including LB\_Yi \citep{YiJF98}, LB\_Kim \citep{KimPC01}, LB\_Keogh
\citep{keogh2005eid}, LB\_PAA \citep{keogh2005eid}, LB\_NewPAA \citep{ZhuS03b},
and LBS \citep{sakurai2005ffs}. It has been widely known that LB\_Keogh
and LBS are among the most efficient lower bounding functions, where
LB\_Keogh has lower time complexity, while LBS has tighter bound.

Beside lower bounding functions, various index structures for DTW
distance have been proposed to guide the search to access only some
parts of the database. In other words, the search result is returned,
while a small portion of the database is accessed for distance calculation,
i.e., when querying, the index structure determines which parts of
the database are likely to contain answers, and then the raw data
on disk are randomly accessed. Generally, this index structure should
be small enough to fit in main memory. Currently, two exact indexing
approaches are typically used, i.e., GEMINI framework with LB\_PAA
\citep{keogh2005eid}, and a more recent approach, FTW indexing \citep{sakurai2005ffs}.
Note that the exact indexing returns a set of querying results with
no false dismissal guaranteed; in the other words, the best answers
must be included in the results. GEMINI framework \citep{FaloutsosRM94}
typically utilizes the multi-dimensional tree, e.g., R{*}-tree \citep{BeckmannKSS90},
as an index structure, while FTW indexing stores indices in a flat
file. However, current indexing techniques are burdened with huge
amount of I/O cost since random access to the database is typically
5 to 10 times slower than the sequential access \citep{WeberSB98}.
Therefore, indexing is efficient when less than 20\% of raw data sequences
are accessed on average. However, current indexing techniques still
consumes large I/O overheads which are not suitable for massive databases.

In this work, we propose a novel index structure and access method
under DTW distance called TWIST (Time Warping in Index Sequential
sTructure). TWIST utilizes advantages from both sequential structure
and index structure, i.e., low I/O and low CPU costs. Instead of randomly
accessing the raw time series data like other indexing techniques,
TWIST separates and stores a collection of time series data in sequential
structures or flat files. For each file, TWIST generates a representative
sequence (called an envelope) and stores this sequence in an index
structure. Therefore, when a query sequence is issued, each envelope
is calculated for a lower bounding distance using our newly proposed
lower bounding function for a group of sequences (LBG). The lower
bounding distance between an envelope and a query sequence guarantees
that all DTW distance between each and every candidate sequence under
this envelope and the query sequence must always be larger than this
lower bounding distance. Additionally, if the lower bounding distance
is larger than the best-so-far distance, no access to the sequences
within the envelope is needed; otherwise, every sequence in the envelope
is sequentially accessed for DTW distance calculation. 

We evaluate our proposed method, TWIST, comparing with the current
best approaches, i.e., FTW indexing and sequential scan with LB\_Keogh
lower bounding function. As will be demonstrated, TWIST prunes off
a large number of candidate sequences and is much faster than the
rival methods by a few orders of magnitude. Furthermore, when the
size of databases exponentially increases, our query processing time
only grows linearly. 

The rest of the paper is organized as follows. Section 2 provides
literature reviews of related work in speeding up similarity search
under DTW distance. In Section 3, our proposed index structure \textendash{}
TWIST, its access method, and novel proposed lower bounding distance
functions, are described. We show the superiority of TWIST over the
best existing method in Section 4. Finally, in Section 5, we conclude
our work and provide the direction of future research.

\section{Related Work}

After Dynamic Time Warping (DTW) distance measure \citep{BerndtC94}
has been introduced in data mining community \citep{KeoghK03,LohKW04,WangSH06,VlachosYCM06,BagnallRKLJ06,LinKWL07},
it shows the superiority of similarity matching over traditional Euclidean
distance due to its great flexibility in sequence alignment since
time series data mining has been long studied. Specifically, DTW distance
utilizes a dynamic programming to find the optimal warping path and
calculate the distance between two time series sequences. Unfortunately,
to calculate DTW distance, exhaustive computation is generally required.
In addition, since DTW distance is not qualified as a distance metric,
neither distance-based \citep{CiacciaPZ97,313789} nor spatial-based
\citep{BerchtoldKK96,Guttman84,BeckmannKSS90} index structure can
be used efficiently in similarity search under DTW distance. 

\begin{figure}
\noindent \begin{centering}
\begin{tabular}{c}
\includegraphics[scale=0.3]{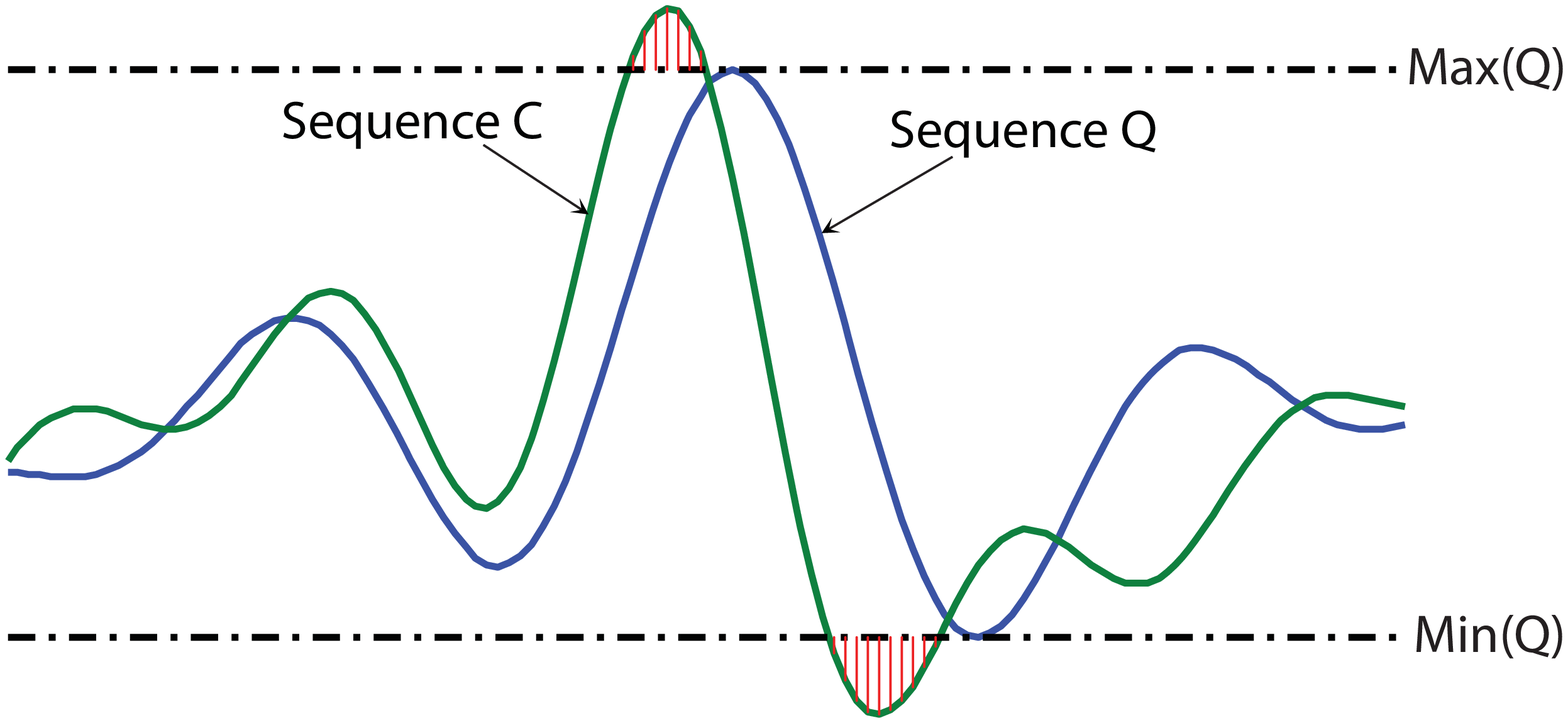}\tabularnewline
(a)\tabularnewline
\tabularnewline
\includegraphics[scale=0.3]{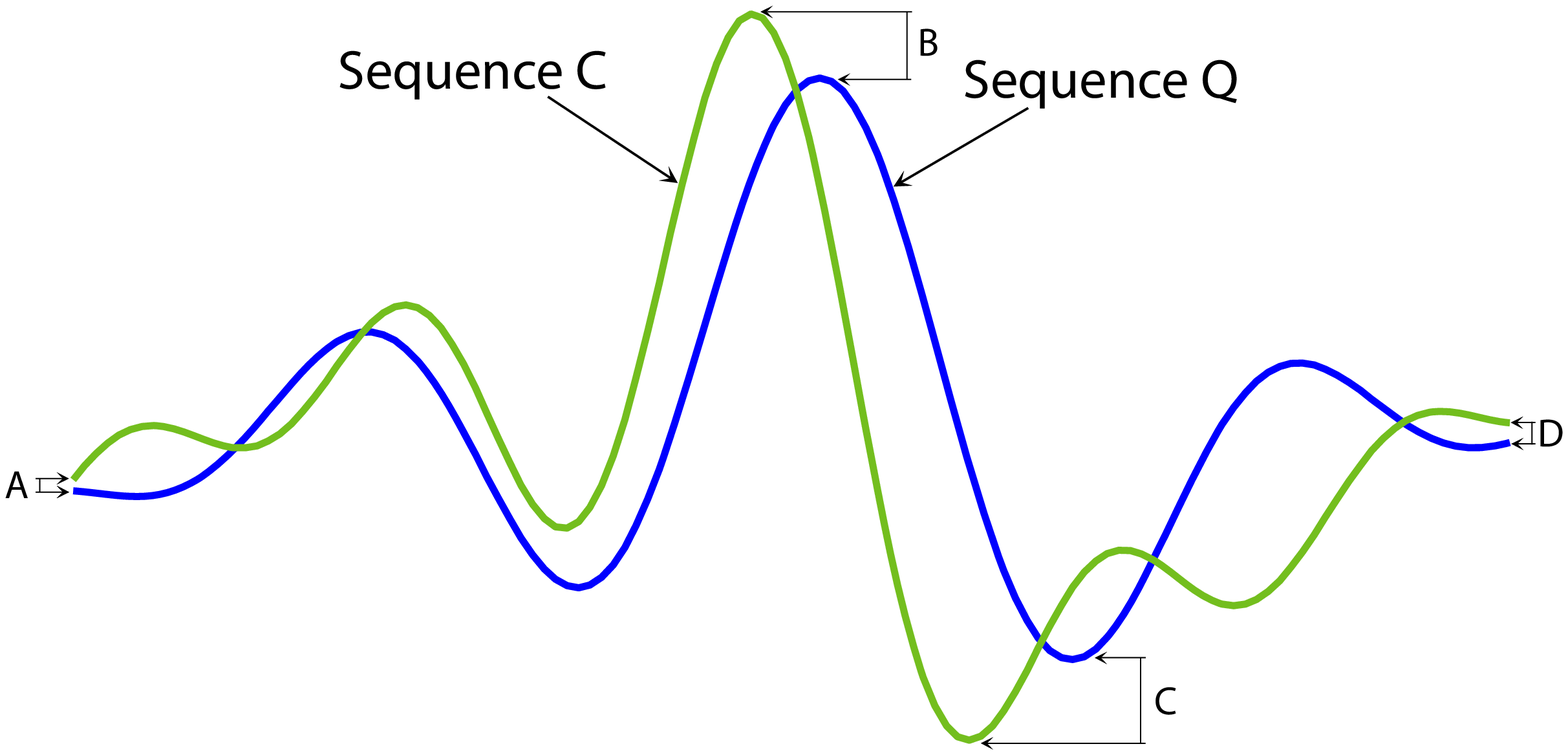}\tabularnewline
(b)\tabularnewline
\tabularnewline
\includegraphics[scale=0.3]{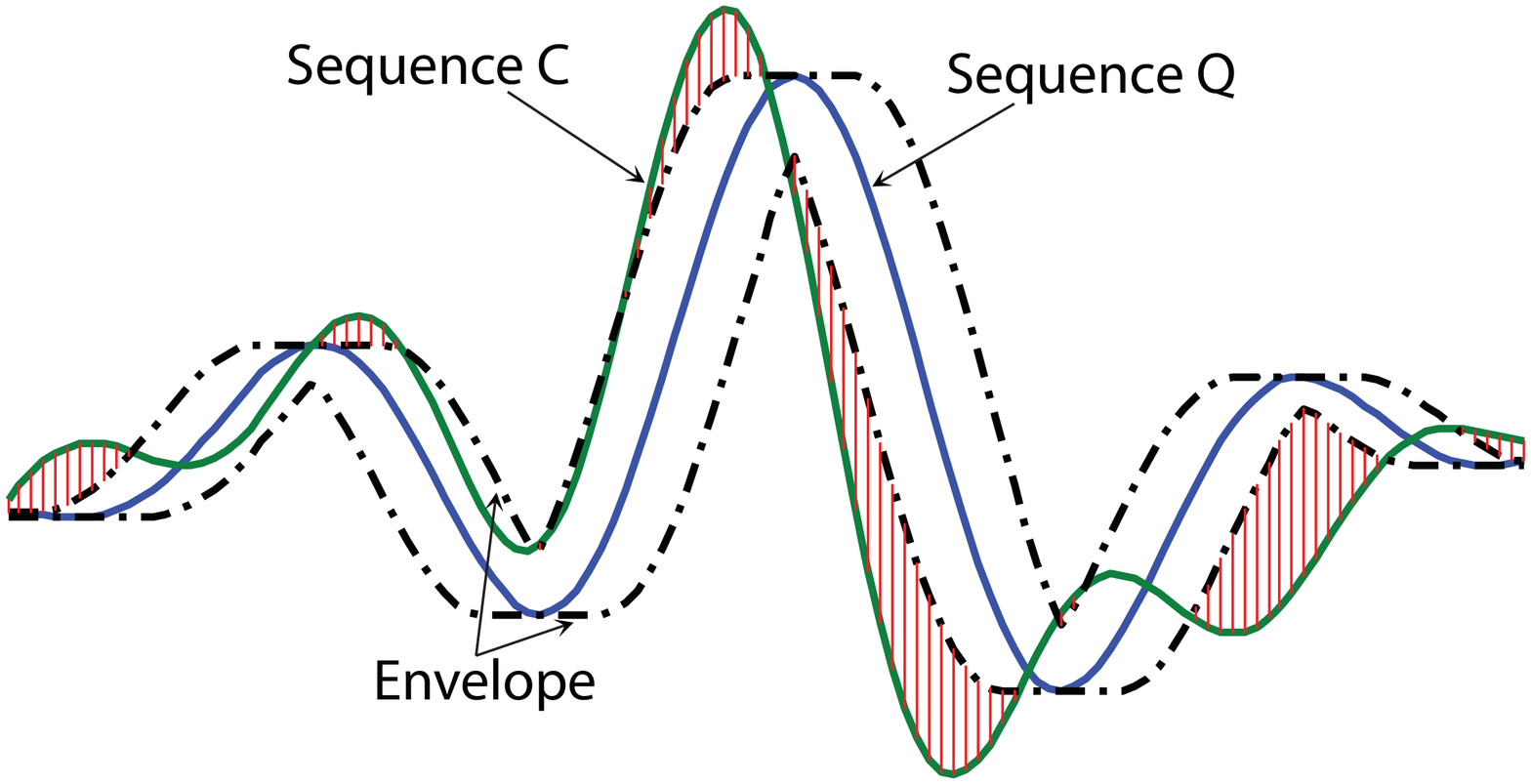}\tabularnewline
(c)\tabularnewline
\end{tabular}
\par\end{centering}

\caption{Illustration of lower bounding distance calculation between a query
sequence and a candidate sequence when using a) LB\_Yi, b) LB\_Kim,
and c) LB\_Keogh}

\label{Flo:lowerbound1}
\end{figure}

Therefore, various lower bounding functions and indexing techniques
for DTW distance have been proposed to resolve these problems. Yi
et al. \citep{YiJF98} first propose a lower bounding function, LB\_Yi,
using two features of a time series sequence, i.e., the minimum and
maximum values. LB\_Yi creates an envelope over a query sequence from
these minimum and maximum values, and then the distance is computed
from the summation of areas between an envelope and a candidate sequence,
as shown in Figure \ref{Flo:lowerbound1}a). Instead of using only
two features, Kim et al. \citep{KimPC01} suggest two additional features,
i.e., the first and the last values of the sequence. LB\_Kim then
calculates distance from the tuples of a query sequence and a candidate
sequence, as shown in Figure \ref{Flo:lowerbound1}b). Although these
two lower bounding functions only require small time complexity, the
uses of LB\_Yi and LB\_Kim is not practical since their lower bounding
distances cannot prune off much of the DTW distance calculations.

\begin{figure}
\noindent \begin{centering}
\begin{tabular}{ccc}
\includegraphics[width=3.5cm]{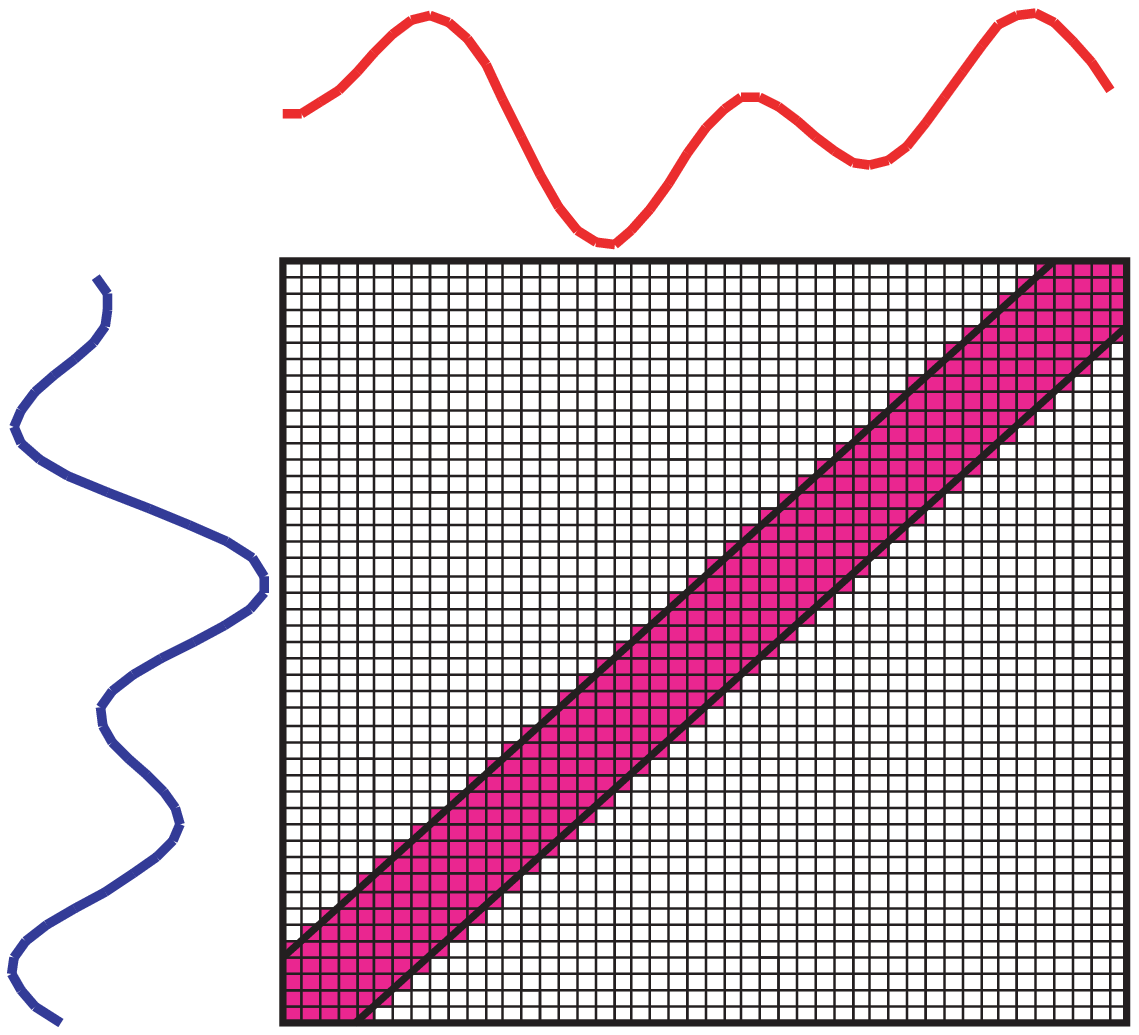} & \includegraphics[width=3.5cm]{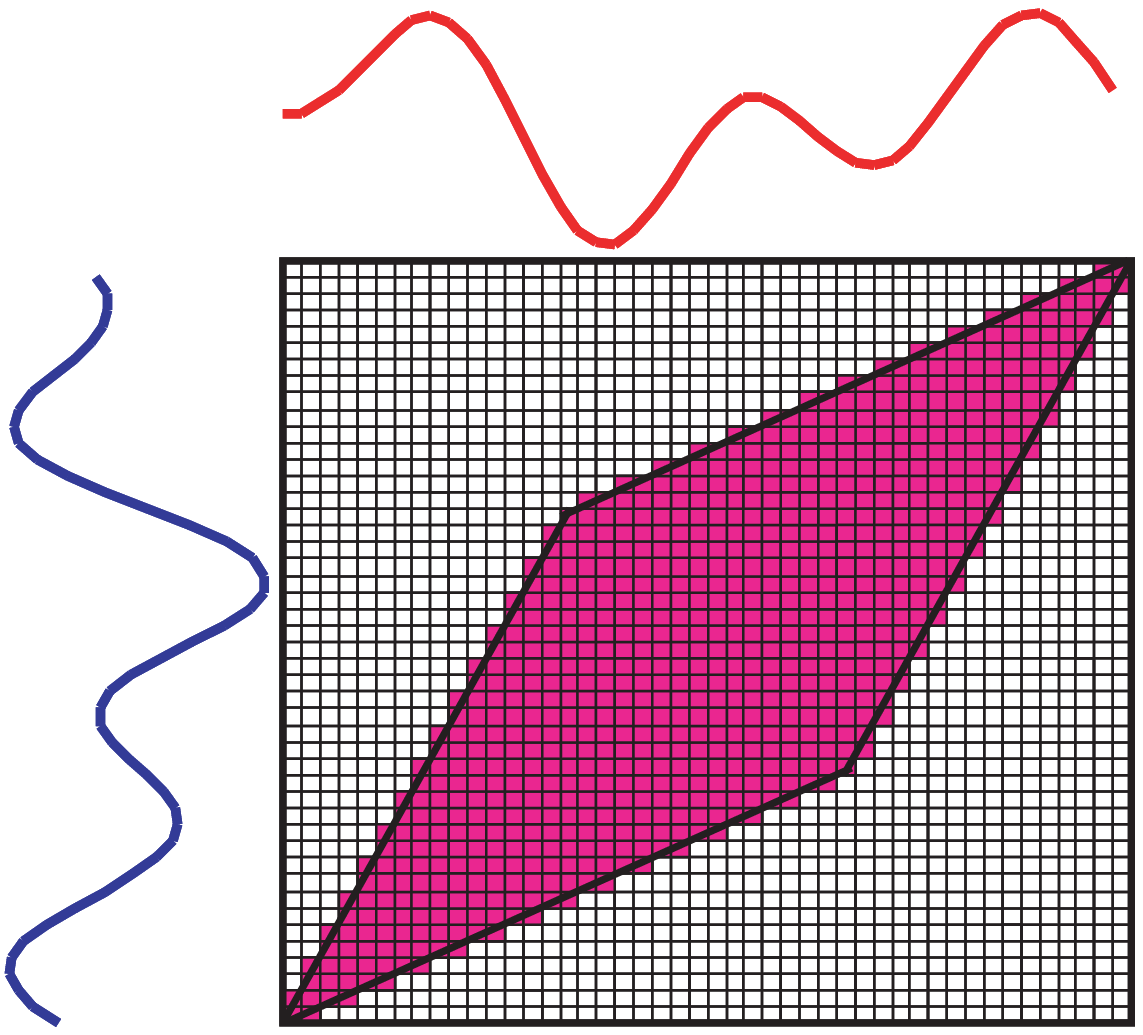} & \includegraphics[width=3.5cm]{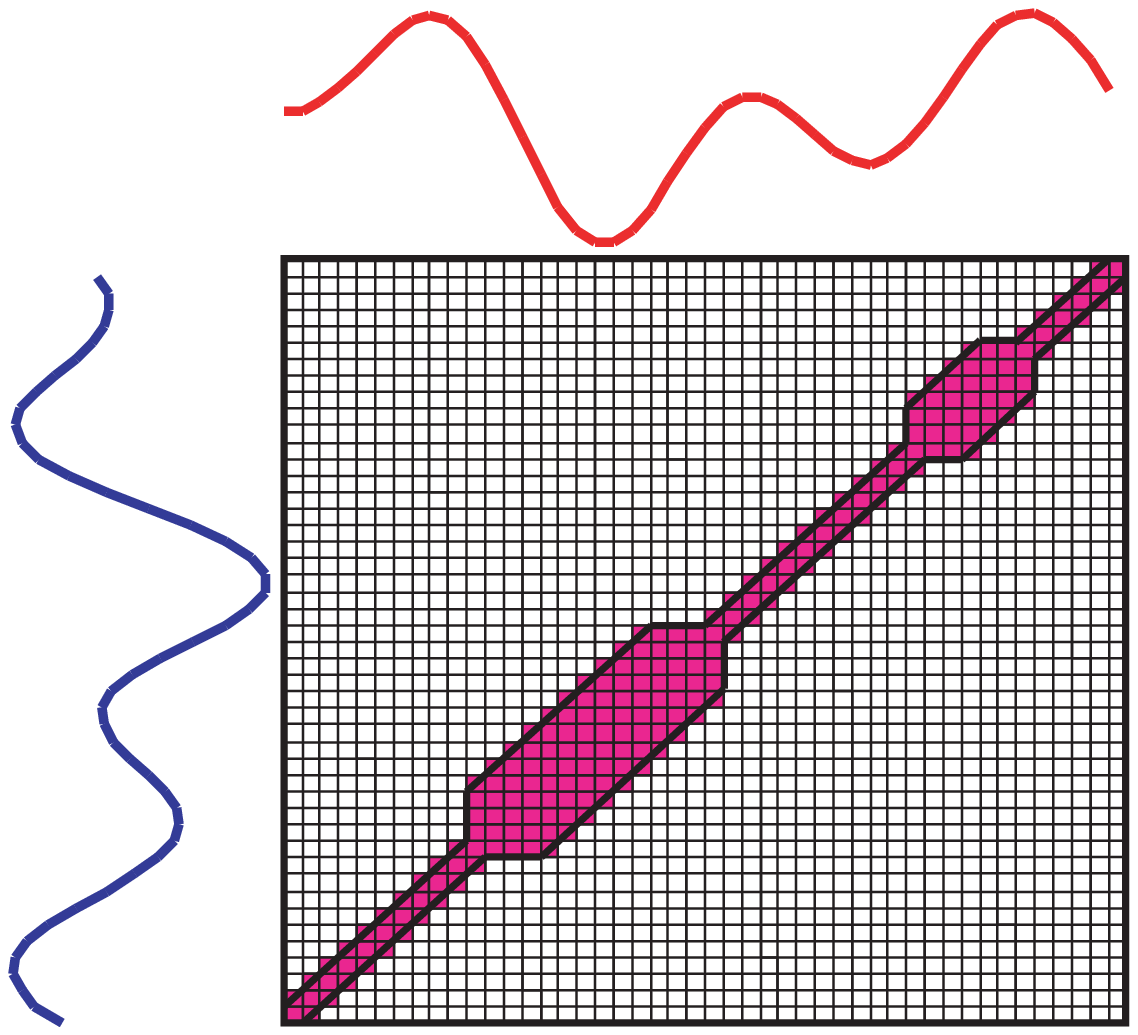}\tabularnewline
(a) & (b) & (c)\tabularnewline
\end{tabular}
\par\end{centering}

\caption{Shapes of a) Sakoe-Chiba band, b) Itakura Parallelogram, and c) Ratanamahatana-Keogh
band}
\label{Flo:globalx}
\end{figure}

Keogh et al. propose a tighter lower bounding function, LB\_Keogh,
utilizing global constraints \citep{108244,1162641,RatanamahatanaK04},
which are generally used to limit the scope of warping in distance
matrix to prevent undesirable paths. In addition, various well-known
global constraints have been proposed, e.g., Sakoe-Chiba band \citep{108244},
Itakura Parallelogram \citep{1162641}, and Ratanamahatana-Keogh (R-K)
band \citep{RatanamahatanaK04}. To be more illustrative, Figure \ref{Flo:globalx}
shows different shapes of global constraints. Note that R-K band is
an arbitrary-shaped constraint which can represent any bands by using
only a single one-dimensional array. LB\_Keogh first creates an envelope
over a query sequence according to the shape and size of the global
constraint. Its lower bounding distance then is an area between the
envelope and a candidate sequence, as shown in Figure \ref{Flo:lowerbound1}c).

In addition, Keogh et al. also propose an indexing technique which
utilizes their discretized version of their lower bounding function,
LB\_PAA. In order to create an index structure, they reduce dimensions
of each time series sequence using Piecewise Average Aggregation (PAA)
technique \citep{KeoghCPM01}, and store the reduced sequence in a
multi-dimensional index structure such as R{*}-tree \citep{BeckmannKSS90}.
Each leaf node of the tree, storing on disk, contains a collection
of segmented sequences, where each sequence points to its raw time
series data. In querying process, an envelope of the query sequence
is created and discretized. Therefore, each MBR (Minimum Bounding
Rectangle) of R{*}-tree is retrieved and is compared with the segmented
query sequence until the leaf node is retrieved in random-access manner.
Then, all discretized candidate sequences in the leaf node are undergone
lower bounding distances calculation using LB\_PAA. If the lower bounding
distance from the LB\_PAA is smaller than the best-so-far distance,
the raw time series sequence is also retrieved by random access, and
the distances are determined using LB\_Keogh and DTW distance, respectively.
It is clear that Keogh et al.'s index structure requires too many
random accesses as the database size slightly increases. Note that
although Zhu et al. later propose a tighter lower bounding function,
LB\_NewPAA \citep{ZhuS03b}, the index structure still consumes high
I/O cost.

\begin{figure}[h]
\noindent \begin{centering}
\begin{tabular}{cc}
\includegraphics[width=5cm]{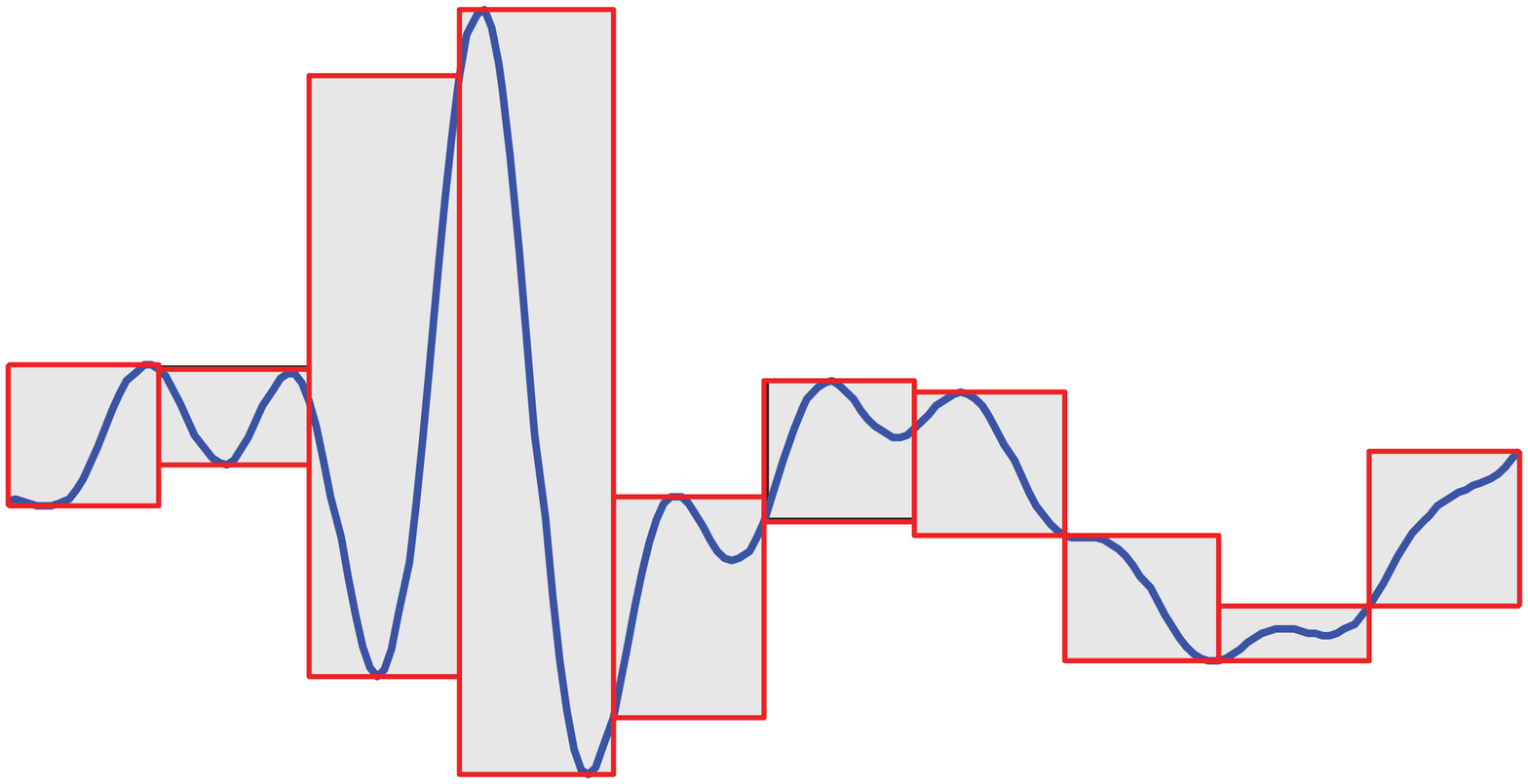} & \includegraphics[width=5cm]{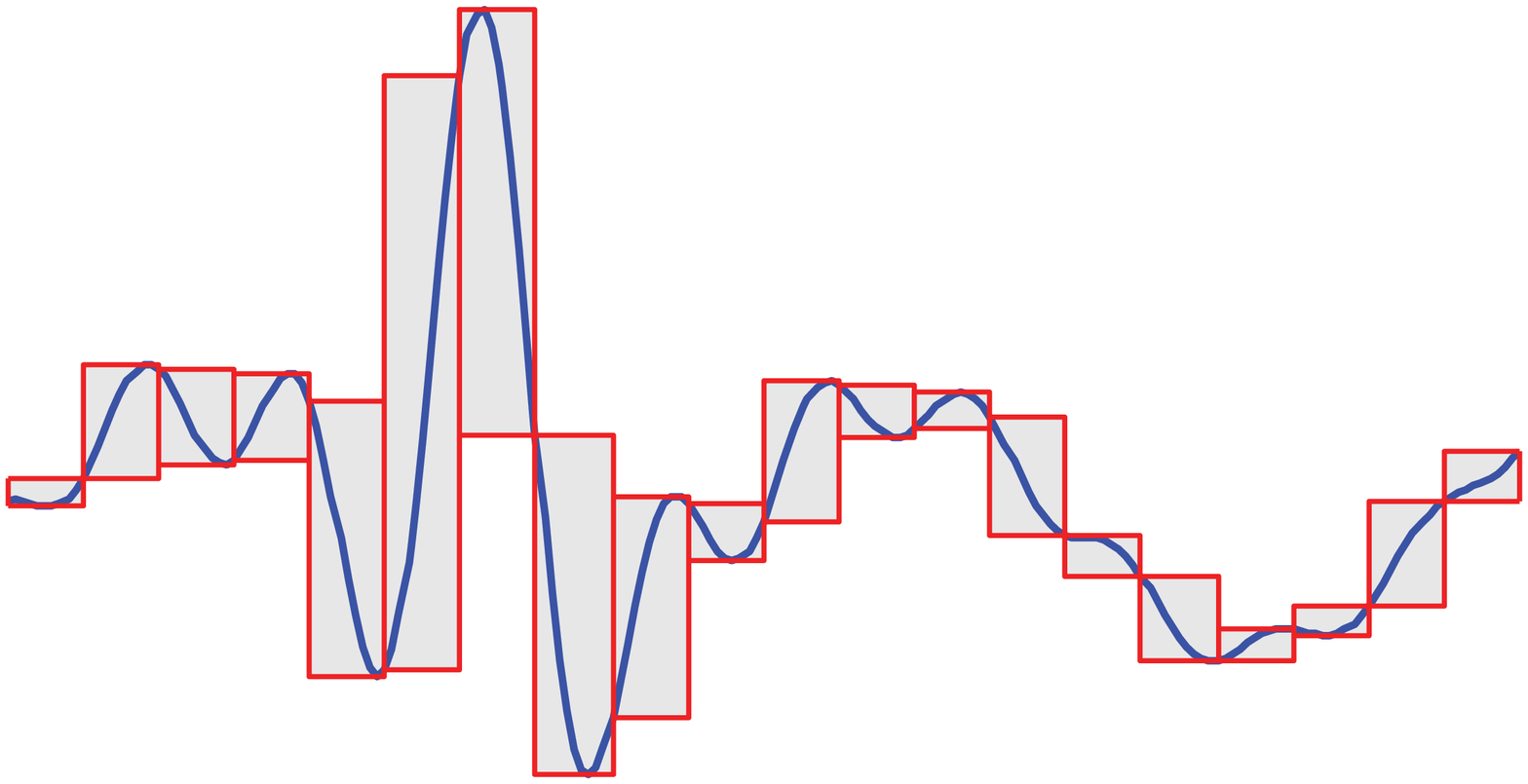}\tabularnewline
\multicolumn{2}{c}{\includegraphics[width=5cm]{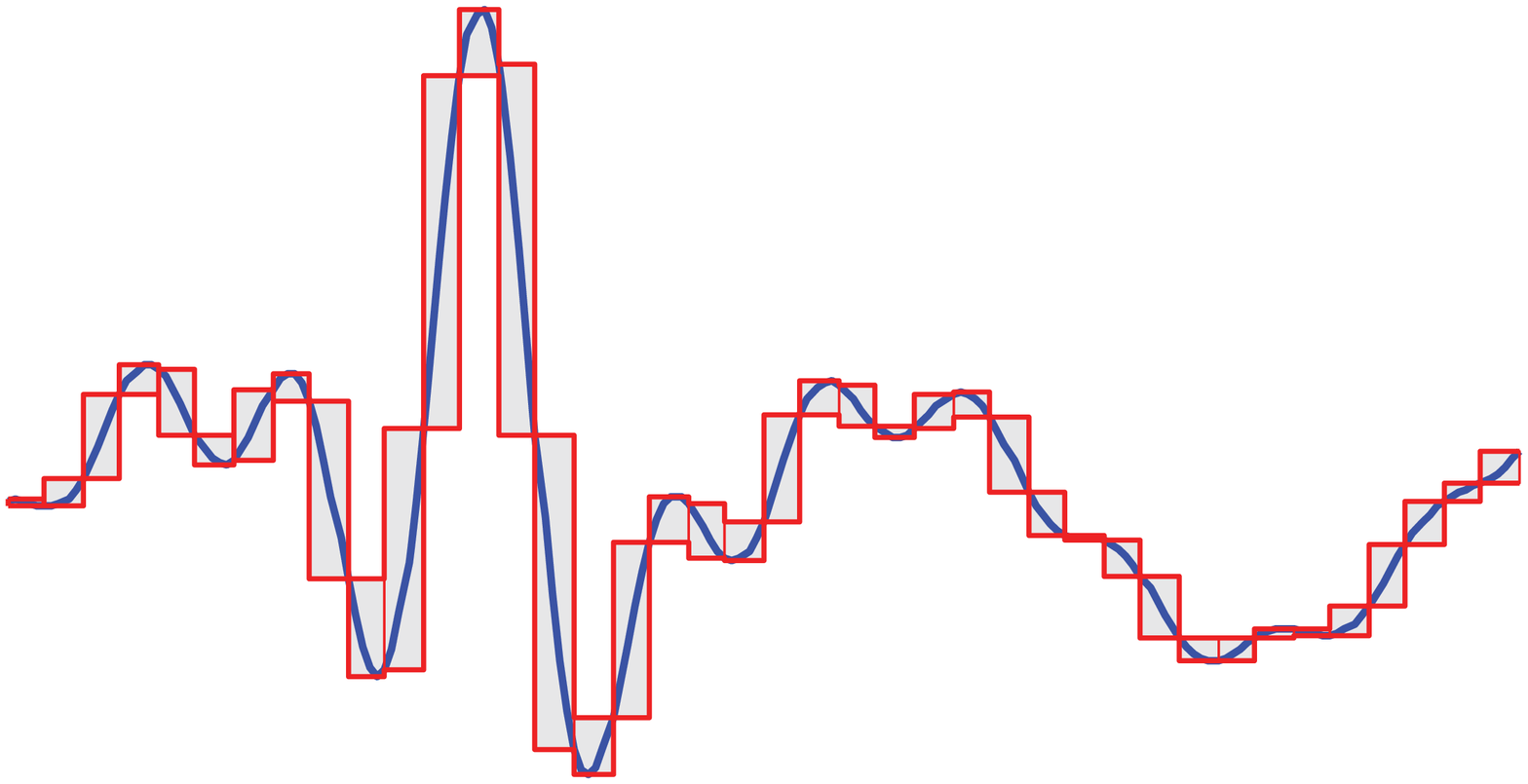}}\tabularnewline
\end{tabular}
\par\end{centering}

\caption{Illustration of segmented sequences with various resolutions}
\label{Flo:segment}
\end{figure}

Sakurai et al. \citep{sakurai2005ffs} propose a new lower bounding
function, LBS (Lower Bounding distance measure with Segmentation),
which requires a quadratic time complexity $O(n^{2}/t^{2})$, where
$n$ is the length of time series and $t$ is the size of a segment.
To calculate lower bounding distance, LBS first quantizes a query
sequence and a candidate sequence into sequences of segments. Each
segment contains two values that indicate the maximum and minimum
among the data points in the segment. Then, dynamic programming is
used to find the optimal distance between these two segmented sequences,
and the resulted distance is determined as a lower bound distance
of DTW distance. Despite the fact that LBS requires larger computational
time and space than those of LB\_PAA at the same resolution, LBS achieves
much tighter lower bounding distance. The example of segmented sequence
is shown in Figure \ref{Flo:segment}.

To use LBS in indexing, Sakurai et al. proposed an index structure
which stores pre-calculated segmented sequences. For each time series
data, a set of segmented sequences is generated by varying segment
sizes from the coarsest to the finest, and the segmented sequence
is stored in a flat file with a pointer to the raw time series data.
In querying process, a query sequence is segmented, and then the index
structure is sequentially accessed and calculated for lower bounding
distance with pre-segmented candidate sequences. If the lower bounding
distance is larger than the best-so-far distance, the raw time series
data is retrieved in random access manner. However, the main drawback
of FTW is that the size of the index structure is approximately twice
the size of the raw time series database. Therefore, this index structure
is definitely impractical for massive time series database since the
entire index file with size larger than the raw data are required
to be read once for every single query causing large I/O overheads.

It is worth to note that the existing index structures are not designed
for massive databases. For example, since LB\_PAA utilizes PAA to
reduce the number of dimensions, as the database size increases, its
pruning power significantly decreases; therefore, a huge number of
sequences must be accessed for distance calculation. Similarly for
FTW indexing, when the database size increases, the index size will
double. In Section 5, our experiments will demonstrate that when the
database exceeds the size of the main memory, our proposed method
significantly outperforms these rival methods.

\section{Background}

Before describing our proposed method, TWIST, we provide some background
knowledge, i.e., Dynamic Time Warping distance (DTW), global constraints,
and lower bounding distance functions including LB\_Keogh and LBS.

\subsection{Dynamic Time Warping Distance}

Dynamic Time Warping (DTW) distance \citep{BerndtC94,RatanamahatanaK05,RatanamahatanaK04}
is a well-known shape-based similarity measure. It uses a dynamic
programming technique to find an optimal warping path between two
time series sequences. To calculate the distance, it first creates
a distance matrix, where each element in the matrix is a cumulative
distance of the minimum of three surrounding neighbors. Suppose we
have two time series, a sequence \emph{$Q=\left\langle q_{1},\ldots,q_{i},\ldots,q_{n}\right\rangle $}
and a sequence \emph{$C=\left\langle c_{1},\ldots,c_{j},\ldots,c_{m}\right\rangle $}.
First, we create an \emph{$n$}-by-\emph{$m$} matrix, and then each
$(i,j)$ element, $\gamma_{i,j}$, of the matrix is defined as:

\begin{equation}
\gamma_{i,j}=\left|q_{i}-c_{j}\right|^{p}+\min\left\{ \gamma_{i-1,j-1},\gamma_{i-1,j},\gamma_{i,j-1}\right\} \label{eq:dtw1}\end{equation}

\noindent where $\gamma_{i,j}$ is the summation of $\left|q_{i}-c_{j}\right|^{p}$
and the minimum cumulative distance of three elements surrounding
the $(i,j)$ element, and \emph{$p$} is the dimension of $L_{p}$-norms.
For time series domain, $p=2$, equipping to Euclidean distance, is
typically used. After we have all distance elements in the matrix,
to find an optimal path, we choose the path $W=\left\langle w_{1},\ldots,w_{k},\ldots,w_{K}\right\rangle $
that yields a minimum cumulative distance at $(n,m)$, where $w_{k}$
is the position $(i,j)$ at $k^{th}$ element of a warping path, $w_{1}=(1,1)$,
and $w_{K}=(n,m)$, which is defined as:

\begin{equation}
DTW(Q,C)=\underset{\forall W\in\mathbb{W}}{\min}\left\{ \sqrt[p]{\overset{K}{\underset{k=1}{\sum}}d_{w_{k}}}\right.\label{eq:dtw2}\end{equation}

\noindent where $d_{w_{k}}$ is the $L_{p}$ distance at the position
$w_{k}$, \emph{$p$} is the dimension of $L_{p}$-norms in Equation
\ref{eq:dtw1}, and \emph{$\mathbb{W}$} is a set of all possible
warping paths. The recursive function are shown in Equation \ref{eq:dtw0}.
Note that, in the original DTW, $p^{th}$ root of the distance must
be computed; however, for fast computation, we usually omit this calculation
since ranking of distance values does not change.

\begin{equation}
DTW(Q,C)=\sqrt[p]{D(n,m)}\label{eq:dtw0}\end{equation}

\begin{equation}
D(i,j)=\left|q_{i}-c_{j}\right|^{p}+\min\left\{ \begin{array}{l}
D(i-1,j-1)\\
D(i-1,j)\\
D(i,j-1)\end{array}\right.\label{eq:dtw00}\end{equation}

\noindent where $D(0,0)=0$, $D(i,0)=D(0,j)=\infty$, $1\leq i\leq n$,
and $1\leq j\leq m$.

\subsection{Global Constraints}

\begin{figure}[h]
\noindent \begin{centering}
\includegraphics[width=4cm]{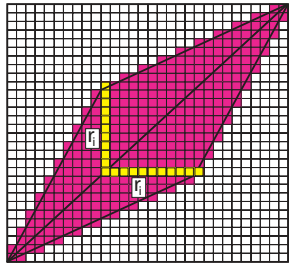}
\par\end{centering}

\caption{Global contraint on DTW distance matrix when applying specific R-K
band}
\label{fig:rk}
\end{figure}

Although unconstrained DTW distance measure gives an optimal distance
between two time series data, an unwanted warping path may be generated.
The global constraint efficiency limits the optimal path to give a
more suitable alignment. Recently, an R-K band \citep{RatanamahatanaK04},
a general model of global constraints, has been proposed. It can be
specified by a one-dimensional array $R$, i.e., $R=\langle r_{1},\ldots,r_{i},\ldots,r_{n}\rangle$,
where $n$ is the length of time series, and $r_{i}$ is the height
above the diagonal in $y$ direction and the width to the right of
the diagonal in $x$ direction, as shown in Figure \ref{fig:rk}.
Each $r_{i}$ value is arbitrary; therefore, R-K band is also an arbitrary-shaped
global constraint. Note that when $r_{i}$ = 0, where $1\leq i\leq n$,
this R-K band represents the well-known Euclidean distance, and when
$r_{i}=n$, $1\leq i\leq n$, this R-K band represents the original
DTW distance with no global constraint. The R-K band can also represent
the S-C band by giving all $r_{i}=c$, where $c$ is the width of
a global constraint. 

\subsection{Lower Bounding Distance Function}

Lower bounding distance function for DTW distance is a function that
is used to calculate a lower bounding distance which must always be
smaller than or equal to the exact DTW distance \citep{YiJF98,KimPC01,keogh2005eid,ZhuS03b,sakurai2005ffs}.
Therefore, in similarity search, the lower bounding function is used
to prune off the candidate sequences that are definitely not the answers.
Typically, lower bounding function consumes much lower computational
time than the DTW distance does. In this work, we consider two lower
bounding functions, i.e., LB\_Keogh \citep{keogh2005eid} proposed
by Keogh et al. and LBS \citep{sakurai2005ffs} proposed by Sakurai
et al. since LB\_Keogh is the best existing lower bounding function
used in sequential search, and LBS is the tightnest lower bounding
function used in indexing. LB\_Keogh creates an envelope from a query
sequence, and then the lower bounding distance is calculated from
areas between the envelope and a candidate sequence. Unlike LB\_Keogh,
LBS creates a segmented query sequence and a segmented candidate sequence,
and then these two segmented sequences are used to determine a lower
bounding distance using dynamic programming.

\subsubsection{LB\_Keogh}

To calculate LB\_Keogh \citep{keogh2005eid}, an envelope $E=\left\langle e_{1},\ldots,e_{i},\ldots,e_{n}\right\rangle $
is generated from a query sequence $Q=\left\langle q_{1},\ldots,q_{i},\ldots,q_{n}\right\rangle $,
where $e_{i}=\left\{ u_{i},l_{i}\right\} $ , and $u_{i}$ and $l_{i}$
are an upper and a lower values of $e_{i}$. With a specified global
constraint $R=\left\langle r_{1},\ldots,r_{i},\ldots,r_{n}\right\rangle $,
elements $u_{i}$ and $l_{i}$ are computed from $u_{i}=\max\left\{ q_{i-r_{i}},\ldots,q_{i+r_{i}}\right\} $
and $l_{i}=\min\left\{ q_{i-r_{i}},\ldots,q_{i+r_{i}}\right\} $,
respectively. The lower bounding distance $LB_{Keogh}(Q,C)$ between
sequences $Q$ and $C$ can be computed by the following equation.

\begin{equation}
LB_{Keogh}(Q,C)=\sqrt[p]{\overset{n}{\underset{i=1}{\sum}}\left\{ \begin{array}{cl}
\left|c_{i}-u_{i}\right|^{p} & \mathrm{if}\, c_{i}>u_{i}\\
\left|l_{i}-c_{i}\right|^{p} & \mathrm{if}\, c_{i}<l_{i}\\
0 & \mathrm{otherwise}\end{array}\right.}\label{eq:lb2}\end{equation}

\noindent where \emph{$p$} is the dimension of $L_{p}$-norms. The
proof of $LB_{Keogh}(Q,C)$\linebreak{}
$\leq DTW(Q,C)$ can be found in the original paper
\citep{keogh2005eid}.

\subsubsection{LBS}

To calculate LBS (Lower bounding distance measure with Segmentation),
a query and a candidate sequences must first be segmented. The segmented
sequence $S^{T}=\left\langle s_{1}^{T},\ldots,s_{b}^{T},\ldots,s_{t}^{T}\right\rangle $
is calculated from the sequence $S=\left\langle s_{1},\ldots,s_{a},\ldots,s_{A}\right\rangle $
with a given segment size $T$, where $s_{b}^{T}=\left\{ us_{b}^{T},ls_{b}^{T}\right\} $,
$us_{b}^{T}=\max\left\{ s_{x},\ldots,s_{y}\right\} $, $ls_{b}^{T}=\min\left\{ s_{x},\ldots s_{y}\right\} $,
$x=\left(a-1\right)\cdot T+1$, $y=b\cdot T$, and $1\leq T\leq A$.
Although LBS has capability to support segments with different lengths,
in this work, we consider each segments with an equal length to demonstrate
maximum performance of LBS. The lower bounding distance $LBS(Q^{T},C^{T})$
between a segmented query sequence $Q^{T}=\left\langle q_{1}^{T},\ldots,q_{i}^{T},\ldots,q_{n}^{T}\right\rangle $
and a segmented candidate sequence $C^{T}=\left\langle c_{1}^{T},\ldots,c_{i}^{T},\ldots,c_{n}^{T}\right\rangle $
can be computed by the following equations.

\begin{equation}
LBS(Q^{T},C^{T})=\sqrt[p]{D(n,m)}\label{eq:ftw1}\end{equation}

\begin{equation}
D(i,j)=T\cdot d(q_{i}^{T},c_{j}^{T})+\min\left\{ \begin{array}{l}
D(i-1,j-1)\\
D(i-1,j)\\
D(i,j-1)\end{array}\right.\label{eq:ftw2}\end{equation}

\begin{equation}
d(q_{i}^{T},c_{j}^{T})=\left\{ \begin{array}{cc}
\left|lq_{i}^{T}-uc_{j}^{T}\right|^{p} & \mathrm{if}\,(lq_{i}^{T}>uc_{j}^{T})\\
\left|lc_{j}^{T}-uq_{i}^{T}\right|^{p} & \mathrm{if}\,(lc_{j}^{T}>uq_{j}^{T})\\
0 & \mathrm{otherwise}\end{array}\right.\label{eq:ftw3}\end{equation}

\noindent where $D(0,0)=0$, $D(i,0)=D(0,j)=\infty$, $1\leq i\leq n$,
$1\leq j\leq m$, $q_{i}^{T}=\left\{ uq_{i}^{T},lq_{i}^{T}\right\} $,
$c_{i}^{T}=\left\{ uc_{i}^{T},lc_{i}^{T}\right\} $, and \emph{$p$}
is the dimension of $L_{p}$-norms. The proof of $LBS(Q^{T},C^{T})\leq DTW(Q,C)$
can be found in the Sakurai et al.'s original paper \citep{sakurai2005ffs}.

\section{Time Warping in Indexed Sequential sTructure (TWIST)}

In this work, we propose a novel index structure called TWIST (\textbf{T}ime
\textbf{W}arping in \textbf{I}ndexed \textbf{S}equential s\textbf{T}ructure)
which consists of both sequential structures and an index structure.
Each sequential structure stores a collection of raw time series sequences,
and the index structure stores a representative and a pointer to its
corresponding sequential structure. The intuitive idea of TWIST is
to minimize the number of random accesses and minimize the number
of distance calculations, giving TWIST a much more suitable choice
for massive database than the existing methods which are not quite
scalable.

\subsection{Problem Definition}

We are interested in a generic top-$k$ querying in this work since
many other mining tasks, e.g., classification and clustering, all
require this best-matched querying as their typical subroutine. Given
a query sequence $Q$, a set $\mathbb{C}$ of equal-length time series
sequences, a global constraint $R$, and an integer $k$, it returns
a set of $k$ nearest-neighbor sequences of $Q$ from $\mathbb{C}$
under DTW distance measure with the constraint $R$.

\subsection{Data Structure}

\begin{figure}[h]
\noindent \begin{centering}
\includegraphics[width=7cm]{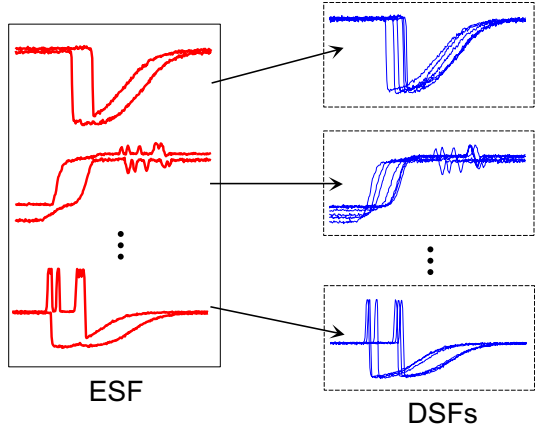}
\par\end{centering}

\caption{A sample data structure of TWIST}
\label{Flo:twist}
\end{figure}

In this section, we describe the data structure of TWIST which is
specially designed to minimize both the I/O and CPU costs in the querying
process. TWIST consists of two main components, i.e., a set of sequential
structures (called Data Sequence File \textendash{} DSF) and an index
structure (called Envelope Sequence File \textendash{} ESF). In addition,
TWIST groups the similar sequences into same sequential structure
so that in the querying process, if this sequential structure greatly
differs from a query sequence, TWIST will simply bypass that structure.
To measure the difference between a query sequence and all the sequences
in a sequential structure, a representative sequence (called an envelope)
is pre-determined and stored in an index structure. The main benefit
of the sequential structure is that, we can access all the data in
the sequential structure much faster than the random access \citep{WeberSB98}.
A sample data structure of TWIST is shown in Figure \ref{Flo:twist}. 

\begin{figure}
\noindent \begin{centering}
\begin{tabular}{cc}
\includegraphics[width=5.5cm]{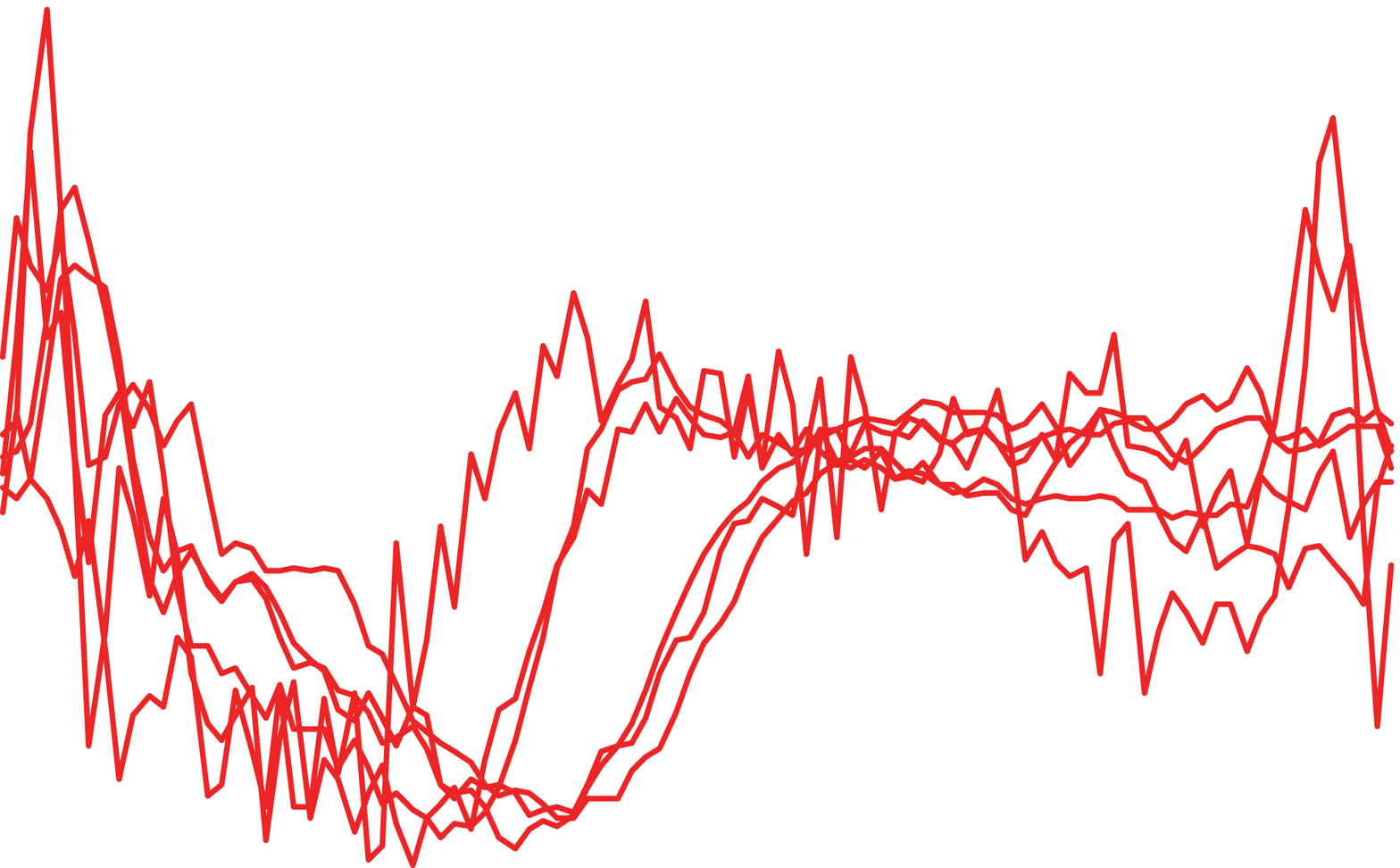} & \includegraphics[width=5.5cm]{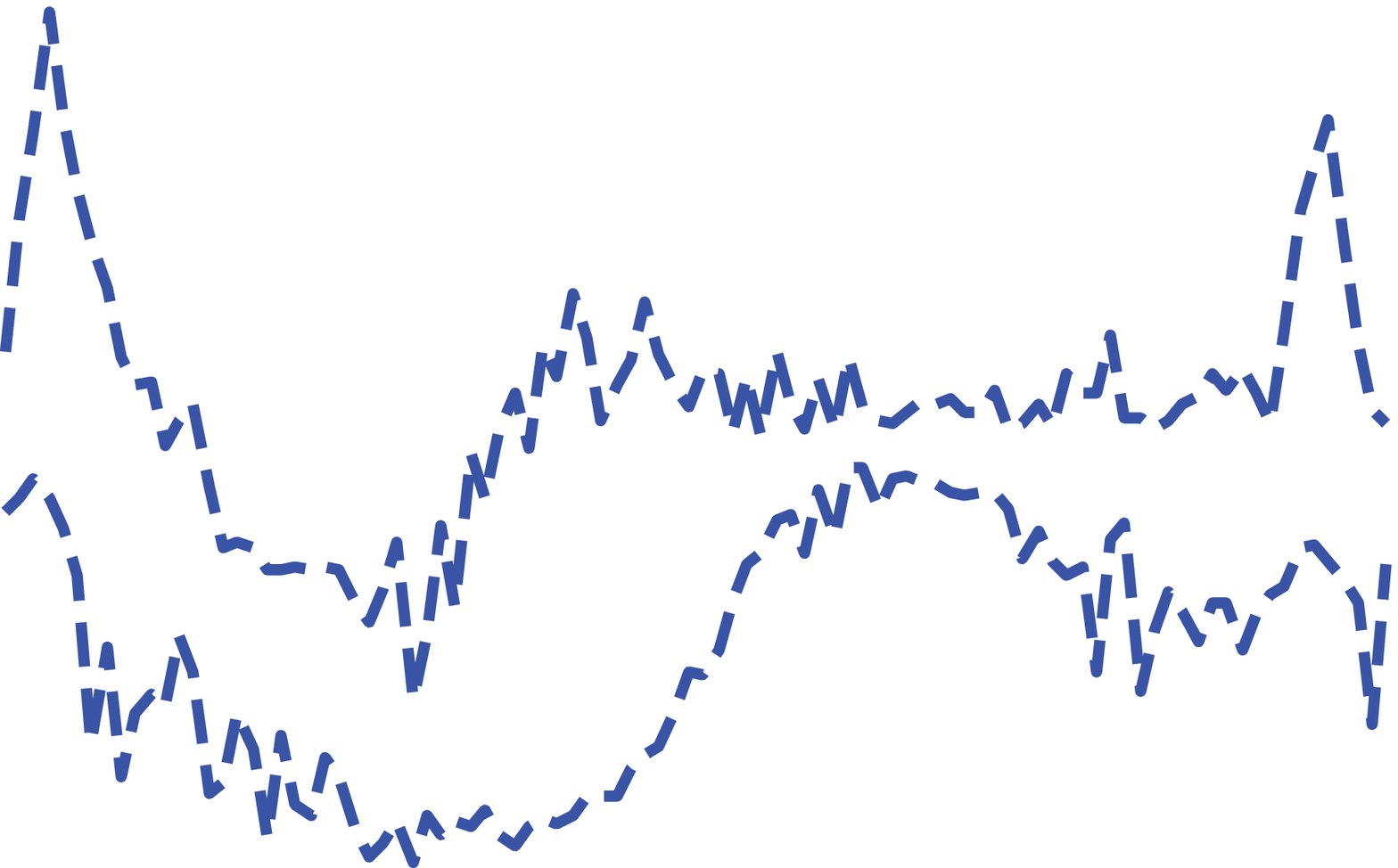}\tabularnewline
\end{tabular}
\par\end{centering}

\caption{An envelope created from a group of sequences}
\label{Flo:twist2}
\end{figure}

Suppose there is a set $\mathbb{S}$ of time series sequences $S=\left\langle s_{1},\ldots,s_{i},\ldots,s_{n}\right\rangle $,
DSF simply stores these sequences sequentially. And for each DSF,
an envelope $EG=\left\langle eg_{1},\ldots,eg_{i},\ldots,eg_{n}\right\rangle $
for a group of time series sequences is generated, where $eg_{i}=\left\{ ueg_{i},leg_{i}\right\} $,
$ueg_{i}=\underset{S\in\mathbb{S}}{\max}\left\{ s_{i}\right\} $,
and $leg_{i}=\underset{S\in\mathbb{S}}{\min}\left\{ s_{i}\right\} $.
In addition, the data structure of ESF is basically an array $A$
of an object $O=\left\{ P,EQ\right\} $ containing a pointer $P$
to DSF and an envelope $EG$. Figure \ref{Flo:twist2} illustrates
an envelope construction for each DSF. The envelope is determined
from an upper bound and a lower bound of a group of sequences.

\subsection{Lower Bounding Distance for a Group of Sequences}

In this work, we propose a novel lower bounding distance function
for a group of sequences called LBG. Instead of calculating lower
bounding distances between a query sequence and a candidate sequence,
LBG returns a lower bounding distance between a query sequence and
a set of candidate sequences; in other words, each DTW distance between
a query sequence and any candidate sequence in the set is always larger
than the lower bounding distance from LBG. Therefore, if the lower
bounding distance is larger than the distance from the best-so-far
distance, LBG can prune off all those candidate sequences since all
the real DTW distances from the candidate sequences are guaranteed
not to be any smaller. More specifically, TWIST utilizes LBG by determining
an LBG for each DFS from an envelope sequence stored in the EFS so
that only some DSFs are accessed which significantly reduces both
CPU and I/O costs.

\begin{figure}[h]
\noindent \begin{centering}
\begin{tabular}{cc}
\includegraphics[width=5.5cm]{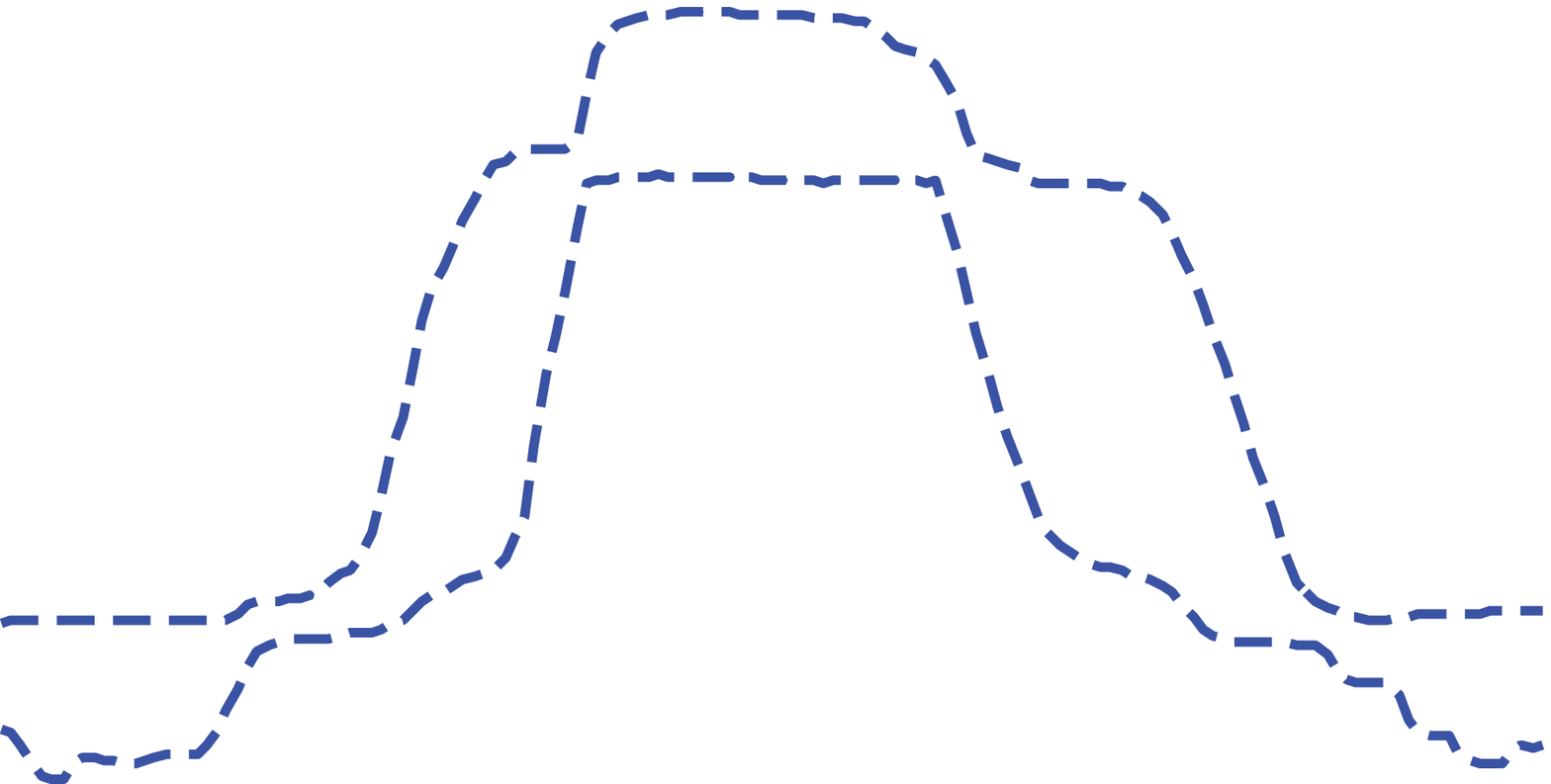} & \includegraphics[width=5cm]{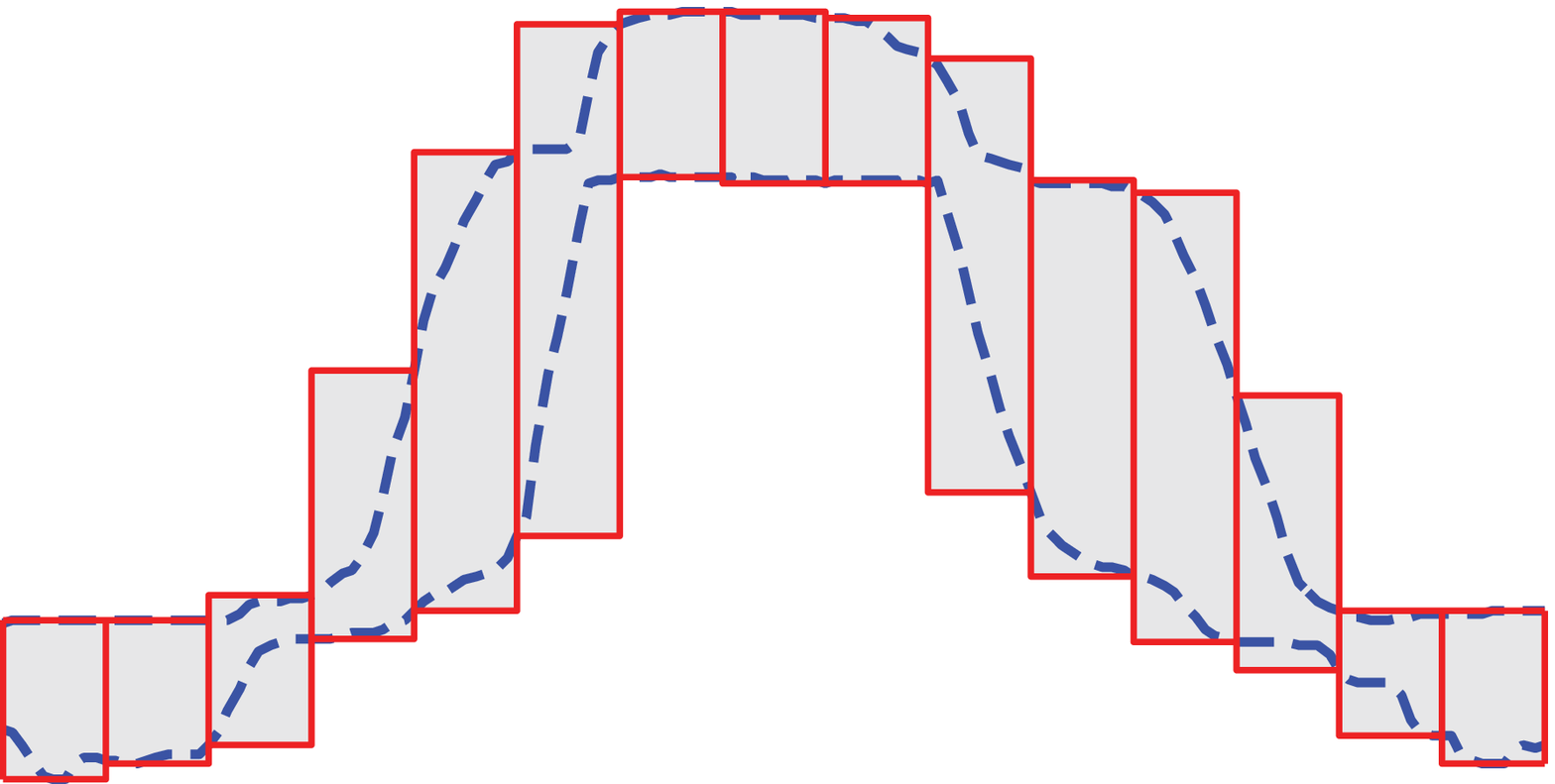}\tabularnewline
(a) & (b)\tabularnewline
\end{tabular}
\par\end{centering}

\caption{Illustration shows a) an envelope used to generate b) a segmented
envelope when calculating LBG}
\label{Flo:lbglbs-1}
\end{figure}

Given a query sequence $Q=\left\langle q_{1},\ldots,q_{a},\ldots,q_{n}\right\rangle $
and an envelope $EG=$\linebreak{}
$\left\langle eg_{1},\ldots,eg_{b},\ldots,eg_{n}\right\rangle $,
where $eg_{b}=\left\{ ueg_{b},leg_{b}\right\} $. LBG first creates
segmented query sequences $Q^{T}=\left\langle q_{1}^{T},\ldots,q_{i}^{T},\ldots,q_{t}^{T}\right\rangle $
and segmented envelope $EG^{T}=$\linebreak{}
$\left\langle eg_{1}^{T},\ldots,eg_{j}^{T},\ldots,eg_{t}^{T}\right\rangle $
with segment size $T$, where $q_{i}^{T}=\left\{ uq_{i}^{T},lq_{i}^{T}\right\} $
and $eg_{j}^{T}=\left\{ ueg_{j}^{T},leg_{j}^{T}\right\} $. An element
$q_{i}^{T}$ of segmented query sequence $Q^{T}$ is computed by $uq_{i}^{T}=\max\left\{ s_{x},\ldots,s_{y}\right\} $
and $lq_{i}^{T}=\min\left\{ s_{x},\ldots s_{y}\right\} $, where $x=\left(a-1\right)\cdot T+1$,
and $y=a\cdot T$. On the other hand, to segment an envelope $EG$,
elements $ueg_{j}^{T}$ and $leg_{j}^{T}$ are created as follows,
$ueg_{j}^{T}=\max\left\{ ueg_{x},\ldots,ueg_{y}\right\} $ and $leg_{j}^{T}=\min\left\{ leg_{x},\ldots leg_{y}\right\} $,
where $x=\left(b-1\right)\cdot T+1$, and $y=b\cdot T$. To be more
illustrative, Figure \ref{Flo:lbglbs-1} shows the segmented envelope
$EG^{T}$ created from an envelope $EG$. 

The lower bounding distance $LBG(Q^{T},EG^{T})$ between a segmented
query sequence $Q^{T}$ and a segmented envelope $EG^{T}$ can be
computed by the following equations.

\begin{equation}
LBG(Q^{T},EG^{T})=\sqrt[p]{D(n,m)}\label{eq:lbg1}\end{equation}

\begin{equation}
D(i,j)=T\cdot d(q_{i}^{T},eg_{j}^{T})+\min\left\{ \begin{array}{l}
D(i-1,j-1)\\
D(i-1,j)\\
D(i,j-1)\end{array}\right.\label{eq:lbg2}\end{equation}

\begin{equation}
d(q_{i}^{T},eg_{j}^{T})=\left\{ \begin{array}{cc}
\left|lq_{i}^{T}-ueg_{j}^{T}\right|^{p} & \mathrm{if}\,(lq_{i}^{T}>ueg_{j}^{T})\\
\left|leg_{j}^{T}-uq_{i}^{T}\right|^{p} & \mathrm{if}\,(leg_{j}^{T}>uq_{i}^{T})\\
0 & \mathrm{otherwise}\end{array}\right.\label{eq:lbg3}\end{equation}

\noindent where $D(0,0)=0$, $D(i,0)=D(0,j)=\infty$, $1\leq i\leq n$,
$1\leq j\leq m$, and \emph{$p$} is the dimension of $L_{p}$-norms.
\begin{theorem}
Let $Q^{T}=\left\langle q_{1}^{T},\ldots,q_{i}^{T},\ldots,q_{t}^{T}\right\rangle $
and $EG^{T}=\left\langle eg_{1}^{T},\ldots,eg_{j}^{T},\ldots,eg_{t}^{T}\right\rangle $
be the approximate segments of sequence $Q$ and envelope $EG$ of
a group of time series sequences $\mathbb{C}=\left\{ C_{1},\ldots,C_{k},\ldots,C_{m}\right\} $,
respectively, where $q_{i}^{T}=\left\{ uq_{i}^{T},lq_{i}^{T}\right\} $
and $eg_{j}^{T}=\left\{ ueg_{j}^{T},leg_{j}^{T}\right\} $, then
\end{theorem}
\begin{equation}
LBG(Q^{T},EG^{T})\leq DTW(Q,C_{opt})\label{eq:lbg5}\end{equation}

\noindent where $C_{opt}$ is a sequence in $\mathbb{C}$ which gives
minimum distance to sequence $Q$, and $C_{opt}^{T}$ is a segmented
sequence of $C_{opt}$.
\begin{proof}
Following from the proof of LBS \citep{sakurai2005ffs}, we have

\begin{equation}
LBS(Q^{T},C_{opt}^{T})\leq DTW(Q,C_{opt})\label{eq:lbg6}\end{equation}

Since $ueg_{j}^{T}\geq uc_{opt_{j}}^{T}$ and $leg_{j}^{T}\leq lc_{opt_{j}}^{T}$
for all $j$

\begin{flushleft}
\begin{equation}
\begin{array}{ccc}
d\left(q_{i}^{T},eg_{j}^{T}\right) & = & \left\{ \begin{array}{cc}
\left|lq_{i}^{T}-ueg_{j}^{T}\right|^{p} & \mathrm{if}\,(lq_{i}^{T}>ueg_{j}^{T})\\
\left|leg_{j}^{T}-uq_{i}^{T}\right|^{p} & \mathrm{if}\,(leg_{j}^{T}>uq_{j}^{T})\\
0 & \mathrm{otherwise}\end{array}\right.\\
 & \leq & \left\{ \begin{array}{cc}
\left|lq_{i}^{T}-uc_{j}^{T}\right|^{p} & \mathrm{if}\,(lq_{i}^{T}>uc_{j}^{T})\\
\left|lc_{j}^{T}-uq_{i}^{T}\right|^{p} & \mathrm{if}\,(lc_{j}^{T}>uq_{j}^{T})\\
0 & \mathrm{otherwise}\end{array}\right.\\
 & \leq & d\left(q_{i}^{T},c_{j}^{T}\right)\end{array}\label{eq:lbg7}\end{equation}

\par\end{flushleft}

Since $d\left(q_{i}^{T},eg_{j}^{T}\right)\leq d\left(q_{i}^{T},c_{j}^{T}\right)$,
then

\begin{equation}
LBG(Q^{T},EG^{T})\leq LBS(Q^{T},C_{opt}^{T})\label{eq:lbg8}\end{equation}

Therefore, from Equation \ref{eq:lbg6}, we have

\begin{equation}
LBG(Q^{T},EG^{T})\leq DTW(Q,C_{opt})\label{eq:lbg9}\end{equation}

Q.E.D.
\end{proof}
Since LBG utilizes the concept of a lower bounding distance calculation
between a query and a group of sequences. We also propose a lower
bounding distance function extended from LB\_Keogh called LBG$_{\text{K}}$.
LBG$_{\text{K}}$ obtains lower bounding distance from a query sequence
$Q=\left\langle q_{1},\ldots,q_{i},\ldots,q_{n}\right\rangle $ and
an envelope $EG=\left\langle eg_{1},\ldots,eg_{i},\ldots,eg_{n}\right\rangle $,
where $eg_{i}=\left\{ ueg_{i},leg_{i}\right\} $. Given a query sequence
$Q$, an envelope $E$, and a global constraint $R=\left\langle r_{1},\ldots,r_{i},\ldots,r_{n}\right\rangle $.
LBG$_{\text{K}}$ first creates an envelope of global constraint $EGC=\left\langle egc_{1},\ldots,egc_{i}\right.$
$\left.,\ldots,egc_{n}\right\rangle $ from $EG$, where $egc_{i}=\left\{ uegc_{i},legc_{i}\right\} $.
Elements $uegc_{i}$ and $legc_{i}$ are calculated by $uegc_{i}=\max\left\{ ueg_{i-r_{i}},\ldots,ueg_{i+r_{i}}\right\} $
and $legc_{i}=\min\left\{ leg_{i-r_{i}},\ldots,\right.$ $\left.leg_{i+r_{i}}\right\} $,
respectively. The lower bounding distance $LBG_{K}(Q,EG)$ between
the query sequence $Q$ and the envelope $EG$ are determined by Equation
\ref{eq:lbg10} along with its proof of correctness. 

\begin{equation}
LBG_{K}(Q,EG)=\sqrt[p]{\overset{n}{\underset{i=1}{\sum}}\left\{ \begin{array}{cl}
\left|q_{i}-uegc_{i}\right|^{p} & \mathrm{if}\, c_{i}>uegc_{i}\\
\left|legc_{i}-q_{i}\right|^{p} & \mathrm{if}\, c_{i}<legc_{i}\\
0 & \mathrm{otherwise}\end{array}\right.}\label{eq:lbg10}\end{equation}

\noindent where \emph{$p$} is the dimension of $L_{p}$-norms. 
\begin{theorem}
Let $Q=\left\langle q_{1},\ldots,q_{i},\ldots,q_{n}\right\rangle $
be a query sequence and $EGC=$\linebreak{}
$\left\langle eg_{1},\ldots,eg_{i},\ldots,eg_{n}\right\rangle $
be an envelope of global constraint created from an envelope $EG$
of a group of sequences $\mathbb{C}=\left\{ C_{1},\ldots,C_{k},\ldots,C_{m}\right\} $,
where $C_{k}=\left\langle c_{k_{1}},\ldots,c_{k_{i}},\ldots,c_{k_{n}}\right\rangle $,
then 
\end{theorem}
\begin{equation}
LBG_{K}(Q,EG)\leq DTW(Q,C_{opt})\label{eq:lbg11}\end{equation}

\noindent where $C_{opt}$ is the sequence which gives the minimum
DTW distance to $Q$ in $\mathbb{C}$.
\begin{proof}
Since

\begin{equation}
DTW(Q,C_{opt})=\sqrt[p]{\overset{K}{\underset{k=1}{\sum}}d_{w_{k}}}\label{eq:proof1}\end{equation}

\noindent where $d_{w_{k}}$ is the $k^{\mathrm{th}}$ distance calculation
of sequence $Q$ and the nearest $C_{opt}$ in the optimal warping
path which calculates distance between $q_{i}$ and $c_{opt_{i}}$.

For $uegc_{i}$ and $legc_{i}$,

\begin{equation}
\begin{array}{ccc}
uegc_{i} & = & \underset{1-r\leq j\leq i+r}{\max}\left\{ ueg_{j}\right\} \\
 & = & \underset{1-r\leq j\leq i+r}{\max}\left\{ \underset{1\leq k\leq n}{\max}\left\{ c_{k_{j}}\right\} \right\} \\
 & \geq & c_{opt_{j}}\end{array}\label{eq:proof14}\end{equation}

\begin{equation}
\begin{array}{ccc}
legc_{i} & = & \underset{1-r\leq j\leq i+r}{\min}\left\{ leg_{j}\right\} \\
 & = & \underset{1-r\leq j\leq i+r}{\min}\left\{ \underset{1\leq k\leq n}{\min}\left\{ c_{k_{j}}\right\} \right\} \\
 & \leq & c_{opt_{j}}\end{array}\label{eq:proof15}\end{equation}

Since

\begin{equation}
\begin{array}{ccc}
LBG_{Keogh}(Q,EG) & \leq & DTW(Q,C_{opt}),\\
\\\sqrt[p]{\overset{n}{\underset{i=1}{\sum}}\left\{ \begin{array}{cl}
\left|q_{i}-uegc_{i}\right|^{p} & \mathrm{if}\, q_{i}>uegc_{i}\\
\left|legc_{i}-q_{i}\right|^{p} & \mathrm{if}\, q_{i}<legc_{i}\\
0 & \mathrm{otherwise}\end{array}\right.} & \leq & \sqrt[p]{\overset{K}{\underset{k=1}{\sum}}d_{w_{k}}}\end{array}\label{eq:lbg13}\end{equation}

Since $K\geq n$ from the DTW's conditions, there are three possible
cases, i.e., $\left|q_{i}-uegc_{i}\right|^{p}\leq d_{w_{k}}$, $\left|legc_{i}-q_{i}\right|^{p}\leq d_{w_{k}}$,
and $0\leq d_{w_{k}}$.

Suppose

\begin{equation}
\left|q_{i}-uegc_{i}\right|^{p}\leq d_{w_{k}},\label{eq:proof5}\end{equation}

\noindent DTW requires that, for $d_{w_{k}}$ and for all $i-r_{i}\leq j\leq i+r_{i}$,
each data point $q_{i}$ must be compared once with $c_{opt_{j}}$ 

\begin{equation}
\left|q_{i}-uegc_{i}\right|^{p}\leq\left|q_{i}-c_{opt_{j}}\right|^{p}\label{eq:proof7}\end{equation}

\begin{equation}
uegc_{i}\geq c_{opt_{j}}\label{eq:proof6}\end{equation}

The case $\left\Vert legc_{i}-q_{i}\right\Vert _{p}\leq d_{w_{k}}$
yields to a similar argument and $0\leq d_{w_{k}}$ always holds since
$d_{w_{k}}$ is nonnegative.

Hence, 

\begin{equation}
LBG_{Keogh}(Q,EG)\leq DTW(Q,C_{opt})\label{eq:lbg14}\end{equation}

Q.E.D.
\end{proof}

\subsection{Querying Process}

When a query sequence is issued, ESF is first accessed and lower bounding
distance from LBG for each envelope is calculated. Therefore, if LBG
for any DSF is larger than the best-so-far distance, all time series
sequences in that DSF are guaranteed not to be the answers. TWIST
could utilize this distance to prune off a significantly large number
of candidate sequences by using only a very small amount of both CPU
and I/O costs.

Instead of calculating only one level of lower bounding distance,
LBG calculates lower bounding distance iteratively. First, the best-so-far
distance is initialized with an LBG distance between the coarsest
segmented sequences of a query sequence and an envelope. Subsequently,
each finer envelope sequence is used by LBG calculation again and
again. If LBG distance is still smaller than the best-so-far distance,
the DSF is accessed, and all data sequences in DSF are then sequentially
searched. But if finer LBG is returned with anything larger than the
best-so-far distance, the next DSF is then considered. The process
is terminated when all envelope sequences in ESF are exhausted. The
pseudo code of TWIST with LBG is described in Table \ref{Flo:lbglbs-1-1}.

Although implementations of LBG and LBG$_{\text{K}}$ over TWIST are
different, we provide solutions for both. The advantages of LBG$_{\text{K}}$
over LBG are that LBG$_{\text{K}}$ requires to access ESF only once,
while LBG requires twice the access, and when the small global constraint
is applied in the querying, LBG$_{\text{K}}$ is faster. However,
LBG achieves a better query performance in terms of query processing
time than LBG$_{\text{K}}$ since LBG returns a tighter lower bounding
distance, independent of the global constraint.

To query with LBG$_{\text{K}}$ under top-$k$ querying, each envelope
sequence is sequentially retrieved, and its lower bounding distance
is calculated. Then LBG$_{\text{K}}$ distances are sorted into a
priority queue. DSF with smallest LBG$_{\text{K}}$ distance will
first be accessed. Then for each candidate sequence in the DSF, sequential
search is utilized to find the best-so-far sequence. Once the DSF
access is completed, the lower bounding distance from LBG$_{\text{K}}$
distance for the next DSF will then be considered. If the lower bounding
distance between the envelope of the next DSF is larger than the best-so-far
distance, the search is terminated, and a set of nearest-neighbor
sequences is returned. The pseudo code is provided in Table \ref{Flo:lbgkeogh}.

\begin{table}
\caption{Top-$k$ querying under TWIST with LBG}

\noindent \centering{}\begin{tabular}{cl}
\toprule 
\multicolumn{2}{l}{\noun{Algorithm}{[}$\mathbb{C}${]} = \noun{LBG Top-k Querying} {[}$Q,k${]}}\tabularnewline
\midrule
1 & Let: \tabularnewline
2 & ~~~$\mathbb{C}$ be a priority queue of answer sequences\tabularnewline
3 & ~~~$P$ be a pointer to DSF\tabularnewline
4 & ~~~$EG$ be an envelope\tabularnewline
5 & ~~~$d_{best}$ = \noun{PositiveInfinite} be the best-so-far distance\tabularnewline
6 & ~~~$T$ be the coarsest resolution\tabularnewline
7 & for all $\left\{ P,EG\right\} $ in ESF // Finding $d_{best}$ from
the coarsest version of ESF\tabularnewline
8 & ~~~$d_{EG}=LBW_{LBS}(Q^{T},EG^{T})$\tabularnewline
9 & ~~~if ($d_{EG}<d_{best}$) $d_{best}=d_{EG}$ endif\tabularnewline
10 & endfor\tabularnewline
11 & for all $\left\{ P,W\right\} $ in ESF\tabularnewline
12 & ~~~while ($T$ is not the finest resolution) // Use $LBW_{LBS}$
to prune $ESF$\tabularnewline
13 & ~~~~~~$d_{W}=LBW_{LBS}(Q^{T},EG^{T})$\tabularnewline
14 & ~~~~~~if ($d_{W}>d_{best}$) Break and go to the next $\left\{ P,EG\right\} $
endif\tabularnewline
15 & ~~~~~~Set $T$ to be a finer resolution\tabularnewline
16 & ~~~endwhile\tabularnewline
17 & ~~~for all $C$ in $DSF_{P}$\tabularnewline
18 & ~~~~~~$d_{lower}=LB(Q,C)$\tabularnewline
19 & ~~~~~~if ($d_{lower}\leq d_{best}$)\tabularnewline
20 & ~~~~~~~~~$d_{true}=DTW(Q,C)$\tabularnewline
21 & ~~~~~~~~~if ($\mathbb{C}.size()<k$)\tabularnewline
22 & ~~~~~~~~~~~~$\mathbb{C}.enqueue\left(\left\{ C,d_{true}\right\} \right)$\tabularnewline
23 & ~~~~~~~~~else\tabularnewline
24 & ~~~~~~~~~~~~if ($d_{true}\leq d_{best}$)\tabularnewline
25 & ~~~~~~~~~~~~~~~$\mathbb{C}.enqueue\left(\left\{ C,d_{true}\right\} \right)$\tabularnewline
26 & ~~~~~~~~~~~~~~~$\mathbb{C}.dequeue()$\tabularnewline
27 & ~~~~~~~~~~~~~~~$d_{best}=\mathbb{C}.peek().d_{true}$\tabularnewline
28 & ~~~~~~~~~~~~endif\tabularnewline
29 & ~~~~~~~~~endif\tabularnewline
30 & ~~~~~~endif\tabularnewline
31 & ~~~endfor\tabularnewline
32 & endfor\tabularnewline
33 & Return $\mathbb{C}$\tabularnewline
\bottomrule
\end{tabular}\label{Flo:lbglbs-1-1}
\end{table}

\clearpage{}

\begin{table}
\caption{Top-$k$ querying under TWIST with LBG$_{\text{K}}$}
\label{Flo:lbgkeogh}

\noindent \centering{}\begin{tabular}{cl}
\toprule 
\multicolumn{2}{l}{\noun{Algorithm}{[}$\mathbb{C}${]} = LBG$_{\text{K}}$\noun{ Top-k
Querying} {[}$Q,k${]}}\tabularnewline
\midrule
1 & Let: \tabularnewline
2 & ~~~$\mathbb{W}$ be a priority queue of envelope distances\tabularnewline
3 & ~~~$\mathbb{C}$ be a priority queue of answer sequences\tabularnewline
4 & ~~~$P$ be a pointer to DSF\tabularnewline
5 & ~~~$EG$ be an envelope\tabularnewline
6 & ~~~$d_{best}$ = \noun{PositiveInfinite} be the best-so-far distance\tabularnewline
7 & Initialize $d_{best}$ = \noun{PositiveInfinite}\tabularnewline
8 & for all $\left\{ P,EG\right\} $ in ESF // Calculate LBG distance
from ESF for all DSF\tabularnewline
9 & ~~~$d_{W}=LBW_{K}(Q,W)$\tabularnewline
10 & ~~~$\mathbb{W}.enqueue\left(\left\{ P,d_{W}\right\} \right)$ \tabularnewline
11 & endfor\tabularnewline
12 & // Dequeue $\left\{ P,d_{W}\right\} $ with smallest $d_{W}$ \tabularnewline
13 & // keep searching for an answer while $d_{W}\leq d_{best}$\tabularnewline
14 & while ($\left\{ P,d_{W}\right\} =\mathbb{W}.dequeue()$ and $d_{W}\leq d_{best}$)\tabularnewline
15 & ~~~for all $C$ in $DSF_{P}$\tabularnewline
16 & ~~~~~~$d_{lower}=LB(Q,C)$\tabularnewline
17 & ~~~~~~if ($d_{lower}\leq d_{best}$)\tabularnewline
18 & ~~~~~~~~~$d_{true}=DTW(Q,C)$\tabularnewline
19 & ~~~~~~~~~if ($\mathbb{C}.size()<k$)\tabularnewline
20 & ~~~~~~~~~~~~$\mathbb{C}.enqueue\left(\left\{ C,d_{true}\right\} \right)$\tabularnewline
21 & ~~~~~~~~~else\tabularnewline
22 & ~~~~~~~~~~~~if ($d_{true}\leq d_{best}$)\tabularnewline
23 & ~~~~~~~~~~~~~~~$\mathbb{C}.enqueue\left(\left\{ C,d_{true}\right\} \right)$\tabularnewline
24 & ~~~~~~~~~~~~~~~$\mathbb{C}.dequeue()$\tabularnewline
25 & ~~~~~~~~~~~~~~~$d_{best}=\mathbb{C}.peek().d_{true}$\tabularnewline
26 & ~~~~~~~~~~~~endif\tabularnewline
27 & ~~~~~~~~~endif\tabularnewline
28 & ~~~~~~endif\tabularnewline
29 & ~~~endfor\tabularnewline
30 & endwhile\tabularnewline
31 & Return $\mathbb{C}$\tabularnewline
\bottomrule
\end{tabular}
\end{table}

Although this paper emphasizes on top-$k$ querying, range query can
simply be adapted. Instead of using the best-so-far distance to prune
off the database, the range distance is used to specify the maximum
distance between a query sequence and a candidate sequence. In addition,
an integer $k$ is set to be positive infinite.

\subsection{Indexing Process}

To maintain a data structure, we also propose a machanism to efficiently
insert and delete data sequences over our proposed index structure
TWIST.

\subsubsection{Data Sequence Insertion}

In case of insertion, suppose there exist DSFs and ESF, cost of insertion
between a new sequence and an envelope is computed for all envelopes
in ESF, the new sequence will be in the minimum cost envelope. After
the minimum-cost envelope has been found, the envelope's DSF is accessed,
and the new sequence is added. The envelope is updated accordingly
to the ESF. Generally, the cost is computed from the size of an envelope
after insertion. If DSF exceeds the maximum number of sequences per
file (maximum page size), TWIST splits this DSF into two DSFs, and
two new envelopes are also generated and stored in the ESF. For clarification,
we provide the insertion algorithm in Table \ref{Flo:insertion}.
Note that the maximum page size is a user-defined parameter which
determines a maximum number of sequences within each DSF.

\begin{table}[h]
\caption{Inserting a new sequence to TWIST}

\noindent \centering{}\begin{tabular}{cl}
\toprule 
\multicolumn{2}{l}{\noun{Algorithm} \noun{Insertion} {[}$C${]}}\tabularnewline
\midrule
1 & // Find the minimum-cost DSF\tabularnewline
2 & Initialize $cost_{min}$ = \noun{PositiveInfinite}, $P_{min}$ = \noun{null}\tabularnewline
3 & for all $\left\{ P,EG_{P}\right\} $ in ESF\tabularnewline
4 & ~~~$cost_{EG}=Cost(EG_{P},C)$\tabularnewline
5 & ~~~if ($cost_{EG}<cost_{min}$)\tabularnewline
6 & ~~~~~~$cost_{min}=cost_{EG}$\tabularnewline
7 & ~~~~~~$P_{min}=P$\tabularnewline
8 & ~~~endif\tabularnewline
9 & endfor\tabularnewline
10 & Add $C$ in $DSF_{P_{min}}$ \tabularnewline
11 & // Check if the size of $DSF_{P_{min}}$ exceeds $\alpha$\tabularnewline
12 & if ($DSF_{P_{min}}.size()>\alpha$)\tabularnewline
13 & ~~~// Split $DSF_{P_{min}}$ into two DSFs, $DSF_{X}$ and $DSF_{Y}$\tabularnewline
14 & ~~~$\left[\left\{ S,EG_{S}\right\} ,\left\{ T,EG_{T}\right\} \right]=SplitDSF(DSF_{P_{min}})$\tabularnewline
15 & ~~~Delete $\left\{ P_{min},EG_{P_{min}}\right\} $ from ESF\tabularnewline
16 & ~~~Add $\left\{ X,EG_{T}\right\} ,\left\{ X,EG_{Y}\right\} $ to
ESF\tabularnewline
17 & else\tabularnewline
18 & ~~~// Update $EG_{P_{min}}$ from $C$\tabularnewline
19 & ~~~$EG_{P_{min}}=UpdateEnvelope(EG_{P_{min}},C)$\tabularnewline
20 & ~~~Update $\left\{ P_{min},EG_{P_{min}}\right\} $ to ESF\tabularnewline
21 & endif\tabularnewline
\bottomrule
\end{tabular}\label{Flo:insertion}
\end{table}
\begin{figure}[h]
\noindent \begin{centering}
\includegraphics[width=7cm]{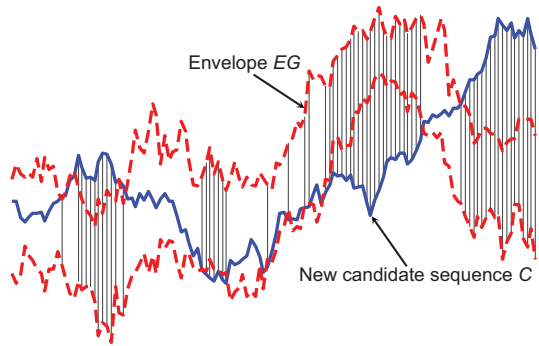}
\par\end{centering}

\caption{Shadowed area represents total cost of insertion between a sequence
$C$ and an envelope $EG$}

\label{Flo:cost-1}
\end{figure}
\begin{table}[h]
\caption{Cost function for an insertion of a sequence $C$ into $EG$}

\noindent \centering{}\begin{tabular}{cl}
\toprule 
\multicolumn{2}{l}{\noun{Algorithm Cost} {[}$EG,C${]}}\tabularnewline
\midrule
1 & Let: \tabularnewline
2 & ~~~$cost_{sum}=0$\tabularnewline
3 & for each $c_{i},ueg_{i},leg_{i}$\tabularnewline
4 & ~~~if ($c_{i}>ueg_{i}$)\tabularnewline
5 & ~~~~~~$cost_{sum}=cost_{sum}+\left|c_{i}-leg_{i}\right|^{p}$\tabularnewline
6 & ~~~else if ($c_{i}<leg_{i}$)\tabularnewline
7 & ~~~~~~$cost_{sum}=cost_{sum}+\left|ueg_{i}-c_{i}\right|^{p}$\tabularnewline
8 & ~~~endif\tabularnewline
9 & Return $cost_{sum}$\tabularnewline
\bottomrule
\end{tabular}\label{Flo:cost}
\end{table}
Generally, the cost function is calculated from total area of an envelope
after a new sequence is inserted. To be more illustrative, the shadowed
areas in Figure \ref{Flo:cost-1} indicate the cost of insertion.
Given a new time series sequence $C=\left\langle c_{1},\ldots,c_{i},\ldots,c_{n}\right\rangle $
and an envelope $EG=\left\langle eg_{1},\ldots,eg_{i},\ldots,eg_{n}\right\rangle $,
where $eg_{i}=\left\{ ueg_{i},leg_{i}\right\} $, the cost function
$Cost(EG,C)$ is defined as (also shown in Table \ref{Flo:cost}).

\begin{equation}
Cost(EG,C)=\sqrt[p]{\overset{n}{\underset{i=1}{\sum}}\left\{ \begin{array}{cc}
\left|c_{i}-leg_{i}\right|^{p} & \mathrm{if}\, c_{i}>ueg_{i}\\
\left|ueg_{i}-c_{i}\right|^{p} & \mathrm{if}\, c_{i}<leg_{i}\\
0 & \mathrm{otherwise}\end{array}\right.}\label{eq:insertion1}\end{equation}

\noindent where \emph{$p$} is the dimension of $L_{p}$-norms. 

If the number of sequence in DSF exceeds the maximum page size, the
DSF needs to split into two DSFs to reduce the envelope size. Generally,
TWIST tries to split sequences into two groups so that each new envelope
sequence is tight and has only small overlaps. In this paper, $k$-means
clustering \citep{kmeans} ($k=2$) with Euclidean distance is adopted
as a heuristic function for separating the data into two appropriate
groups. However, other algorithms such as splitting algorithms in
R-tree \citep{Guttman84} and R{*}-tree \citep{BeckmannKSS90} can
be used in place of $k$-means clustering algorithm since splitting
algorithms are also designed to separate and minimize Minimum Bounding
Rectangle (MBR); however, these splitting algorithms require relatively
large time complexity. Pseudo code of the splitting algorithm is provide
in Table \ref{Flo:split}.

\begin{table}[h]
\caption{Splitting algorithm, separating a DSF into two DSFs}

\noindent \centering{}\begin{tabular}{cl}
\toprule 
\multicolumn{2}{l}{\noun{Algorithm SplitDSF} {[}$DSF${]}}\tabularnewline
\midrule
1 & // Run $k$-means clustering algorithm \tabularnewline
2 & // Fix $k=2$\tabularnewline
3 & $\left[DSF_{X},DSF_{Y}\right]=KMeans(DSF)$\tabularnewline
4 & // Create $EG_{X}$ and $EG_{Y}$\tabularnewline
5 & $EG_{X}=CreateEnvelope(DSF_{X})$\tabularnewline
6 & $EG_{Y}=CreateEnvelope(DSF_{Y})$\tabularnewline
7 & Return $\left[\left\{ X,EG_{X}\right\} ,\left\{ Y,EG_{Y}\right\} \right]$\tabularnewline
\bottomrule
\end{tabular}\label{Flo:split}
\end{table}

\begin{table}[h]
\caption{An envelope construction algorithm}

\noindent \centering{}\begin{tabular}{cl}
\toprule 
\multicolumn{2}{l}{\noun{Algorithm CreateEnvelope} {[}$DSF${]}}\tabularnewline
\midrule
1 & Let: \tabularnewline
2 & ~~~$EG$ be an envelope\tabularnewline
3 & for each sequence $C$ in $DSF$\tabularnewline
4 & ~~~for each $c_{i},ueg_{i},leg_{i}$\tabularnewline
5 & ~~~~~~$ueg_{i}=\max\left\{ ueg_{i},c_{i}\right\} $\tabularnewline
6 & ~~~~~~$leg_{i}=\min\left\{ leg_{i},c_{i}\right\} $\tabularnewline
7 & ~~~endfor\tabularnewline
8 & endfor\tabularnewline
9 & Return $EG$\tabularnewline
\bottomrule
\end{tabular}\label{Flo:create}
\end{table}

\begin{table}[h]
\caption{An envelope sequence update algorithm after a new sequence insertion}

\noindent \centering{}\begin{tabular}{cl}
\toprule 
\multicolumn{2}{l}{\noun{Algorithm UpdateEnvelope} {[}$EG,C${]}}\tabularnewline
\midrule
1 & for each $c_{i},ueg_{i},leg_{i}$\tabularnewline
2 & ~~~$ueg_{i}=\max\left\{ ueg_{i},c_{i}\right\} $\tabularnewline
3 & ~~~$leg_{i}=\min\left\{ leg_{i},c_{i}\right\} $\tabularnewline
4 & endfor\tabularnewline
5 & Return $EG$\tabularnewline
\bottomrule
\end{tabular}\label{Flo:update}
\end{table}

After new DSFs are created in the insertion step, new envelopes are
generated by an algorithm described in Table \ref{Flo:create} by
finding the maximum and minimum values for each DSF. If the number
of sequences in DSF exceeds the maximum allowed, the envelope in ESF
is simply updated using the existing envelope and a new sequence.
To update the existing envelope $EG=\left\langle eg_{1},\ldots,eg_{i},\ldots,eg_{n}\right\rangle $
from a new time series sequence $C=\left\langle c_{1},\ldots,c_{i},\ldots,c_{n}\right\rangle $,
elements are updated by $ueg_{i}=\max\left\{ ueg_{i},c_{i}\right\} $
and $leg_{i}=\min\left\{ leg_{i},c_{i}\right\} $, where $eg_{i}=\left\{ ueg_{i},leg_{i}\right\} $.
The updating algorithm is described in Table \ref{Flo:update}.

\subsubsection{Data Sequence Deletion}

To delete a data sequence, coresponding DSF is accessed and the sequence
is simply deleted. However, when DSF changes, ESF needs to be updated
as well. In particular, we provide two deletion policies, i.e., eager
deletion and lazy deletion. For eager deletion, after each sequence
deletion, TWIST immediately recalculates a new envelope from the entire
set of sequences in that DSF, and updates the changes into the ESF.
On the other hand, lazy deletion simply deletes a sequence from DSF
without the need of ESF update since TWIST guarantees that false dismissals
will never occur in the lower bounding calculation of LBG. The treadeoffs
are, of course, a deletion time and the tightness of an envelope between
these two deletion policies. If eager deletion is applied, the deletion
time increases but its envelope sequence is tighter, while the deletion
time is very fast in lazy deletion, but the envelope sequence is not
as tight. We provide a pseudo code for the deletion algorithm in Table
\ref{Flo:deletion}.

\begin{table}[h]
\caption{Delete an existing sequence from TWIST}

\noindent \centering{}\begin{tabular}{cl}
\toprule 
\multicolumn{2}{l}{\noun{Algorithm Deletion} {[}$C${]}}\tabularnewline
\midrule
1 & Select $DSF_{P}$ which contains $C$\tabularnewline
2 & Delete $C$ from $DSF_{P}$\tabularnewline
3 & if (\noun{IsEager})\tabularnewline
4 & ~~~$EG_{P}=CreateEnvelope(DSF_{P})$\tabularnewline
5 & ~~~Update $\left\{ P,EG_{P}\right\} $ to ESF\tabularnewline
6 & endif\tabularnewline
\bottomrule
\end{tabular}\label{Flo:deletion}
\end{table}

\section{Experimental Evaluation}

In experimental evaluation, we compare our proposed method, TWIST,
with the best existing indexing method, FTW \citep{sakurai2005ffs},
and the best naïve method, sequential search with LB\_Keogh \citep{keogh2005eid},
in many evaluation metrics, i.e., querying time, indexing time, the
number of page accesses, and storage requirement. In addition, two
solutions of our proposed method are evaluated, i.e., TWIST with LBG
and TWIST with LBG$_{\text{K}}$. Although FTW indexing outperforms
R{*}-tree with LB\_PAA \citep{keogh2005eid}, our method shows superiority
over FTW by few orders of magnitude. In addition, sequential search
with LB\_Keogh is also evaluated to show the best performance of naïve
method when no indexing structure is utilized. It is important to
note that we make our best effort in tuning the rival methods to run
at their best performances by applying early abandon \citep{keogh2005eid}
and early stopping \citep{sakurai2005ffs} techniques; however, as
will be demonstrated, our proposed method still outperforms them in
all terms. 

To verify that our proposed method is scalable for massive time series
database, we use a database with the size exceeding the main memory.
Otherwise, the operating system is likely to cache the data into the
main memory. Therefore, our database size ranges from 256MB to 4 GB.
We perform our experiments on a Windows-XP computer with Intel Core
2 Duo 2.77 GHz, 2GB of RAM, and 80 GB of 5400 rpm internal hard drive.
All codes in our experiments are implemented with Java 1.5.

\begin{figure}
\noindent \begin{centering}
\begin{tabular}{cc}
\includegraphics[width=5.5cm]{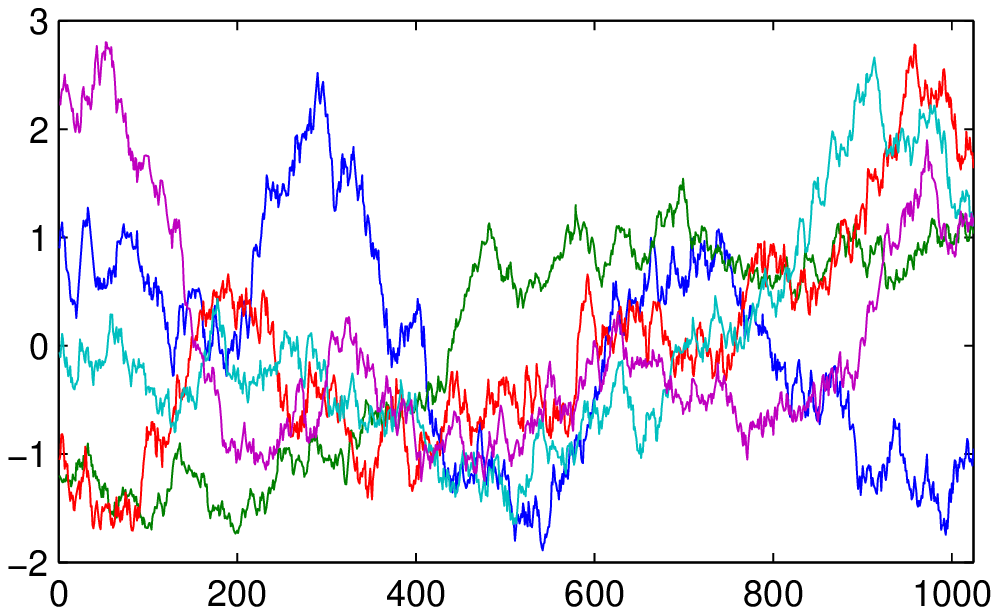} & \includegraphics[width=5.5cm]{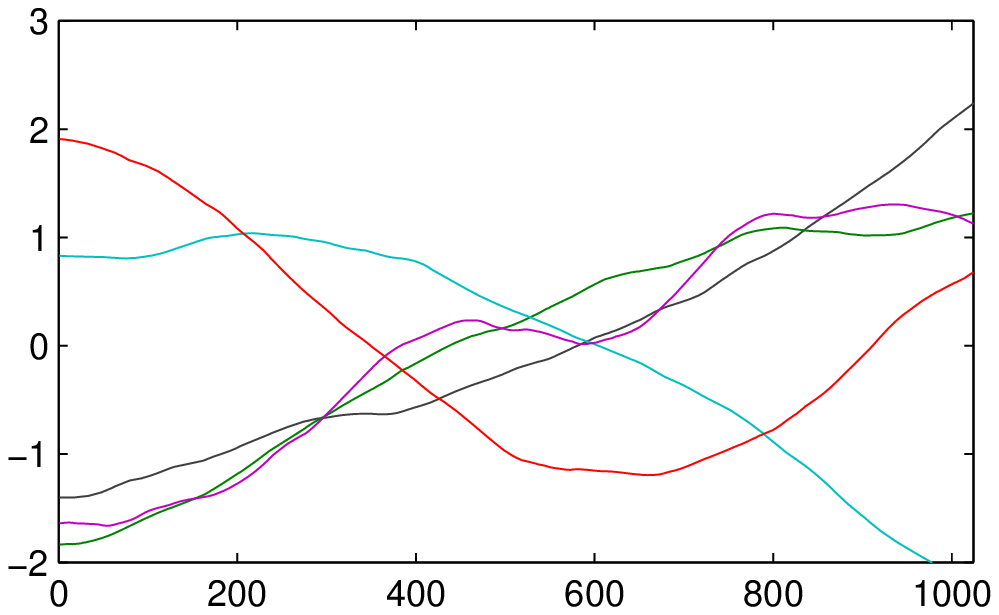}\tabularnewline
(a) & (b)\tabularnewline
 & \tabularnewline
\multicolumn{2}{c}{\includegraphics[width=5.5cm]{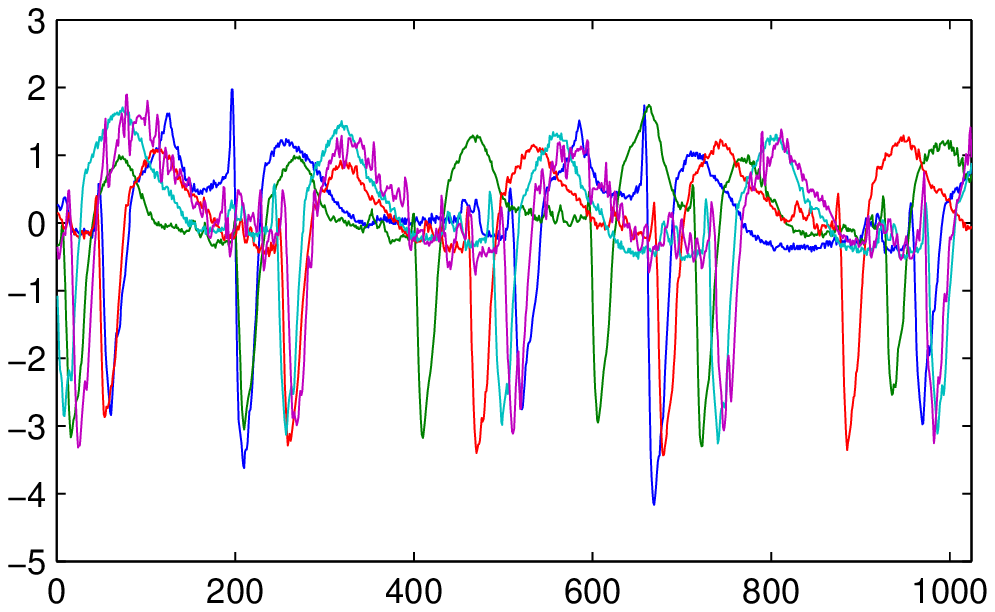}}\tabularnewline
\multicolumn{2}{c}{(c)}\tabularnewline
\end{tabular}
\par\end{centering}

\caption{Example sequences for datasets a) Random Walk I, b) Random Walk II,
and c) Electrocardiogram}
\label{Flo:datasets}
\end{figure}

\subsection{Datasets}

To visualize the performance in various dimensions, many different
datasets listed below are generated by varying the numbers of sequences
in the databases ($2^{16}=65536$, $2^{17}=131072$, $2^{18}=262144$,
and $2^{19}=524288$ sequences) and the sequence lengths (512, 1024,
and 2048 data points). All data sequences are Z-normalized; some examples
for each dataset are shown in Figure \ref{Flo:datasets}.
\begin{enumerate}
\item Random Walk I \citep{sakurai2005ffs,AssentKAS08}: To demonstrate
the scalability of our proposed method, a large amount of sequences
are generated by a following equation: $t_{i+1}=t_{i}+N(0,1)$, where
$N(0,1)$ is a random value drawn from a normal distribution.
\item Random Walk II \citep{AssentKAS08}: We generate a set of random walk
sequences from a following equation: $t_{i+1}=2t_{i}-t_{i-1}+N(0,1)$,
where $N(0,1)$ is a random value drawn from a normal distribution.
\item Electrocardiogram \citep{electro}: This dataset is recorded from
human subjects with atrial fibrillation which has 250 samples per
second. In addition, this dataset was made at Boston's Beth Israel
Hospital and revised for MIT-BIH Arrhythmia Database. To build the
dataset, we segment all the original sequences into small subsequences.
\end{enumerate}

\subsection{Querying Time\label{sub:Querying-Time}}

\begin{figure}
\noindent \begin{centering}
\begin{tabular}{cc}
\includegraphics[width=4.8cm]{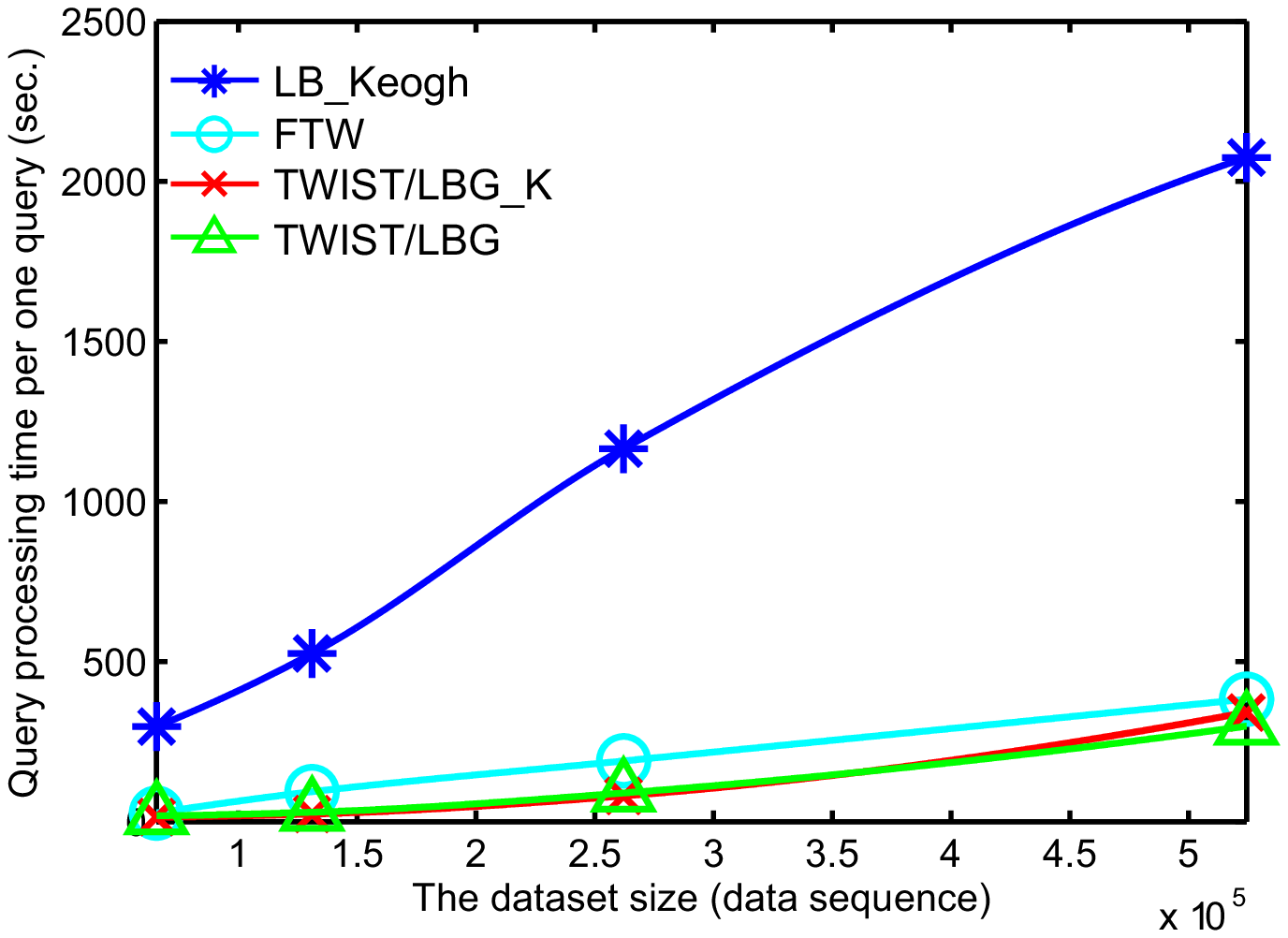} & \includegraphics[width=4.8cm]{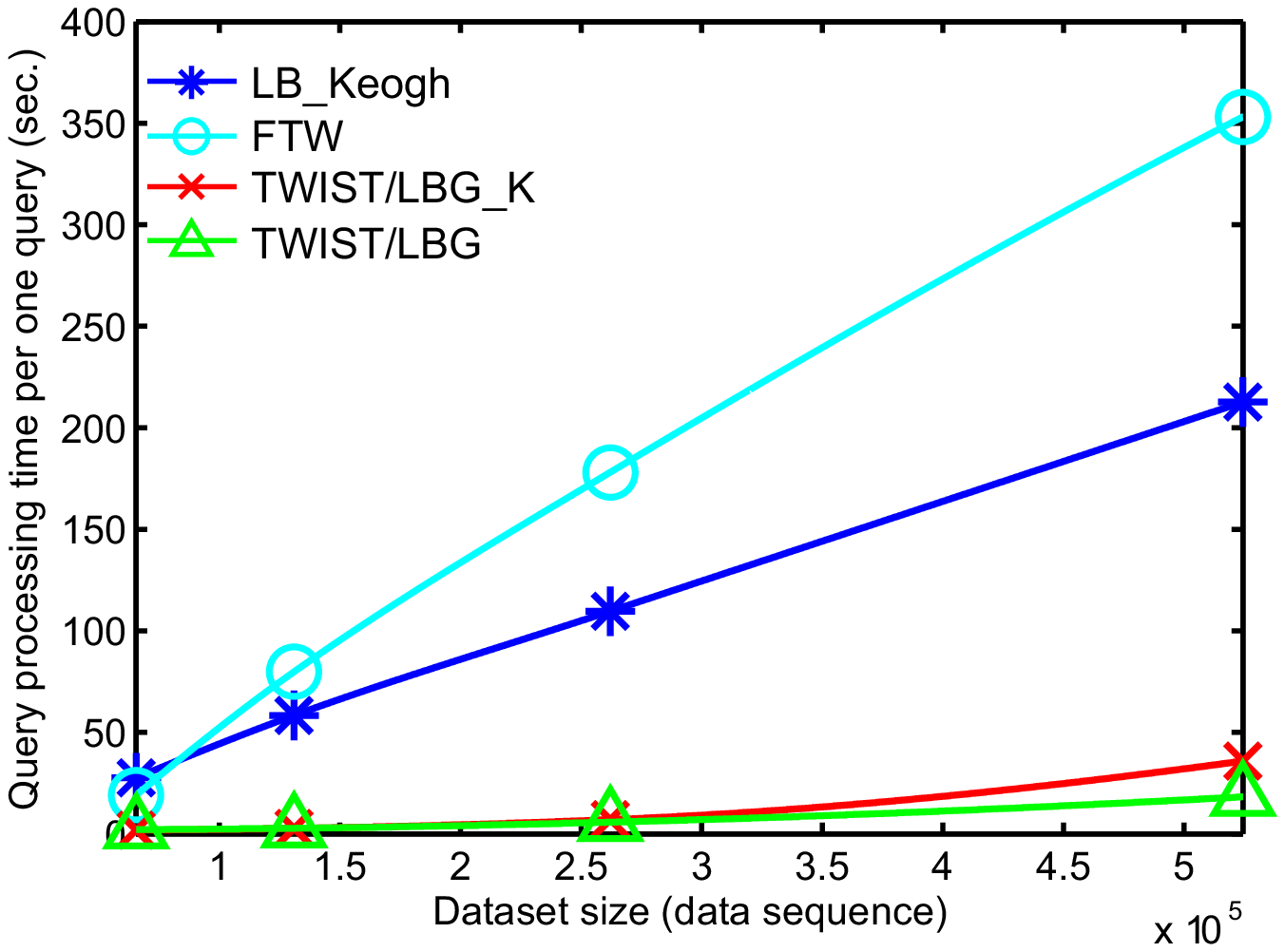}\tabularnewline
(a) Random Walk I & (b) Random Walk II\tabularnewline
\multicolumn{2}{c}{\includegraphics[width=4.8cm]{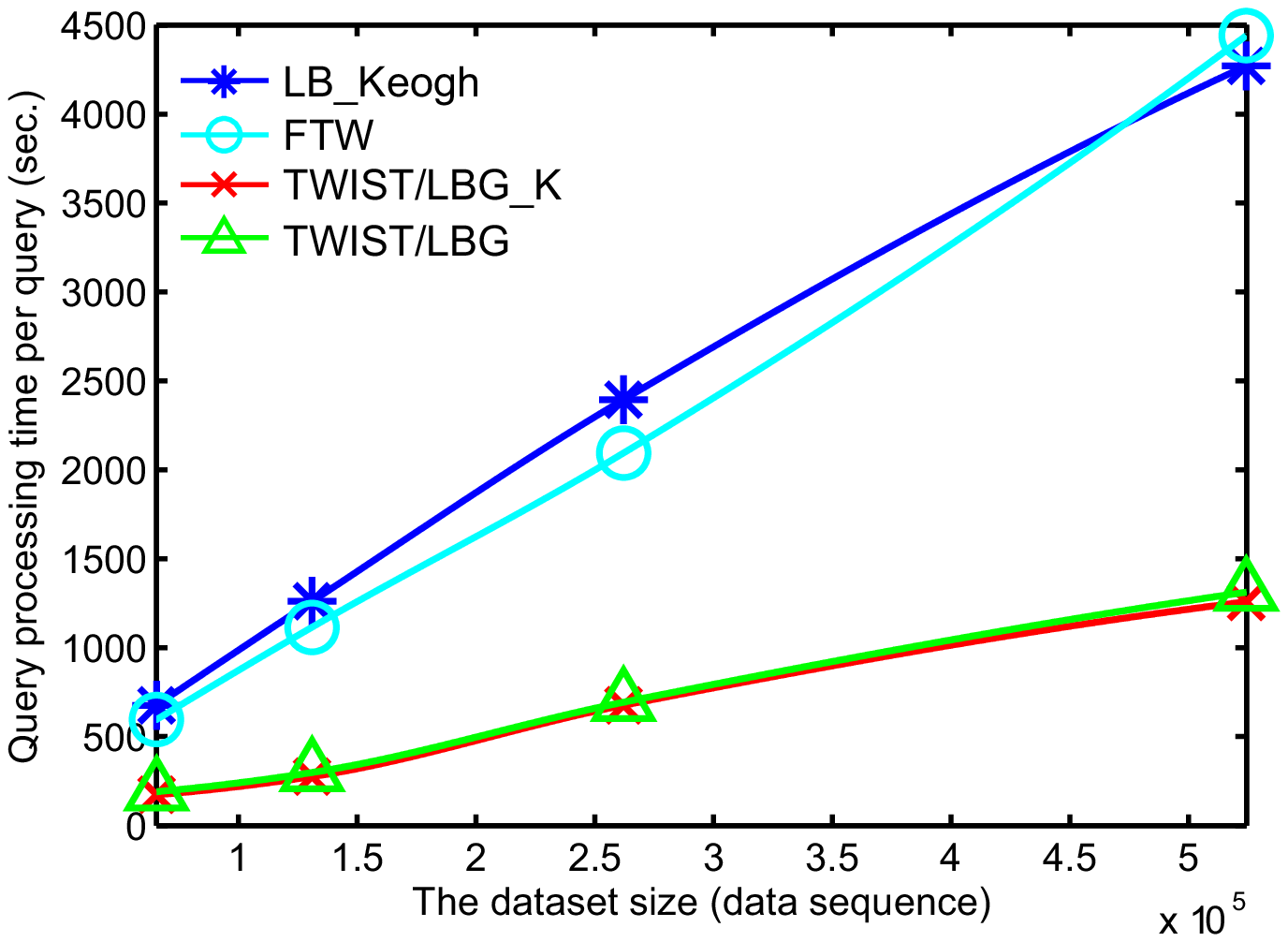}}\tabularnewline
\multicolumn{2}{c}{(c) Electrocardiogram}\tabularnewline
\end{tabular}
\par\end{centering}

\caption{TWIST outperforms the rival methods, and is slightly affected by an
increase in the dataset size, where sequence length, global constraint,
an integer $k$, and page size are set to 2048, 10\%, 1, and 128,
respectively}
\label{Flo:exp1}
\end{figure}

\begin{figure}
\noindent \begin{centering}
\begin{tabular}{cc}
\includegraphics[width=4.8cm]{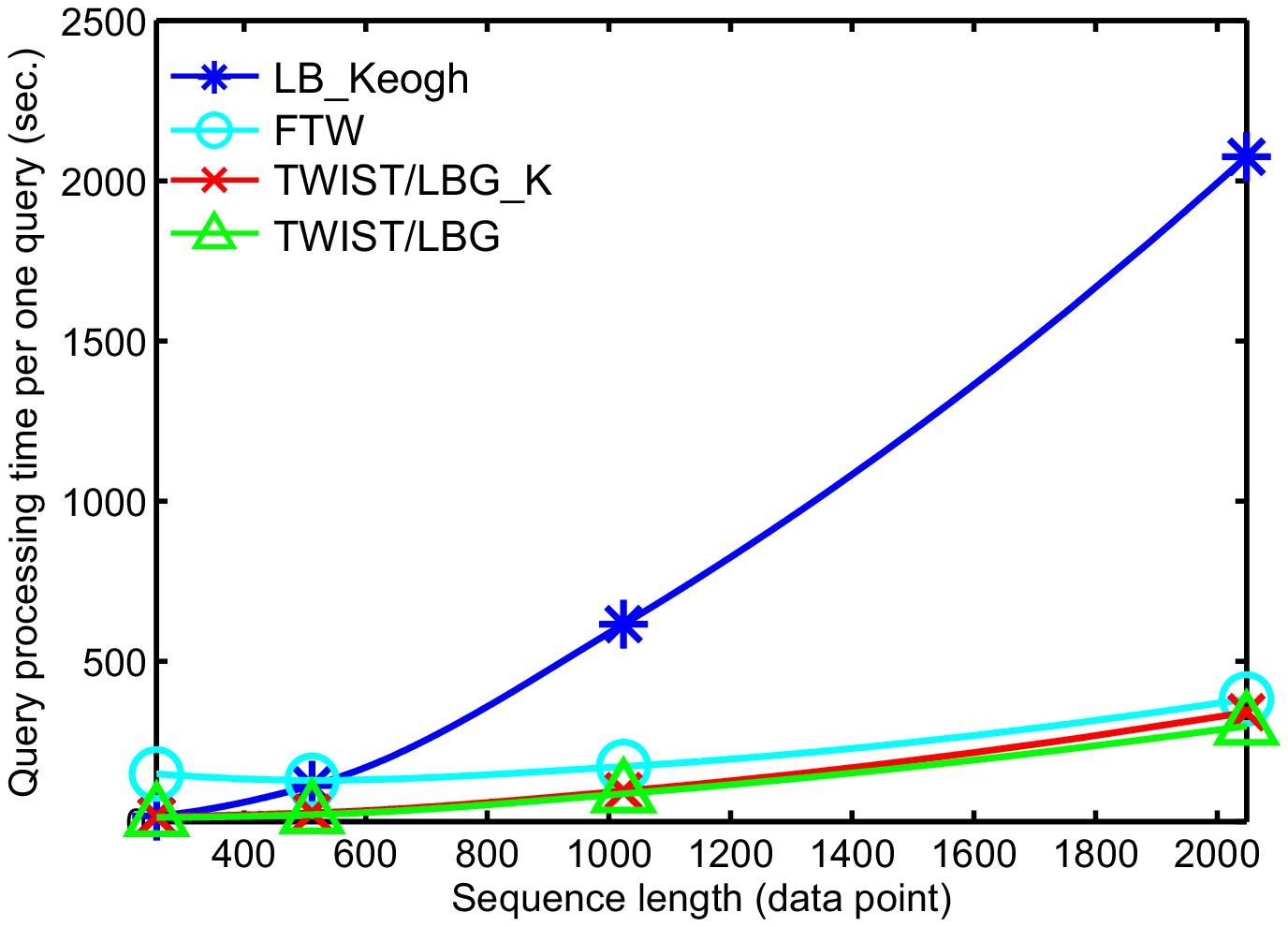} & \includegraphics[width=4.8cm]{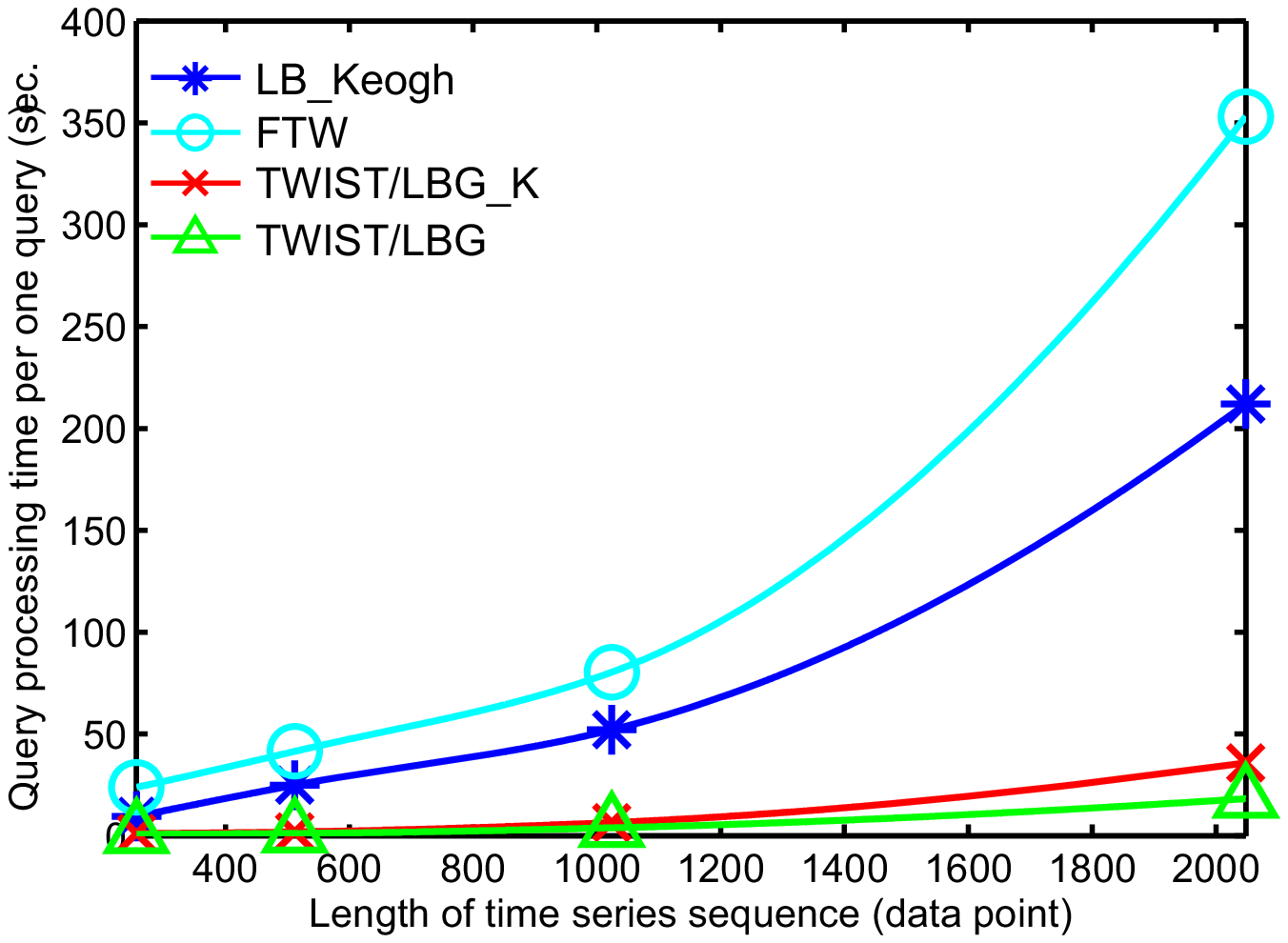}\tabularnewline
(a) Random Walk I & (b) Random Walk II\tabularnewline
\multicolumn{2}{c}{\includegraphics[width=4.8cm]{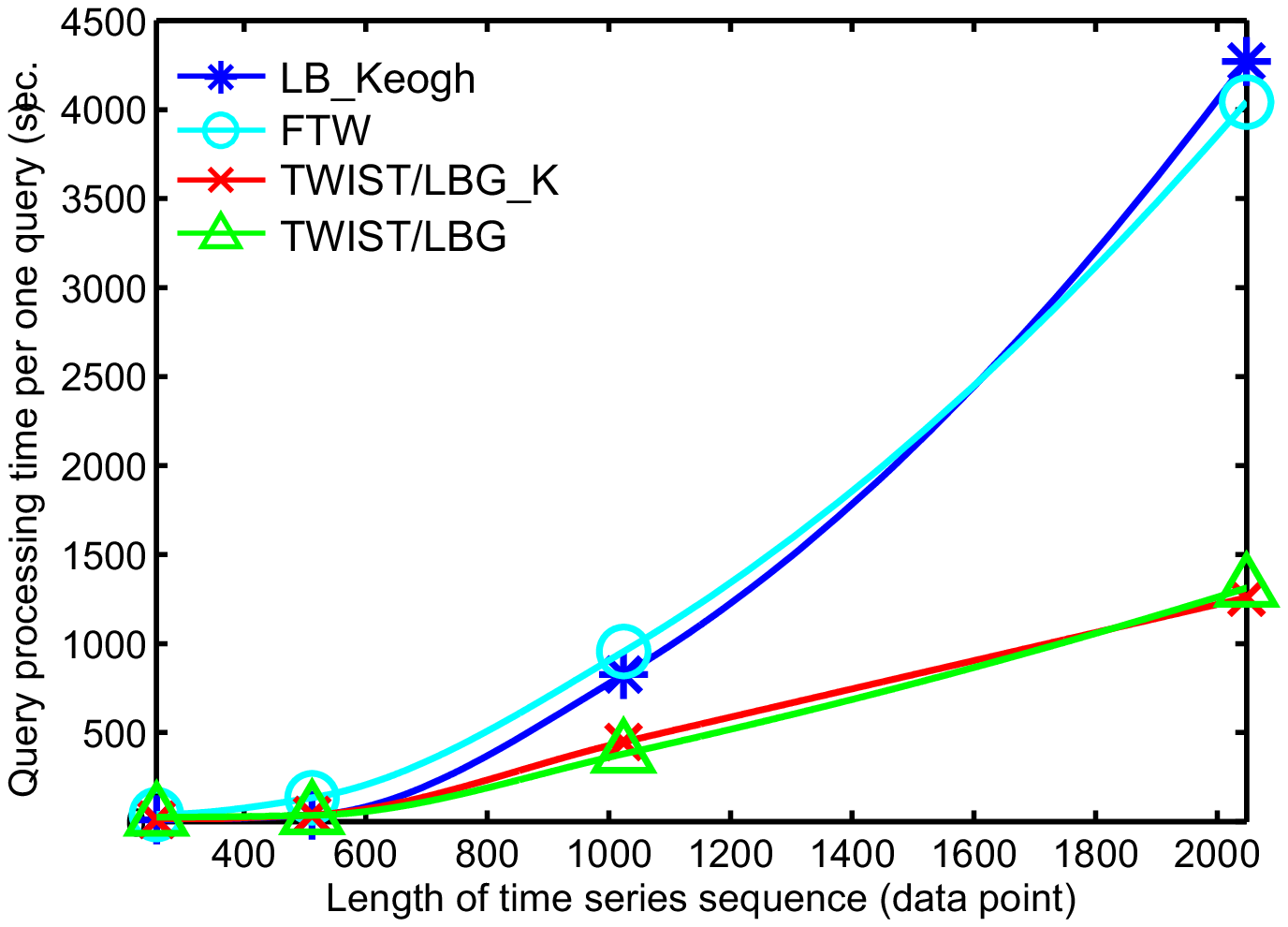}}\tabularnewline
\multicolumn{2}{c}{(c) Electrocardiogram}\tabularnewline
\end{tabular}
\par\end{centering}

\caption{Although sequence length increases, TWIST requires only small query
processing time comparing with FTW and LB\_Keogh, where database size,
global constraint, an integer $k$, and page size are set to 524288,
10\%, 1, and 128, respectively}
\label{Flo:exp2}
\end{figure}

\begin{figure}
\noindent \begin{centering}
\begin{tabular}{cc}
\includegraphics[width=4.8cm]{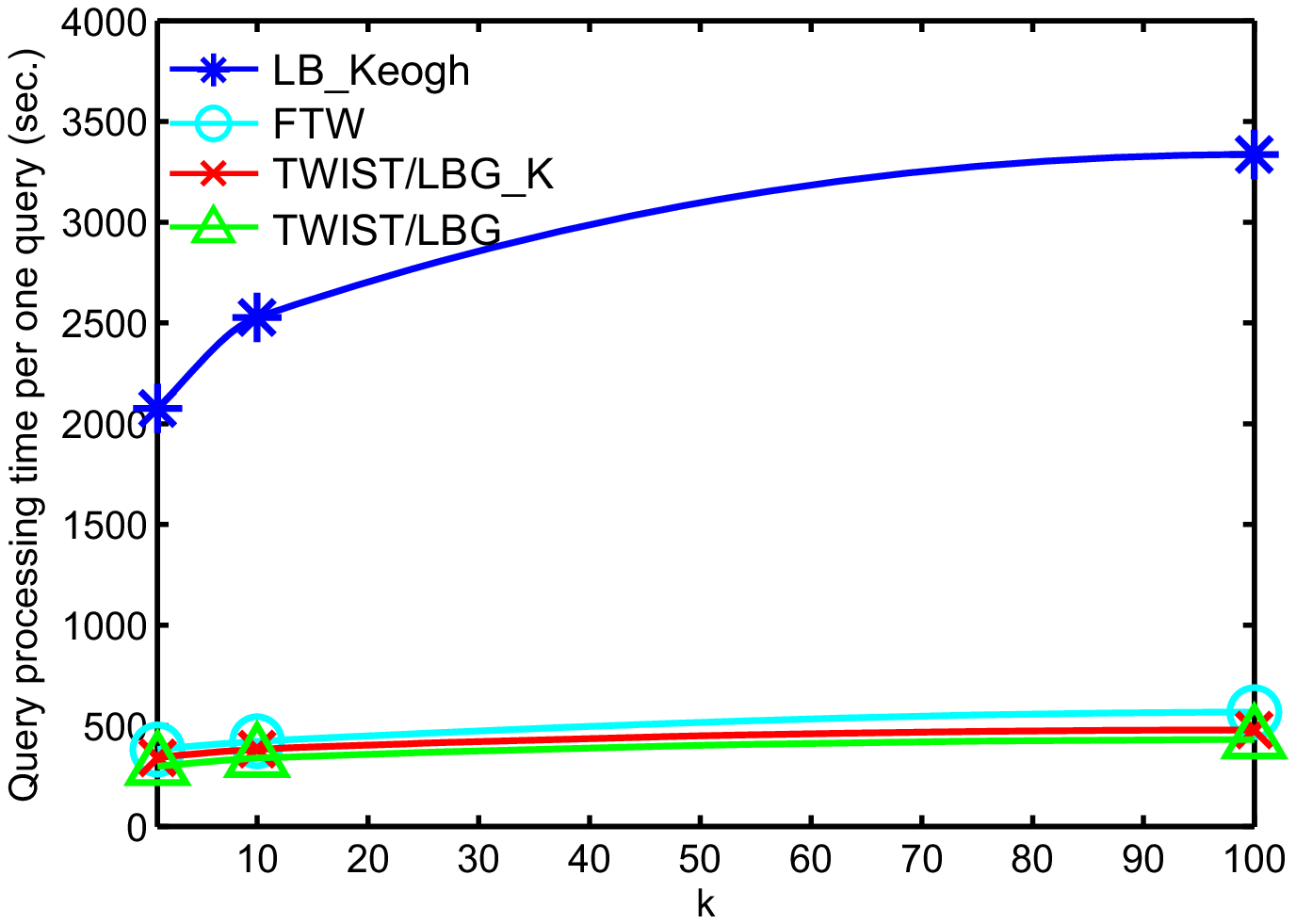} & \includegraphics[width=4.8cm]{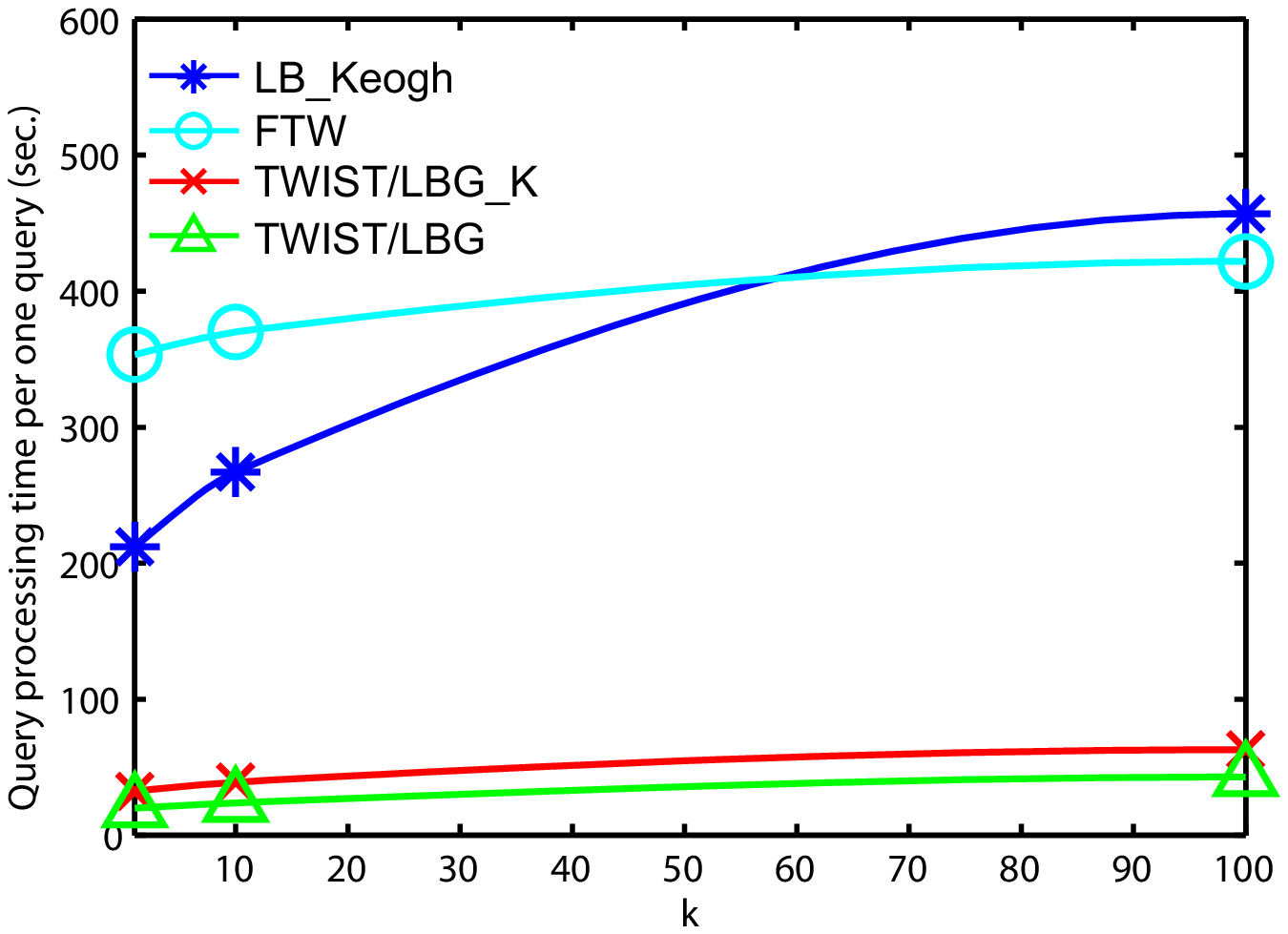}\tabularnewline
(a) Random Walk I & (b) Random Walk II\tabularnewline
\multicolumn{2}{c}{\includegraphics[width=4.8cm]{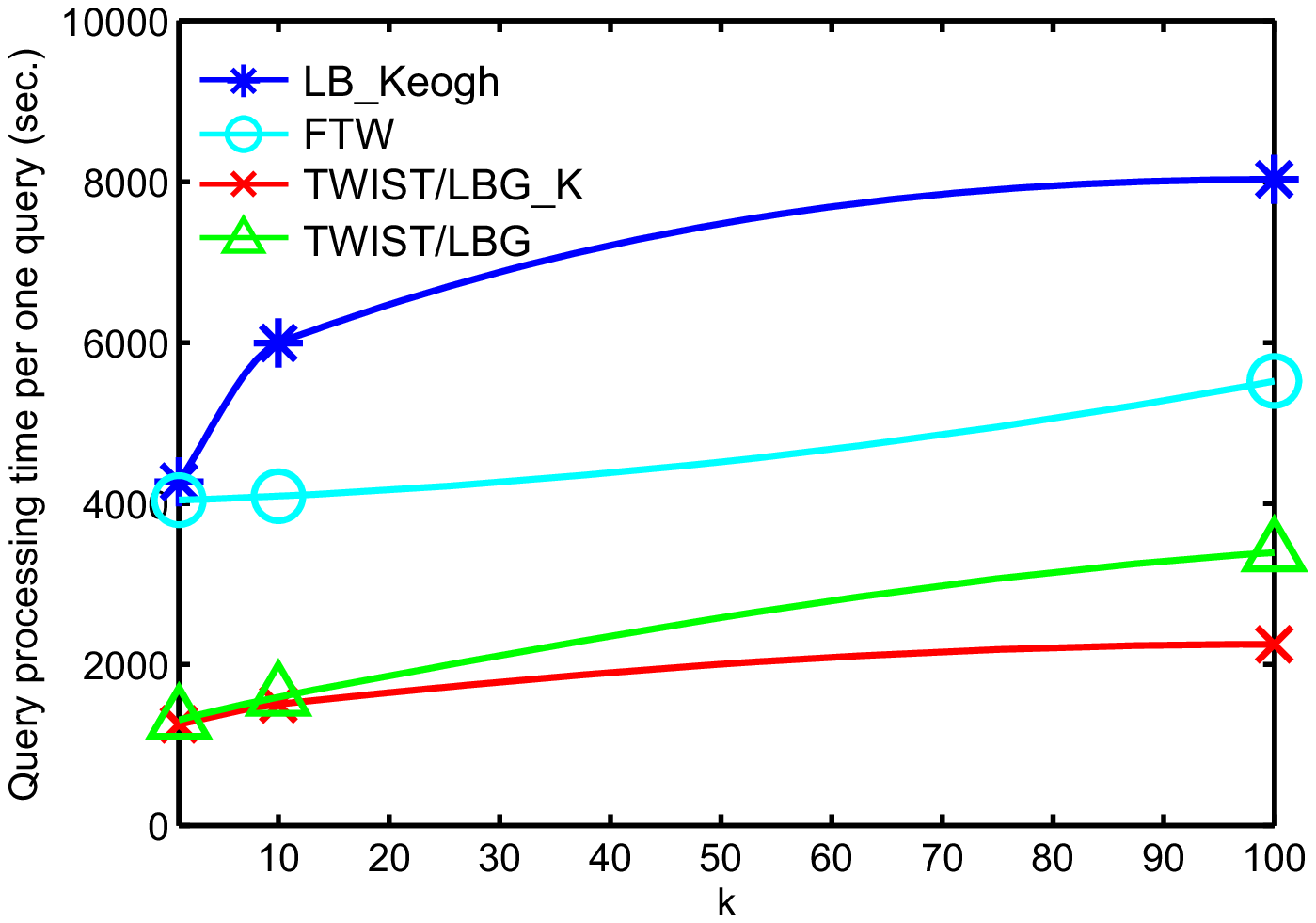}}\tabularnewline
\multicolumn{2}{c}{(c) Electrocardiogram}\tabularnewline
\end{tabular}
\par\end{centering}

\caption{TWIST is faster than FTW and LB\_Keogh for all values of $k$, where
database size, sequence length, global constraint size, and maximum
page size are set to 524288, 2048, 10\%, and 128, respectively}
\label{Flo:exp3}
\end{figure}

\begin{figure}
\noindent \begin{centering}
\begin{tabular}{cc}
\includegraphics[width=4.8cm]{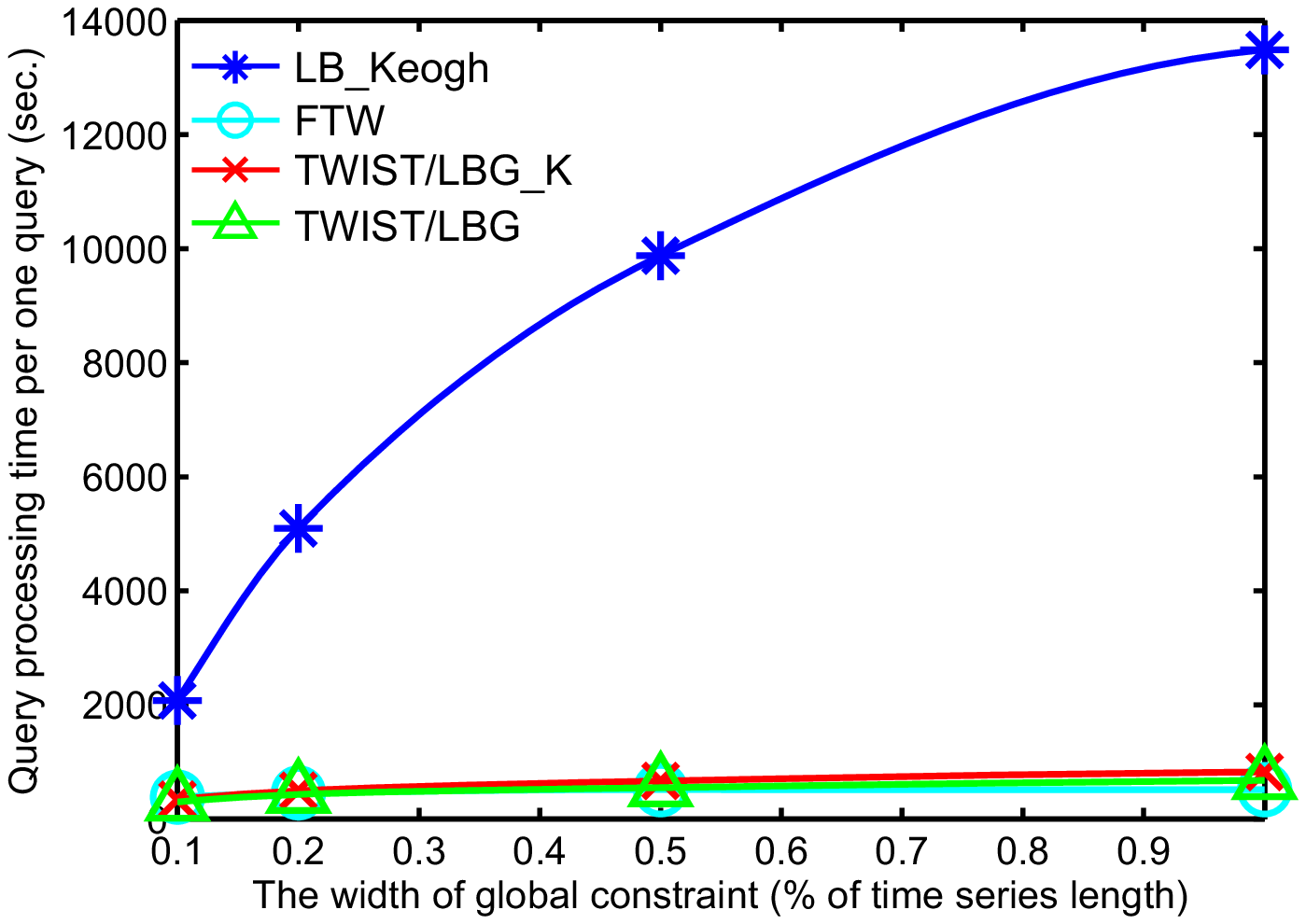} & \includegraphics[width=4.8cm]{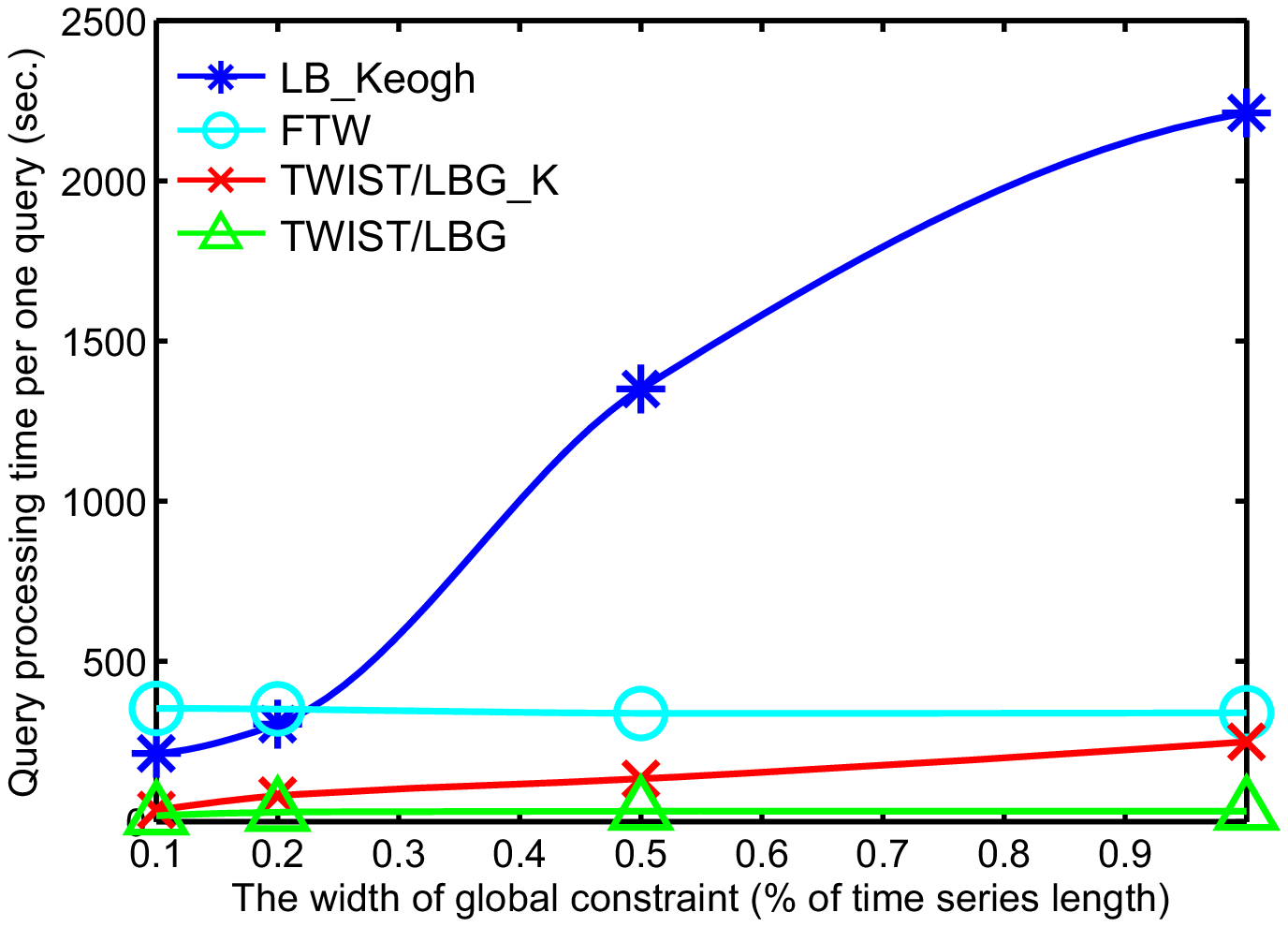}\tabularnewline
(a) Random Walk I & (b) Random Walk II\tabularnewline
\multicolumn{2}{c}{\includegraphics[width=4.8cm]{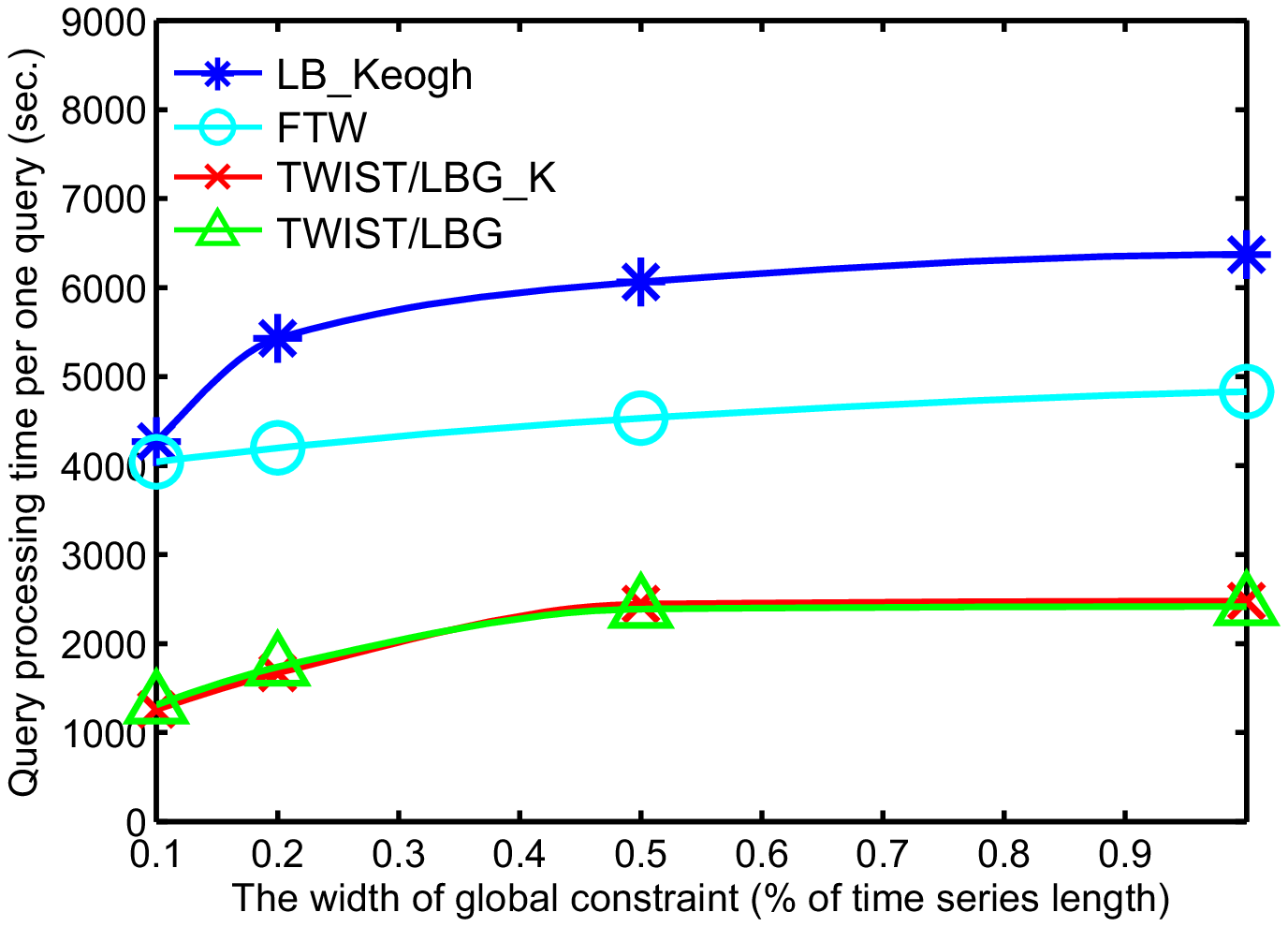}}\tabularnewline
\multicolumn{2}{c}{(c) Electrocardiogram}\tabularnewline
\end{tabular}
\par\end{centering}

\caption{TWIST and FTW are not affected by the increment of the global constraint\textquoteright{}s
width; however, TWIST outperforms both FTW and LB\_Keogh, where database
size, sequence length, an integer $k$, and page size are set to 524288,
2048, 1, and 128, respectively}
\label{Flo:exp3-2}
\end{figure}

\begin{figure}
\noindent \begin{centering}
\begin{tabular}{cc}
\includegraphics[width=4.8cm]{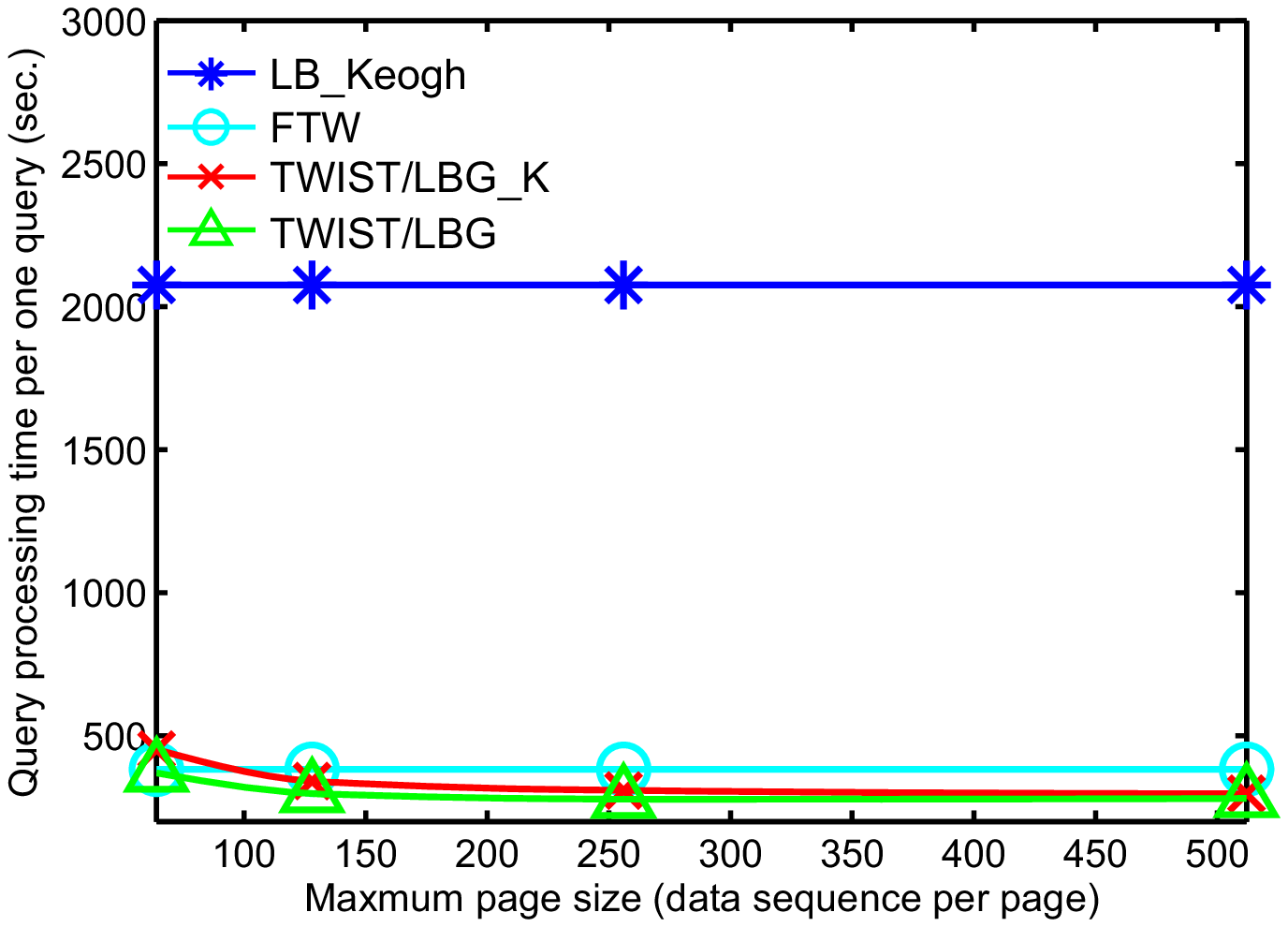} & \includegraphics[width=4.8cm]{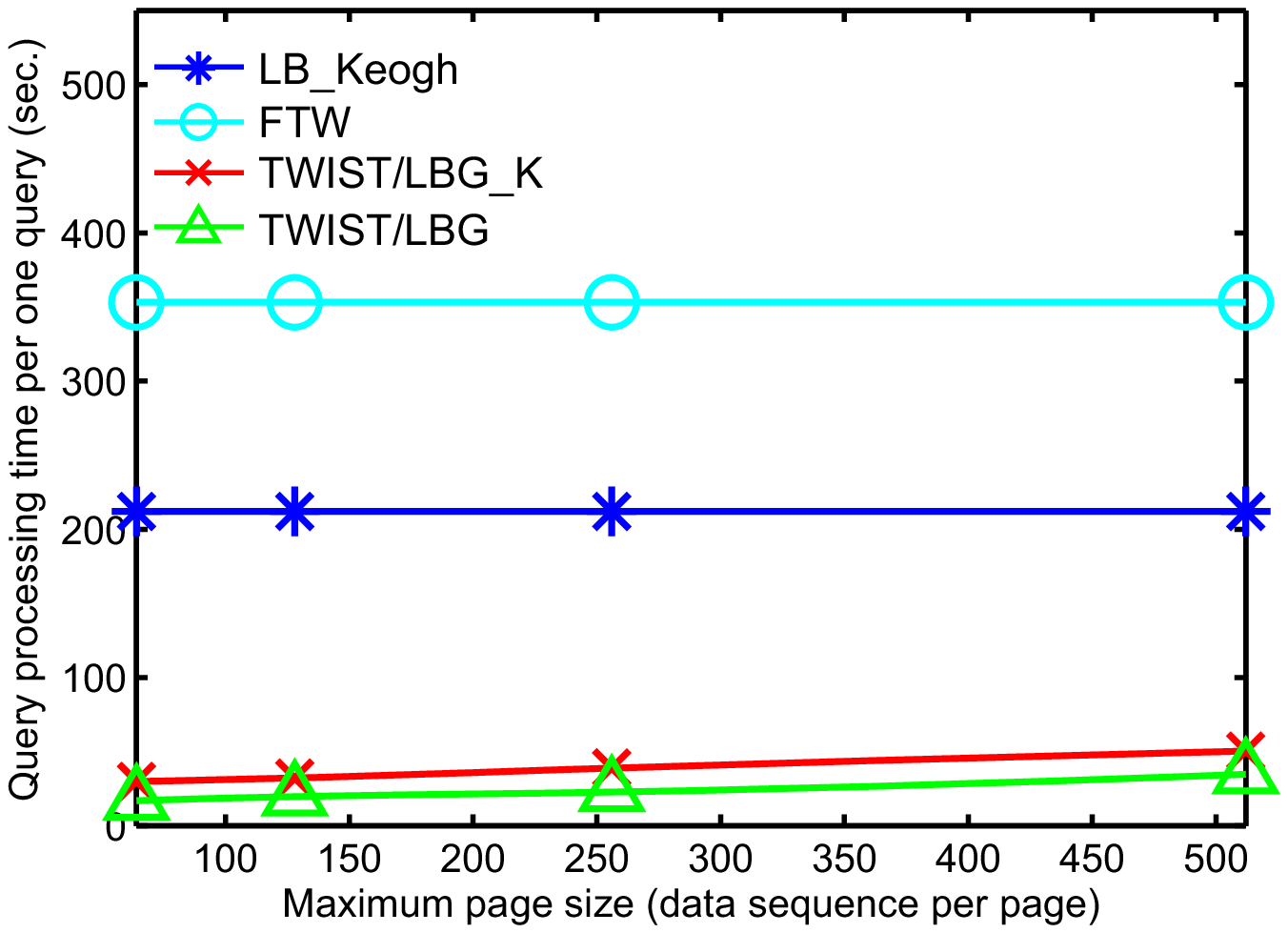}\tabularnewline
(a) Random Walk I & (b) Random Walk II\tabularnewline
\multicolumn{2}{c}{\includegraphics[width=4.8cm]{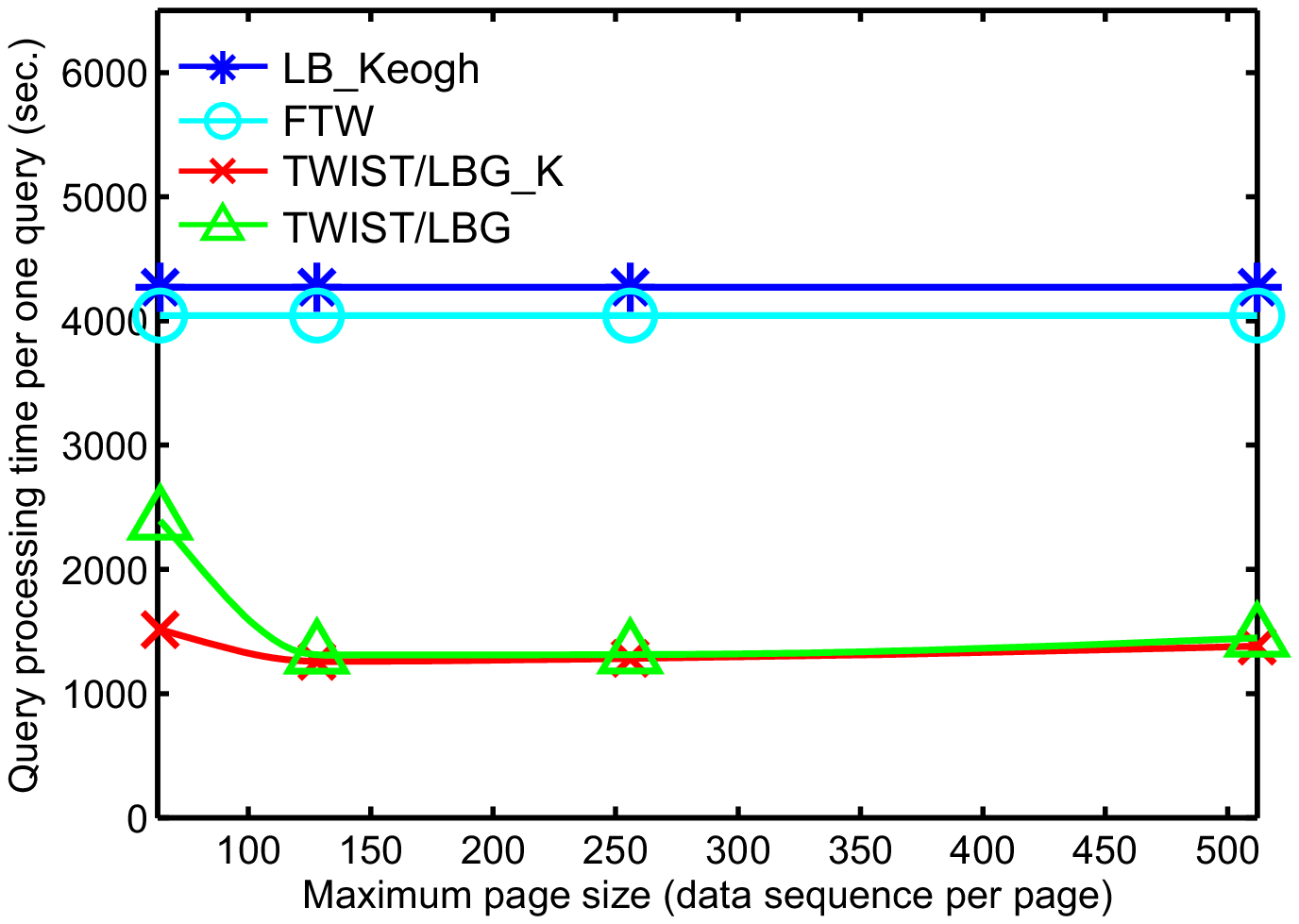}}\tabularnewline
\multicolumn{2}{c}{(c) Electrocardiogram}\tabularnewline
\end{tabular}
\par\end{centering}

\caption{When maximum page size changes, TWIST still outperforms the rival
methods, where database size, sequence length, global constraint size,
and an integer $k$ are set to 524288, 2048, 10\%, and 1, respectively}
\label{Flo:exp3-1}
\end{figure}

In this experiment, query processing times are averaged over 100 runs,
and are compared in the best-matched problem by varying four parameters,
i.e., the number of time series sequences, the dataset size, the width
of global constraint, an integer $k$, and the maximum page size (only
for TWIST). In order to observe the trend for each parameter, the
default values are fixed as follows, the dataset size as 524288 ($2^{19}$)
sequences, the length of time series sequence as 2048 data points,
the default width of global constraint as 10\% of sequence length,
an integer $k$ in top-$k$ querying as 1, and the maximum number
of sequences in DSF as 128 sequences. In addition, a dataset of 524288
sequences with length 2048, giving approximately 4 GB in size, and
10\% constraint width of global constraint is typically used in time
series data mining community \citep{RatanamahatanaK05}. Note that
for LBS, the default segment size proposed in the original paper is
used, i.e., 1024, 256, 64, and 16, and LBG uses the same segment size
to that of LBS. In sequential search in DSF, we implement LBS to reduce
the DTW distance calculation. However, the segmented sequence is generated
online; in other words, no index structure is stored on DSF.

Figures \ref{Flo:exp1}, \ref{Flo:exp2}, \ref{Flo:exp3}, \ref{Flo:exp3-2},
and \ref{Flo:exp3-1} illustrate the performance of TWIST, comparing
in terms of querying time against two rival methods by varying the
dataset size, sequence length, the width of global constraint, an
integer $k$, and maximum number of sequences in DSF. As expected,
TWIST greatly outperforms sequential search with LB\_Keogh and FTW
indexing.

\subsection{Indexing Time}

\begin{figure}
\noindent \begin{centering}
\begin{tabular}{cc}
\includegraphics[width=5cm]{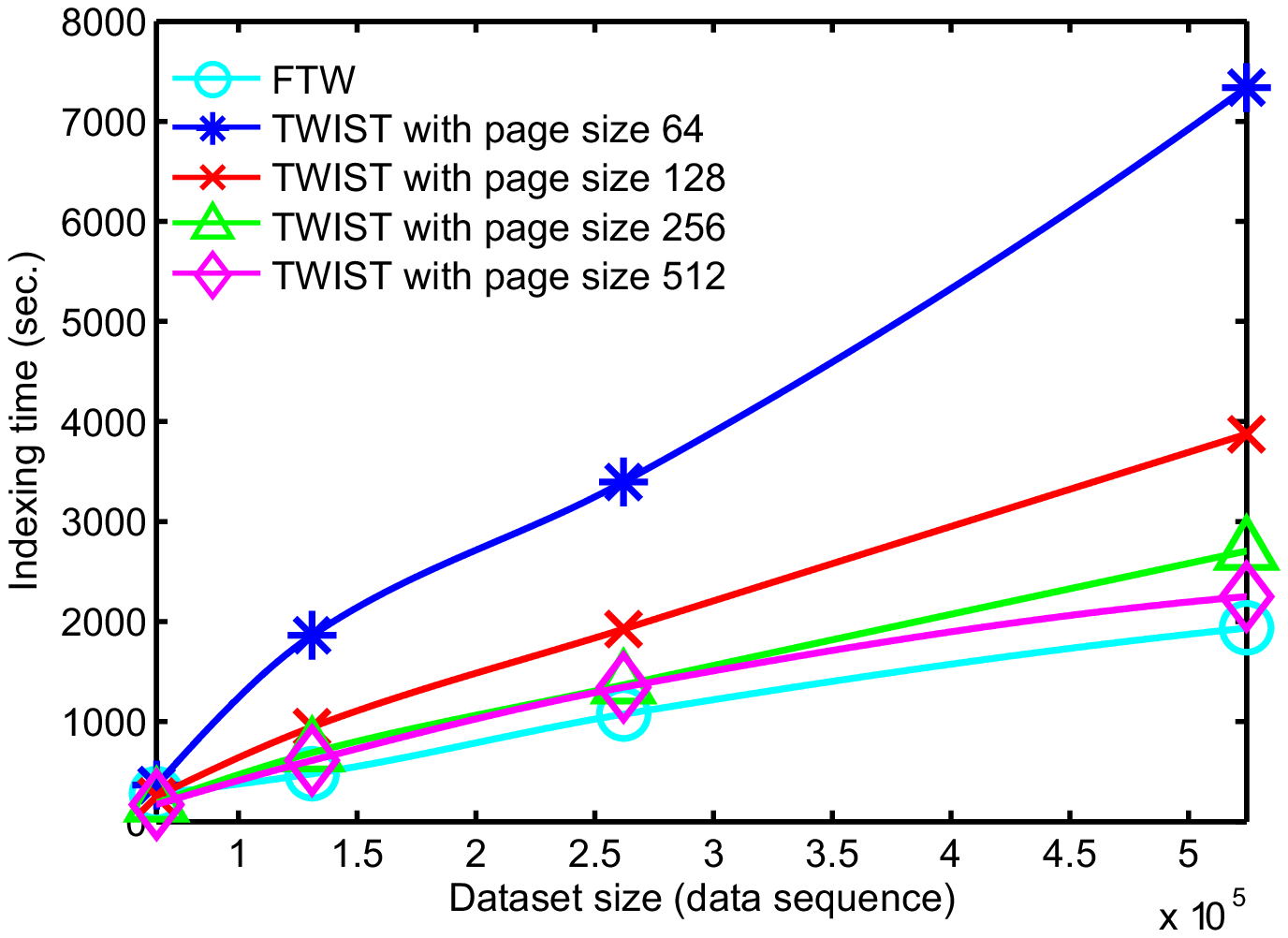} & \includegraphics[width=5cm]{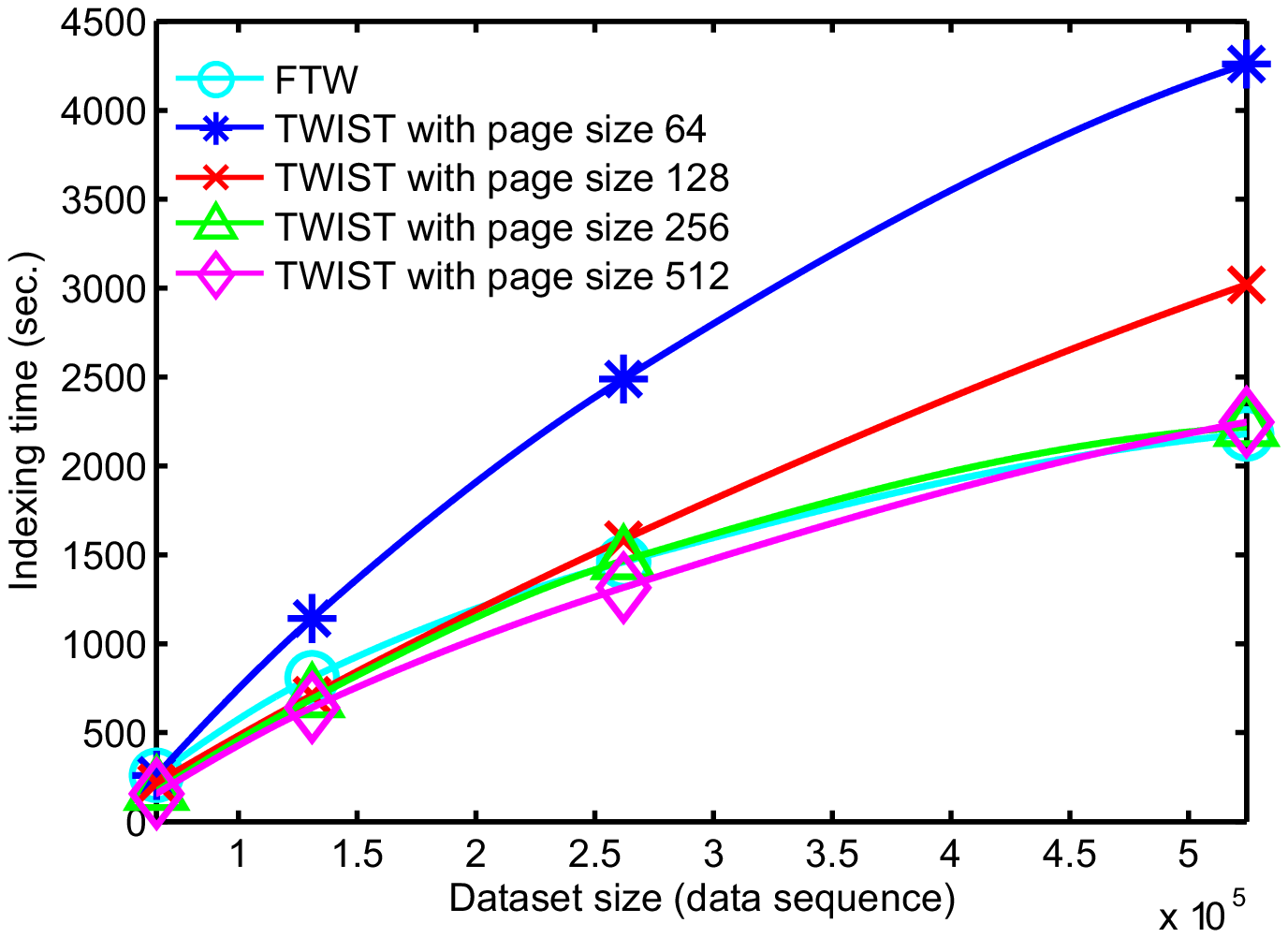}\tabularnewline
(a) Random Walk I & (b) Random Walk II\tabularnewline
\multicolumn{2}{c}{\includegraphics[width=5cm]{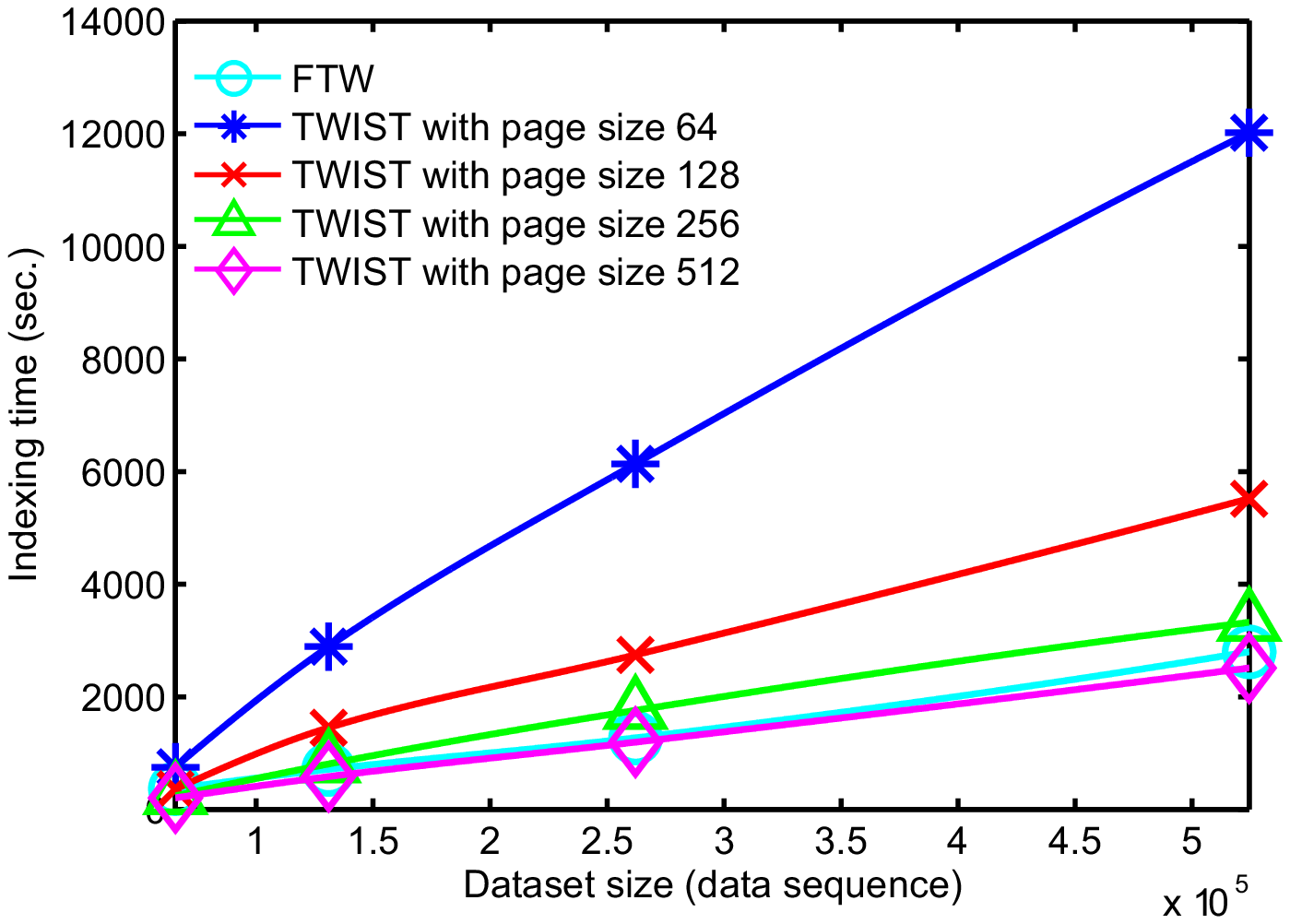}}\tabularnewline
\multicolumn{2}{c}{(c) Electrocardiogram}\tabularnewline
\end{tabular}
\par\end{centering}

\caption{As page size increases, the indexing time of TWIST significantly reduces
and is comparable to FTW\textquoteright{}s. Note that TWIST still
queries faster than FTW for all page sizes (see Figure \ref{Flo:exp3-1}).}
\label{Flo:exp2-1}
\end{figure}

Indexing time is a wall clock time that an algorithm consumes to build
the index structure. In this experiment, we only compare the indexing
time with FTW indexing since the sequential search with LB\_Keogh
does not need an index structure. From an experiment shown in Figure
\ref{Flo:exp2-1}, our indexing time is comparable to FTW's; however,
if the maximum page size is larger, TWIST can greatly reduce indexing
time, but it may trade off with querying time (see Figure \ref{Flo:exp3-1}).
The parameters used in this experiment are set to be the same as the
default parameters from the example in the previous experiments. Although
the indexing time is comparable to the FTW indexing, TWIST requires
very small storage space comparing with FTW indexing (as will be shown
in Section \ref{sub:Storage-Requirement}).

\begin{figure}
\noindent \begin{centering}
\begin{tabular}{cc}
\includegraphics[width=4.8cm]{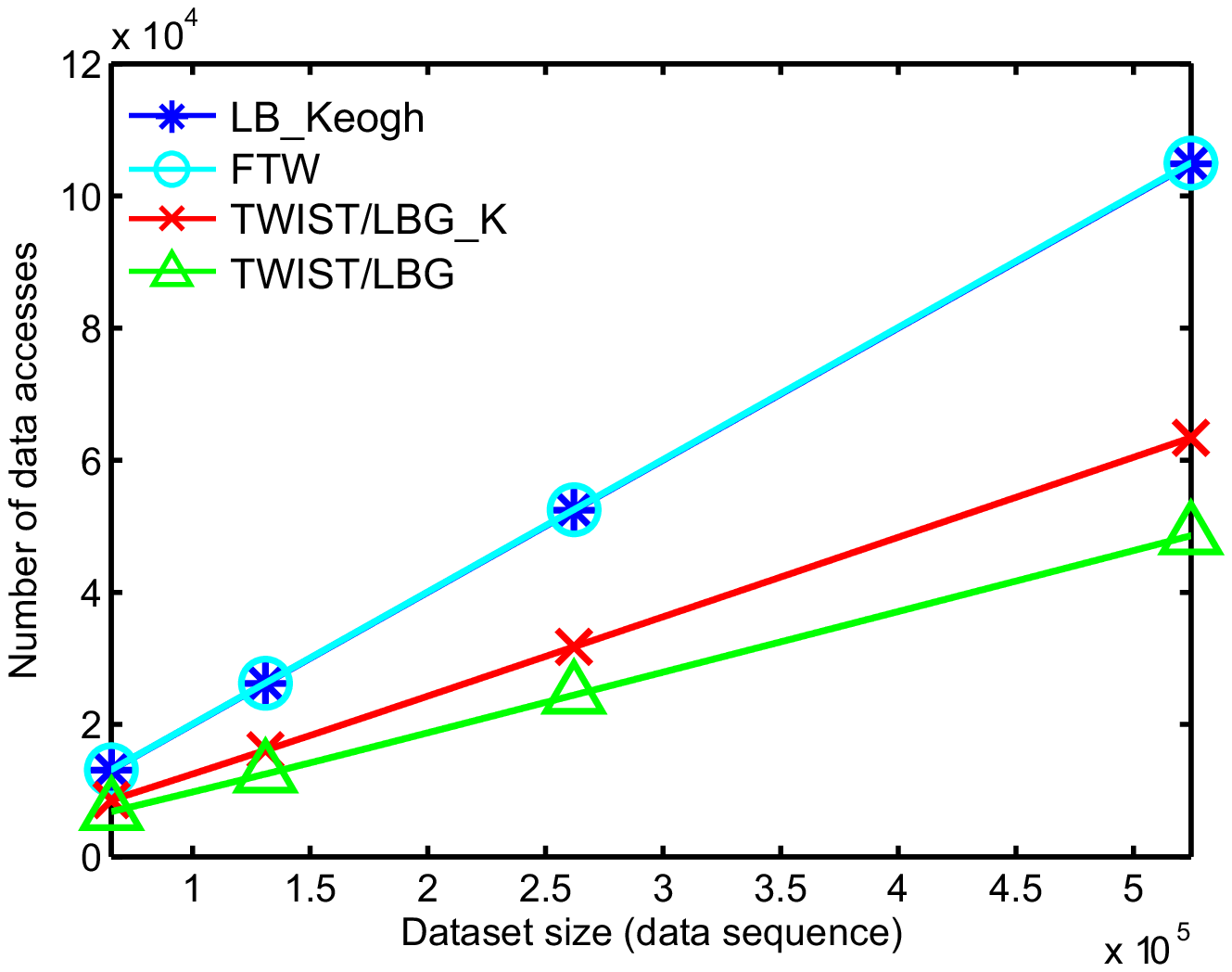} & \includegraphics[width=4.8cm]{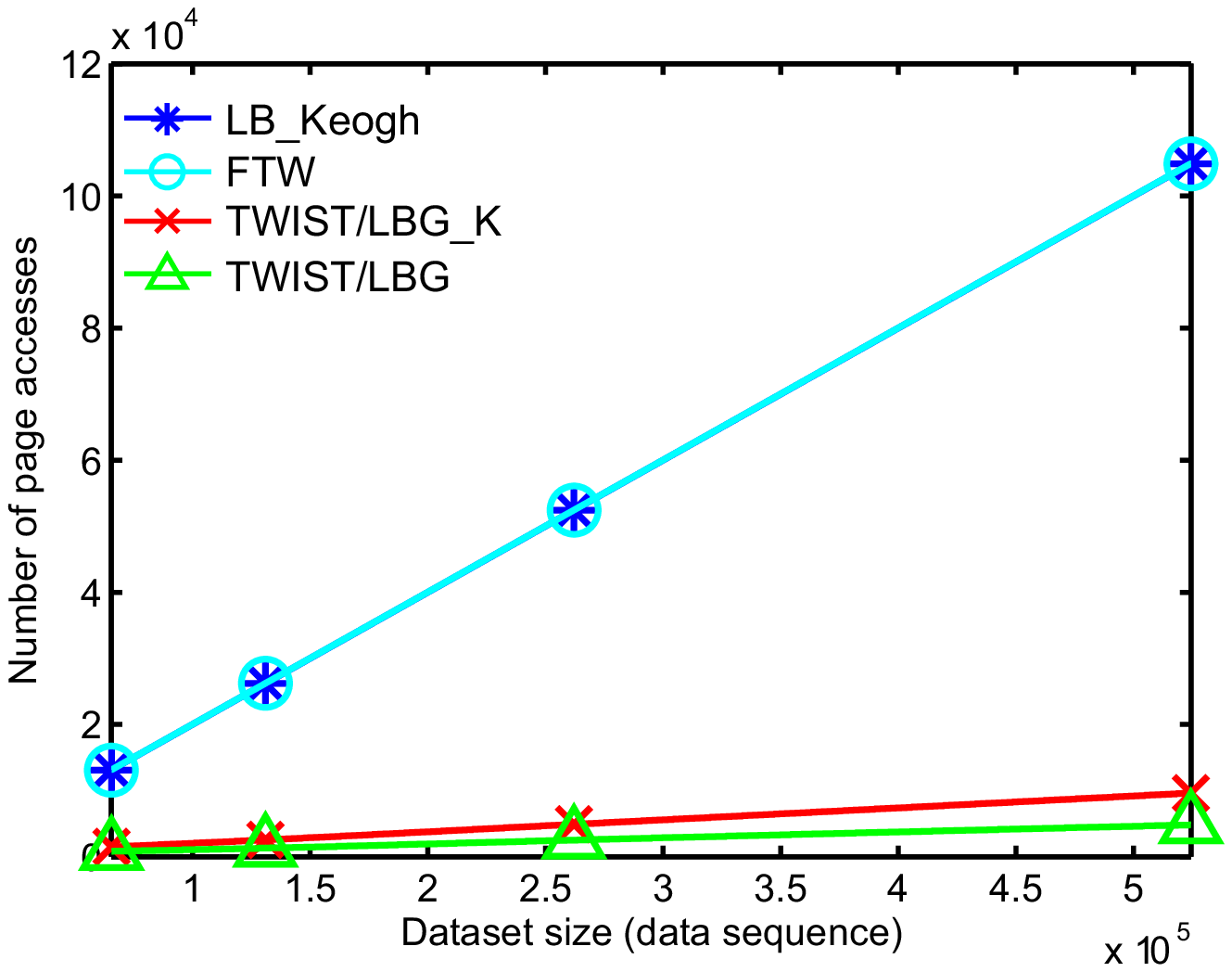}\tabularnewline
(a) Random Walk I & (b) Random Walk II\tabularnewline
\multicolumn{2}{c}{\includegraphics[width=4.8cm]{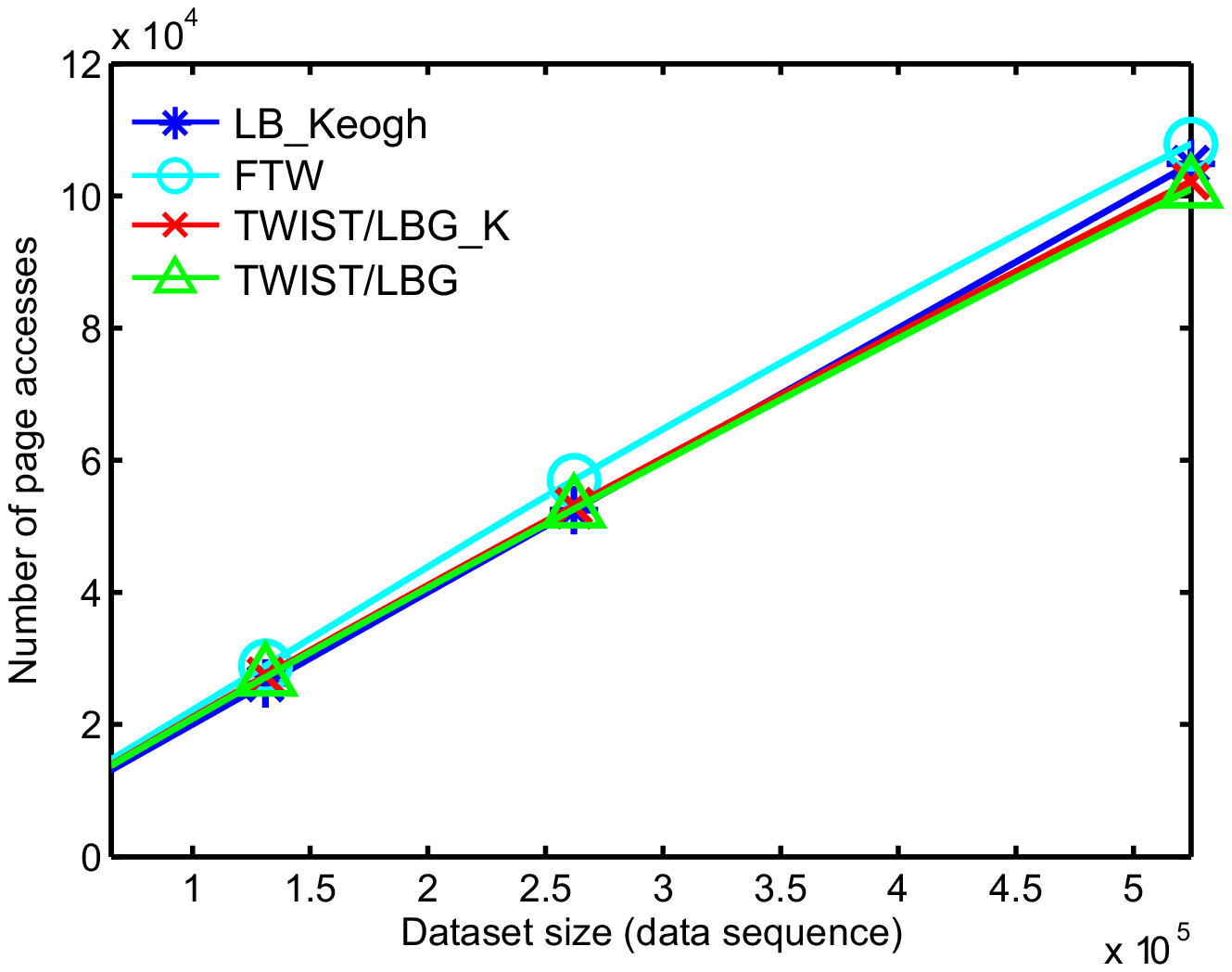}}\tabularnewline
\multicolumn{2}{c}{(c) Electrocardiogram}\tabularnewline
\end{tabular}
\par\end{centering}

\caption{Number of page accesses of TWIST is smaller than other rival methods,
especially in Random Walk I and Random Walk II, when speedup factor
is 5.}
\label{Flo:access}
\end{figure}

\begin{figure}
\noindent \begin{centering}
\begin{tabular}{cc}
\includegraphics[width=4.8cm]{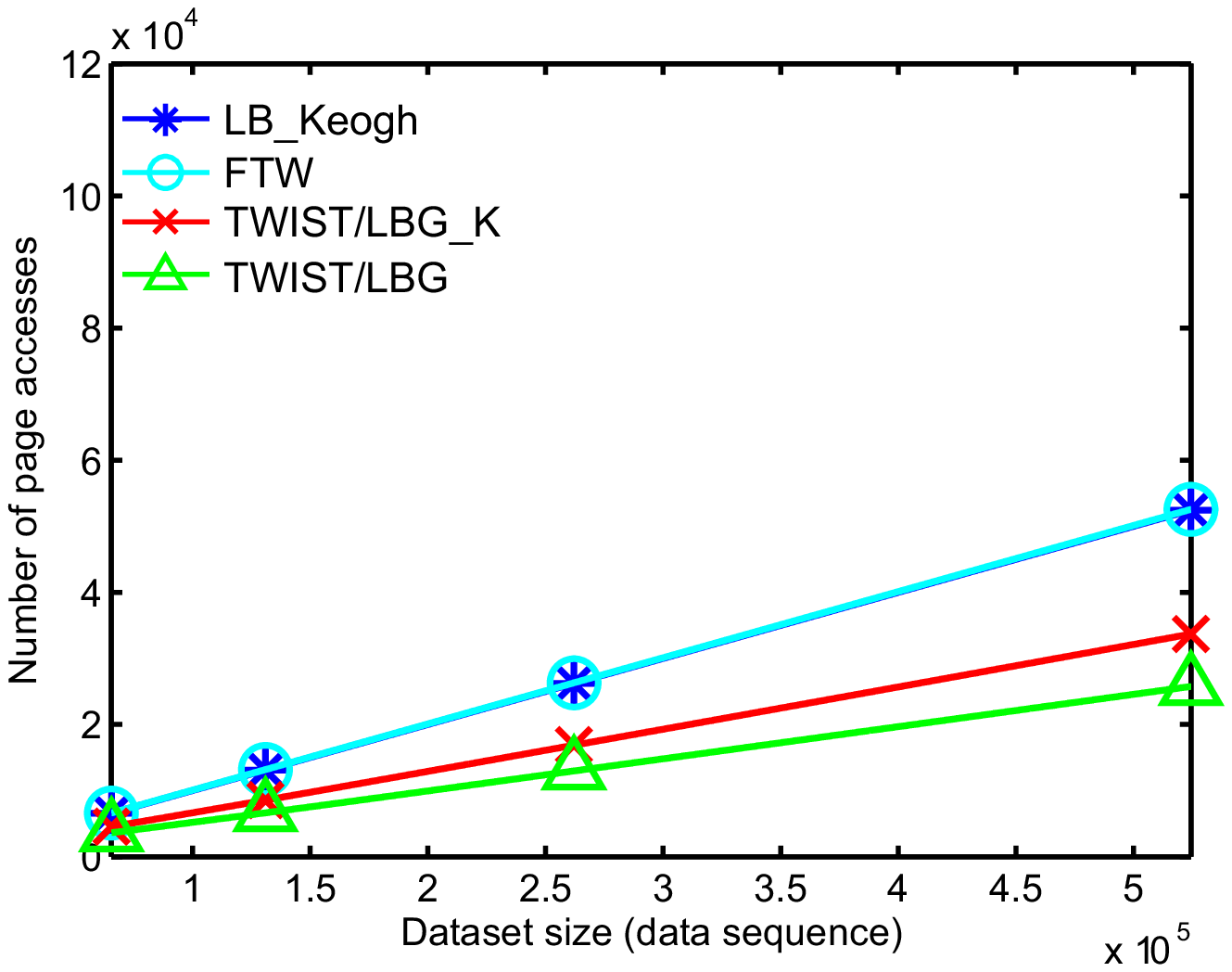} & \includegraphics[width=4.8cm]{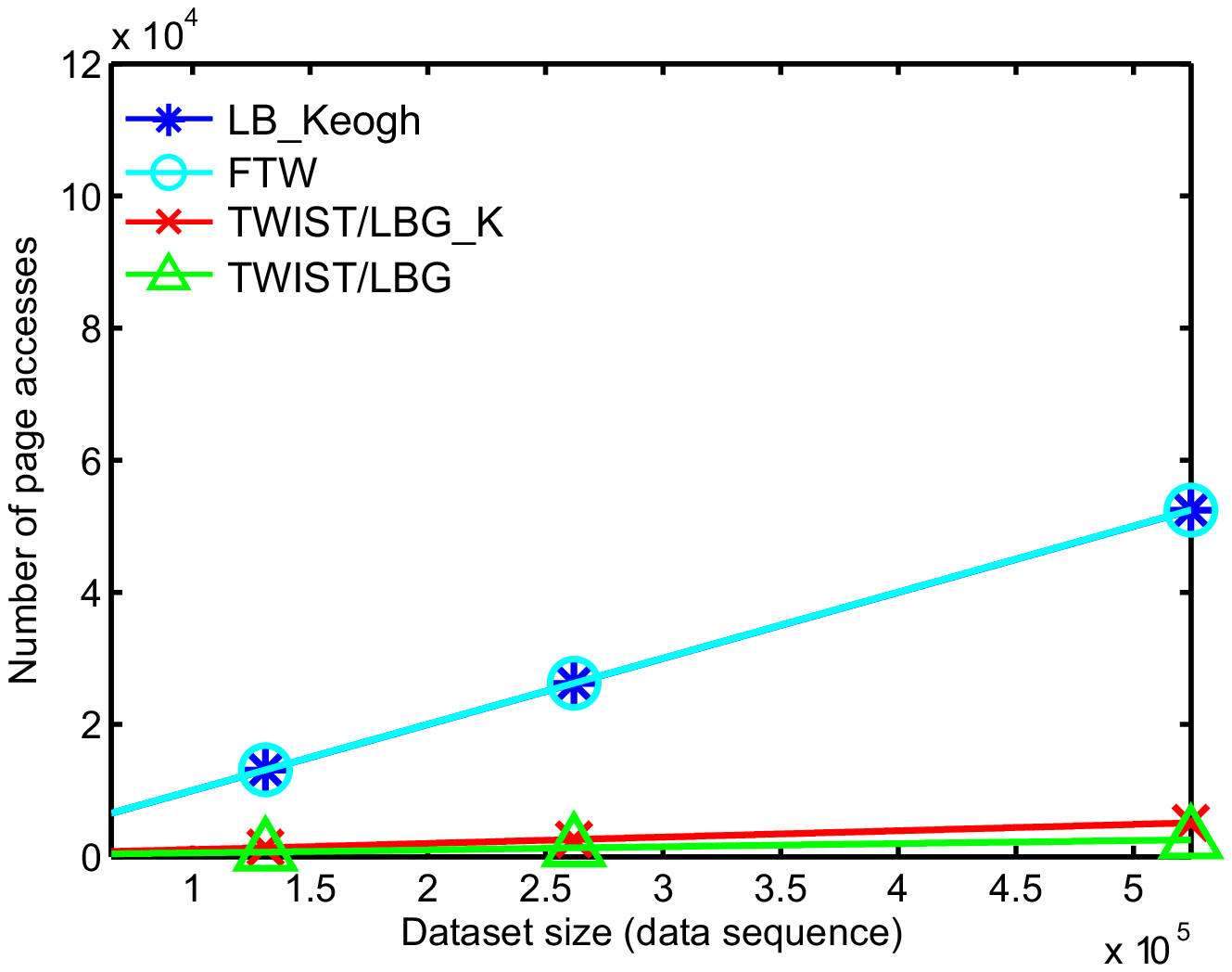}\tabularnewline
(a) Random Walk I & (b) Random Walk II\tabularnewline
\multicolumn{2}{c}{\includegraphics[width=4.8cm]{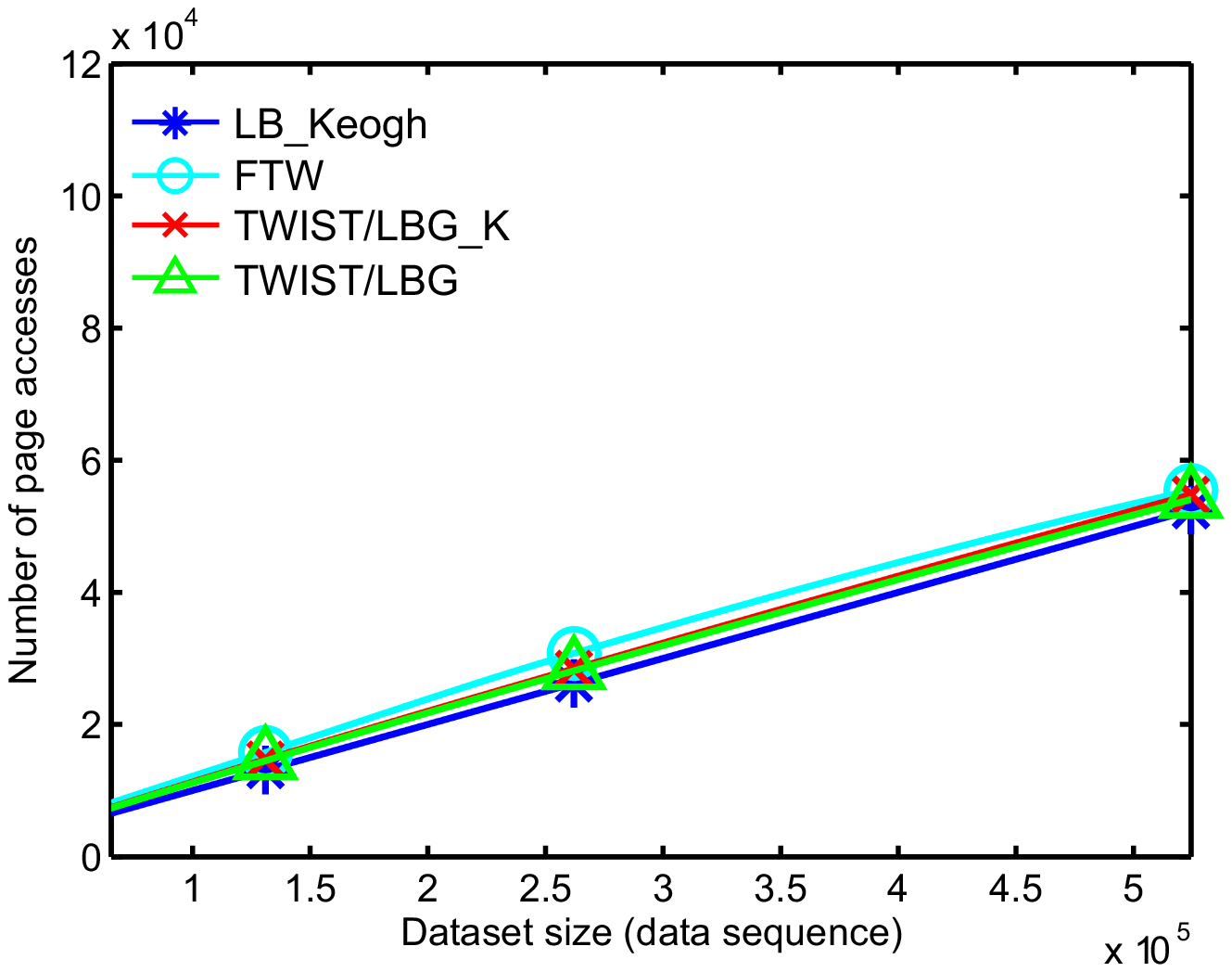}}\tabularnewline
\multicolumn{2}{c}{(c) Electrocardiogram}\tabularnewline
\end{tabular}
\par\end{centering}

\caption{Number of page accesses of TWIST is smaller than other rival methods,
especially in Random Walk I and Random Walk II, when speedup factor
is 10.}
\label{Flo:access-1}
\end{figure}

\subsection{Number of Page Accesses}

The number of page accesses ($\eta$) is generally evaluated in order
to estimate the I/O cost. We calculate the number of page accesses
for TWIST with LBG and TWIST with LBG$_{\text{K}}$ according to the
following equations.

\begin{equation}
\eta_{LBG}=\frac{2\alpha+\beta}{SF}+\delta\label{eq:page2}\end{equation}
\begin{equation}
\eta_{LBG_{K}}=\frac{\alpha+\beta}{SF}+\delta\label{eq:page1}\end{equation}

\noindent where $\alpha$ is a number of envelopes in ESF, $\beta$
is a number of accessed candidate sequences, $\delta$ is a number
of random accesses to DSFs, $SF$ is Speedup Factor proposed by Weber
et al. \citep{WeberSB98} stating that the sequential access is much
faster than random access up to 5 to 10 times. Generally, two values
of SFs are considered, i.e., 5 and 10, which represent traditional
and practical speedup factor of sequential access over random access.

Since sequential scan accesses the entire database, it can therefore
be considered as an upper bound. Surprisingly, as shown in Figures
\ref{Flo:access} and \ref{Flo:access-1}, the number of page accesses
of FTW indexing is approximately equal to that of the sequential scan,
and is very large when comparing with our proposed method TWIST because
FTW retrieves the entire index structure which has database size nearly
doubled. On the other hand, in average cases, TWIST can reduce a great
number of data accesses since it tries to minimize the number of DSF
accesses and the number of accessed candidate. For experimental parameters,
dataset size, sequence length, maximum page size, global constraint,
and $k$, are set to 524288, 2048, 128, 10\%, and 1, respectively.

\subsection{Storage Requirement\label{sub:Storage-Requirement}}

\begin{figure}[h]
\noindent \begin{centering}
\begin{tabular}{cc}
\includegraphics[width=4.8cm]{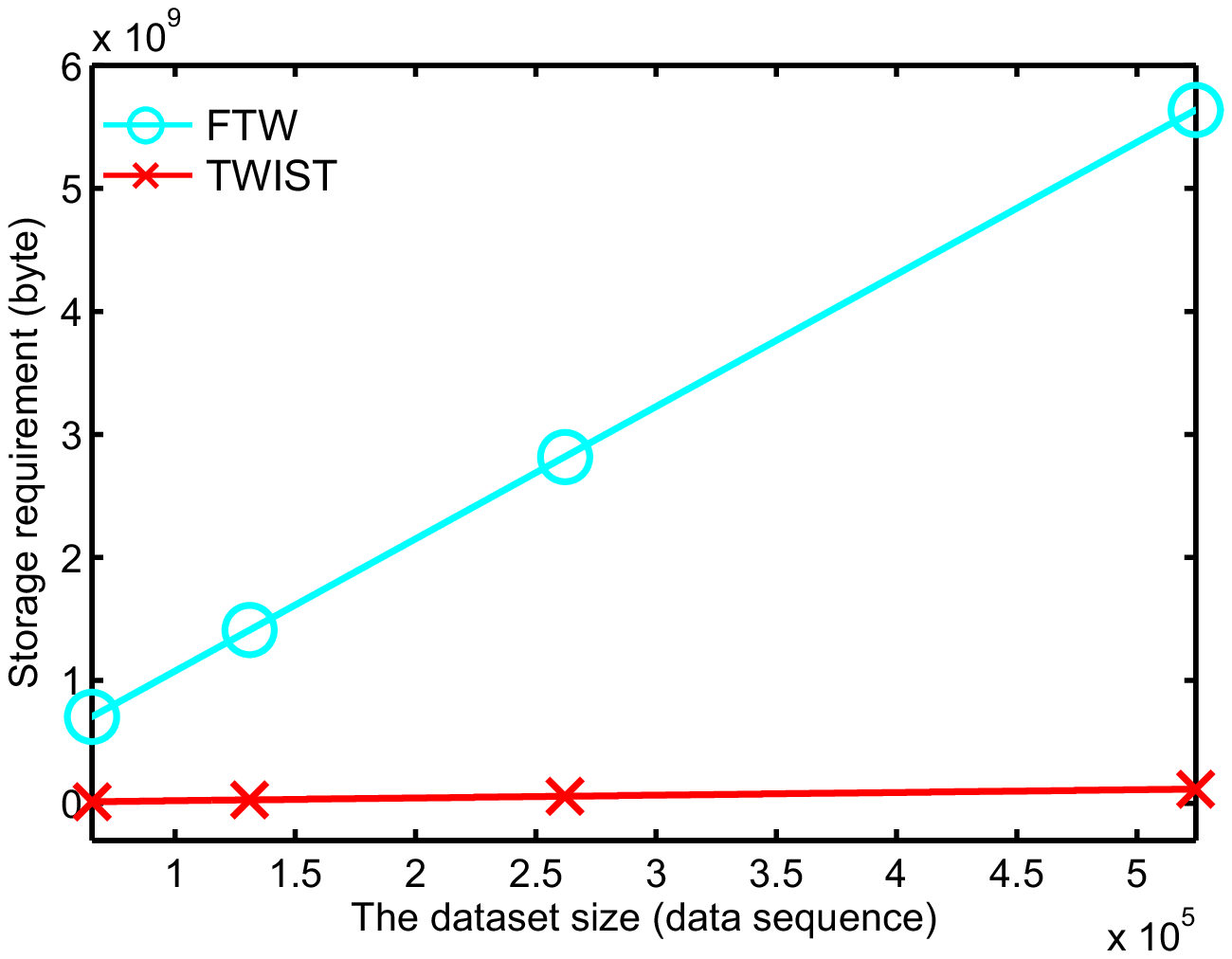} & \includegraphics[width=4.8cm]{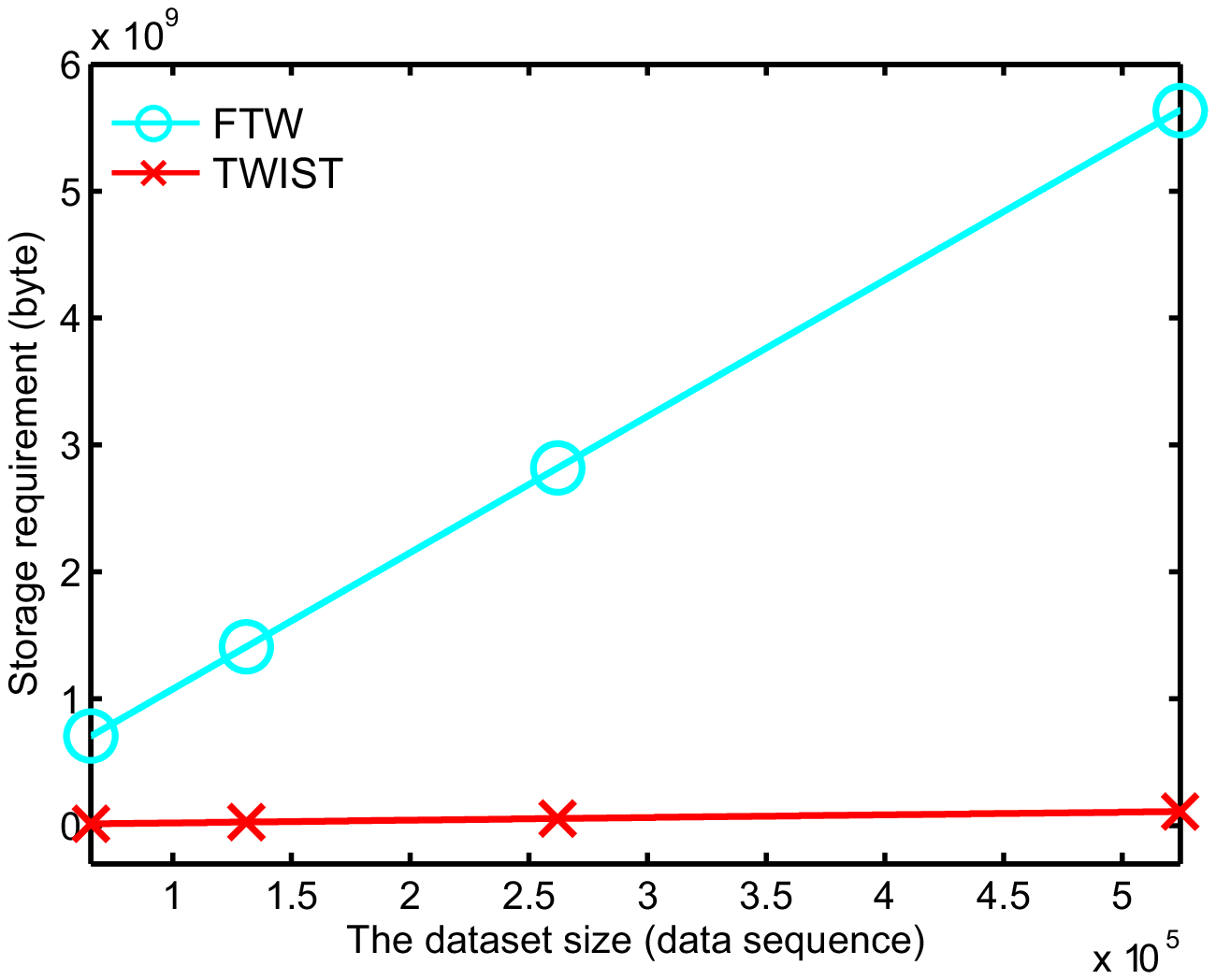}\tabularnewline
(a) Random Walk I & (b) Random Walk II\tabularnewline
\multicolumn{2}{c}{\includegraphics[width=4.8cm]{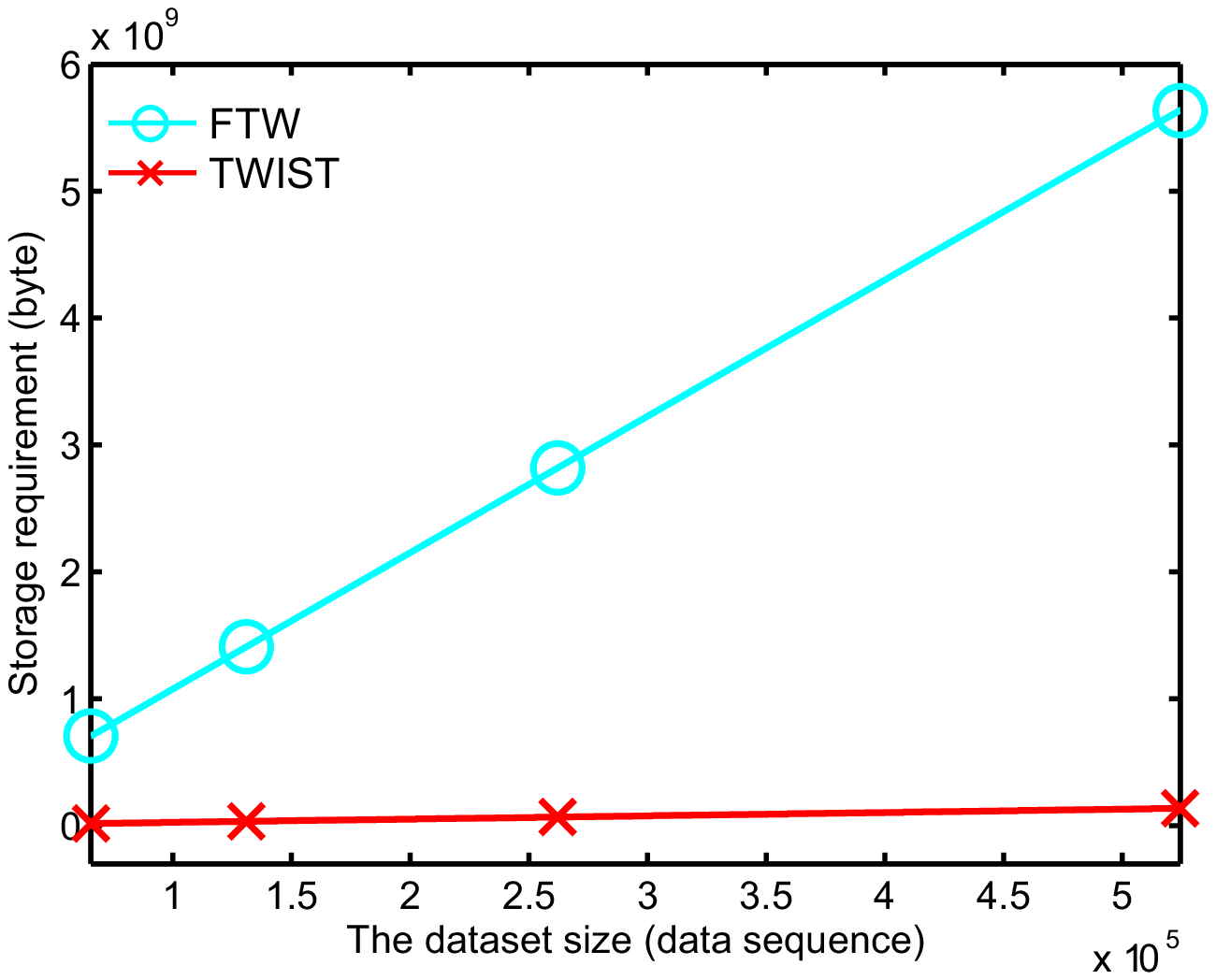}}\tabularnewline
\multicolumn{2}{c}{(c) Electrocardiogram}\tabularnewline
\end{tabular}
\par\end{centering}

\caption{Illustration of storage requirement comparison showing that TWIST\textquoteright{}s
index file requires only small amount of storage when comparing with
FTW\textquoteright{}s, where dataset size, sequence length, and maximum
page size are set to 524288, 2048, and 64, respectively.}
\label{Flo:exp2-1-1}
\end{figure}

In this section, we demonstrate the storage requirement for storing
an index file comparing with the rival method, FTW. Since FTW creates
a set of segmented sequences for each candidate sequence, the index
file\textquoteright{}s size is larger than the data file\textquoteright{}s.
Therefore, FTW index structure is not practical in real world application.
Unlike FTW, TWIST\textquoteright{}s index file requires only small
amount of storage, i.e., only the envelopes from all groups of data
sequences are stored. Figure \ref{Flo:exp2-1-1} shows the comparison
of storage requirement between TWIST and FTW. When the dataset size
is 2$^{\text{19}}$ sequences or 4 GB, FTW requires nearly 5 GB, but
as expected TWIST requires only 110 MB; in other words, TWIST requires
approximately 51 times less storage space than FTW, while still outperforming
in terms of querying processing time.

\subsection{Discussion}

As expected, query processing time increases when the dataset size
and the sequence length are larger for all approaches. However, from
Figures \ref{Flo:exp1} and \ref{Flo:exp2}, we can see that FTW indexing
and naïve method requires much longer time for a single query than
TWIST with LBG and LBG$_{\text{K}}$, and when database size increases,
the query processing time is also much larger. In Figure \ref{Flo:exp3},
if the global constraint changes, only naïve method with LB\_Keogh
and TWIST with LBG$_{\text{K}}$ are affected since the LB\_Keogh
and LBG$_{\text{K}}$ lose their tightness when the width of the global
constraint increases. Although the best-matched querying ($k=1$)
is typically used in several domains, we also evaluate TWIST when
varying $k$ as shown in Figure \ref{Flo:exp3-2}. Obviously, when
$k$ increases, the query processing time also increases since for
a large value of $k$ the best-so-far distance is also large. If the
best-so-far is large, the search cannot use the lower bounding distance
to prune off the database. However, from Figure \ref{Flo:exp3-1},
TWIST still efficiently retrieves an answer comparing with other methods.
The maximum page size is also another important parameter that must
be considered because TWIST uses it to balance the number of pages
in the database and the number of sequences in each page. In other
words, if the maximum page size is small, the number of random access
increases; otherwise, the number of sequential access will increase.
However, from the experiment, when the maximum page number changes,
TWIST still outperforms FTW and sequential search with LB\_Keogh.
Note that when we set the maximum page size to one, TWIST is identical
to FTW, but when the maximum page size is set to infinite, TWIST is
similar to the naïve method, i.e., sequential scan. Therefore, both
FTW indexing and the naïve method are special cases of TWIST. 

To evaluate the indexing time, we compare TWIST with FTW indexing
by varying the database size and the maximum page size in Figure \ref{Flo:exp2-1}.
From our insertion algorithm, if the number of sequences exceeds the
maximum page size, TWIST splits DSF into two DSFs. Therefore, if the
maximum page size is large, TWIST reduces a number of splitting function
calls; this therefore reduces a number of indexing time since splitting
algorithm requires $k$-means clustering algorithm which has linear
time complexity to a number of page size. Although the large maximum
page size reduces the indexing time, the performance when querying
is a tradeoff. 

Although we provide the evaluation in terms of query processing time
in Section \ref{sub:Querying-Time}, the number of page accesses needs
to be evaluated since the number of page accesses reflects the I/O
cost for each approach. The number of page accesses is formulized
and calculated according to \citep{sakurai2005ffs,WeberSB98} which
state that the sequential access is faster than the random access
five to ten times. From Figures \ref{Flo:access} and \ref{Flo:access-1},
the number of page accesses of FTW indexing must always larger than
the naïve approach since FTW indexing reads all segmented sequences
in the index file which are equal to the number of sequences in the
database. Obviously, TWIST consumes only small amount of page accesses
because TWIST is designed to reduce both sequential and random accesses. 

For the size of an index structure, TWIST utilizes only small amount
of spaces comparing with FTW indexing which always requires the space
twice the database size. In Figure \ref{Flo:exp2-1-1}, we demonstrate
TWIST's storage requirement by varying the database sizes and the
maximum page number since the size of ESF solely depends of the number
of DSF in the database.

\section{Conclusion}

In this work, we propose a novel index sequential structure called
TWIST (Time Warping in Index Sequential sTructure) which significantly
reduces querying time up to 50 times comparing with the best existing
methods, i.e., FTW indexing and sequential scan with LB\_Keogh. More
specifically, TWIST groups similar time series sequences together
in the same file, and then the representative of a group of sequences
is calculated and stored in the index structure. When a query sequence
is issued, a lower bounding distance for a group of sequences is determined
from the query sequence and a representative is retrieved from the
index file. Therefore, if the lower bounding distance for a group
of sequences is larger than the best-so-far distance, all candidate
sequences in the group does not need to be accessed. This can prune
off an impressively large amount of candidate sequences and makes
TWIST feasible for massive time series database.

\section*{Acknowledgement}

This research is partially supported by the Thailand Research Fund
(Grant No. MRG5080246), the Thailand Research Fund given through the
Royal Golden Jubilee Ph.D. Program (PHD/0141/2549 to V. Niennattrakul),
and the Chulalongkorn University Graduate Scholarship to Commemorate
the 72$^{nd}$ Anniversary of His Majesty King Bhumibol Adulyadej.

\end{document}